\documentclass[twocolumn]{aastex61}

\usepackage{amsmath}
\usepackage{tikz}
\usepackage{color}
\usepackage{soul}
\usepackage{array}
\usepackage{multirow}
\usepackage{graphicx}
\usepackage{longtable}
\usepackage{rotating}
\usepackage{natbib}
\usepackage{geometry}

\newcommand{\hilightwhite}[1]{\setlength{\fboxsep}{1pt}\colorbox{white}{#1}\setlength{\fboxsep}{0pt}}

\newcommand{\xmm}{{\it XMM-Newton}}
\newcommand{\cxo}{{\it Chandra}}

\newcommand{\ROSAT}{ROSAT}

\newcommand{\SII}{[S\,{\sc ii}]}
\newcommand{\OIII}{[O\,{\sc iii}]}

\newcommand{\sigmaD}{$\Sigma - D$}

\def\confcount{59}
\def\candcount{15}
\def\totalcount{74}
\def\spccount{41}
\def\spccountsn{40} 
\def\lumcount{41}
\def\typeiacount{16}
\def\typecccount{23}

\def\p0{\phantom{0}}
\def\ND{$N(<D)$}

\def\msun{$M_{\odot}$}
\def\HI{\hbox{H\,{\sc i}}}
\def\HII{\hbox{H\,{\sc ii}}}

\shorttitle{Statistics of SNRs in the LMC}
\shortauthors{Bozzetto et al.}

\begin{document}


\title{Statistical Analysis of Supernova Remnants \\
    in the Large Magellanic Cloud}


\correspondingauthor{Miroslav D. Filipovi\'c}
\email{m.filipovic@westernsydney.edu.au}

\author{Luke M. Bozzetto}
\affil{Western Sydney University, Locked Bag 1797, Penrith South DC, NSW 1797, Australia}

\author{Miroslav D. Filipovi\'c}
\affil{Western Sydney University, Locked Bag 1797, Penrith South DC, NSW 1797, Australia}

\author{Branislav Vukoti{\'c}}
\affil{Astronomical Observatory, Volgina 7, 11060 Belgrade 38, Serbia}

\author{Marko Z. Pavlovi{\'c}}
\affil{Department of Astronomy, Faculty of Mathematics, University of Belgrade, Studentski trg 16, 11000 Belgrade, Serbia}

\author{Dejan Uro{\v s}evi{\'c}}
\affil{Department of Astronomy, Faculty of Mathematics, University of Belgrade, Studentski trg 16, 11000 Belgrade, Serbia}
\affil{Isaac Newton Institute of Chile, Yugoslavia Branch}

\author{Patrick J. Kavanagh}
\affil{School of Cosmic Physics, Dublin Institute for Advanced Studies, 31 Fitzwilliam Place, Dublin 2, Ireland}

\author{Bojan Arbutina}
\affil{Department of Astronomy, Faculty of Mathematics, University of Belgrade, Studentski trg 16, 11000 Belgrade, Serbia}

\author{Pierre Maggi}
\affil{Laboratoire AIM, IRFU/Service d'Astrophysique - CEA/DRF - CNRS - Universit{\'e} Paris Diderot, Bat. 709, CEA-Saclay, 91191 Gif-sur-Yvette Cedex, France}

\author{Manami Sasaki}
\affil{Dr. Karl Remeis-Sternwarte, Erlangen Centre for Astroparticle Physics, Friedrich-Alexander-Universit\"at Erlangen-N\"urnberg, Sternwartstra{\ss}e 7, D-96049 Bamberg, Germany}

\author{Frank Haberl}
\affil{Max-Planck-Institut f\"{u}r extraterrestrische Physik, Giessenbachstra\ss e, D-85748 Garching, Germany}

\author{Evan J. Crawford}
\affil{Western Sydney University, Locked Bag 1797, Penrith South DC, NSW 1797, Australia}

\author{Quentin Roper}
\affil{Western Sydney University, Locked Bag 1797, Penrith South DC, NSW 1797, Australia}

\author{Kevin Grieve}
\affil{Western Sydney University, Locked Bag 1797, Penrith South DC, NSW 1797, Australia}

\author{S. D. Points}
\affil{Cerro Tololo Inter-American Observatory, Casilla 603, La Serena, Chile}





\begin{abstract}
We construct the most complete sample of supernova remnants (SNRs) in {any galaxy -- the Large Magellanic Cloud (LMC) SNR sample. We} study their various properties such as spectral index ($\alpha$), size and surface-brightness. We suggest an association between the spatial distribution, environment density of LMC SNRs and their tendency to be located around supergiant shells. We find evidence that the \typeiacount\ known type~Ia LMC SNRs are expanding in a lower density environment compared to the Core-Collapse (CC) type. The mean diameter of our entire population (\totalcount) is {41}~pc, which is comparable to nearby galaxies. We didn't find any correlation between the type of {SN} explosion, ovality or age. The \ND\ relationship of $a={0.96}$ implies that the randomised diameters are readily mimicking such an exponent. The rate of SNe occurring in the LMC is estimated to be $\sim$1 per 200~yr. The mean $\alpha$ of the entire LMC SNR population is $\alpha$=--{0.52}, which is typical of most SNRs. However, our estimates show a clear flattening of the synchrotron $\alpha$ as the remnants age. As predicted, our CC SNRs sample are significantly brighter radio emitters than the type~Ia remnants. We also estimate the \sigmaD\ relation for the LMC to have a slope {$\sim$3.8} which is comparable with other nearby galaxies. We also find the residency time of electrons in the galaxy (\hilightwhite{$4.0-14.3$~Myr}), implying that SNRs should be the dominant mechanism for the production and acceleration of {CRs}.
\end{abstract}


\section{Introduction}

Observational facts that we have gathered over the past decades together with statistical analysis of the {Large Magellanic Cloud (LMC)} {Supernova Remnant (SNR)} population are essential for our understanding of the processes in these violent celestial objects. The LMC is a galaxy that provides one of the rare opportunities to gather information on a population of {SNRs} that is close enough to be resolved spatially. Moderate-to-high resolution images ($<$1\arcmin) are available at all wavelengths, allowing for more stringent classification of this class of objects, yielding a more credible population. Apart from the {Small Magellanic Cloud (SMC)}, the only other galaxy for which such resolution is attainable is our own, the {Milky Way (MW)}, where the highest resolution observations are possible. However, the {MW} sample is not without its own drawbacks. It is affected by the Malmquist bias\footnote{The distance dependent, volume selection effect: brighter objects are favoured in flux density limited surveys.}, making the sample somewhat incomplete, as well as suffering from distance uncertainties and the absence of uniform coverage. Such challenges are not as pronounced when observing {SNRs} in the {LMC} and we can assume that the intrinsic LMC objects are located at approximately the same distance. We acknowledge that the inclination of the LMC towards the line of sight \citep{2010A&A...520A..24S} may introduce up to $\sim$10\% error in distance when the SNR is positioned closer to the southern end, compared with objects at the northern end of the {LMC}.

The first {SNR} candidates in the {LMC} were presented by \citet{1964IAUS...20..283M}, and later confirmed by \citet{1966MNRAS.131..371W} through follow-up radio and optical observations. These three remnants (N\,49, N\,63A and N\,132D) were, in fact, the first extragalactic {SNRs} ever discovered. Since then, there have been a considerable number of additions to that population, with notable samples from \citet{1973ApJ...180..725M,1980MNRAS.191..469M,1981ApJ...248..925L,1983ApJS...51..345M,1984ApJS...55..189M,1984AuJPh..37..321M} and \citet{1985ApJS...58..197M}. Numerous surveys have also been undertaken in various {electromagnetic (EM)} spectral bands. For example, \citet{1988AJ.....96.1874C} looked into the population and environments of {SNRs} in the {LMC}. In the radio-continuum, \citet{1998A&AS..130..421F} carried out a study of the {Magellanic Clouds (MCs)}. \citet[hereafter HP99]{1999A&AS..139..277H} compiled an X-ray catalogue of {LMC} sources, adding numerous {SNR} candidates to the population. \citet{1999ApJS..123..467W} produced an X-ray atlas of {LMC} {SNRs}, while \citet{2000A&AS..143..391S} compiled a \ROSAT\ HRI catalogue of X-ray sources in the {LMC} region. \citet{2006ApJS..165..480B} and \citet{2015ApJ...799...50L} surveyed {SNRs} in the {MCs} at far {UV} wavelengths. \citet{2013ApJ...779..134S} presented a survey of infrared {SNRs} in the {LMC}. Most recently, \cite{maggi16} compiled the LMC SNR population seen with \xmm.

A well-established (predominantly) non-thermal continuum emission is one of the distinguishing characteristics of SNRs at radio frequencies. The majority of SNRs have a radio spectral index of $\alpha\sim-0.5$ (defined as $S\propto\nu^\alpha$), although there is a large scatter because of the wide variety of SNRs and different environments and stages of evolution \citep{1998A&AS..130..421F,2016arXiv160401458F}. On one side, younger and very old remnants can have a steeper spectral index of $\alpha\sim-0.8$, while mid-to-older aged remnants tend to have radio spectra with $\alpha \sim -0.5$. SNRs that harbour a Pulsar Wind Nebula (PWN) exhibit flatter radio spectra with $\alpha \sim -0.1$. As one of the most energetic class of sources in the Universe, SN/SNRs greatly impact the structure, physical properties and evolution of the interstellar medium (ISM) of the host galaxy. Conversely, the interstellar environments in which SNRs reside will heavily affect the remnants' evolution.

A complete sample of SNRs in any galaxy provides the opportunity to study the global properties of SNRs, in addition to carrying out detailed analysis on the subclasses (e.g., sorted by X-ray and radio morphology or by progenitor SN type). Towards this goal, we have been identifying new LMC SNRs using combined optical, radio, IR and X-ray observations. Apart from the above mentioned LMC SNR survey papers, there are a number of studies focusing on particular LMC SNRs. Some recent studies include: \citet{2007MNRAS.378.1237B}, \citet{2009SerAJ.179...55C}, \citet{2008SerAJ.177...61C,2010A&A...518A..35C,2014AJ....148...99C}, \citet{2010SerAJ.181...43B}, \citet{grondin12}, \citet{2012MNRAS.420.2588B,2012RMxAA..48...41B,2012SerAJ.184...69B,2012SerAJ.185...25B}, \citet{2012A&A...540A..25D,2014AJ....147..162D}, \citet{kavanagh13}, \citet{2013MNRAS.432.2177B}, \citet{2014ApJ...780...50B}, \citet{2014MNRAS.439.1110B,2014MNRAS.440.3220B,2014Ap&SS.351..207B}, \citet{warth14}, \citet{Maggi14}, \citet{2015MNRAS.454..991R}, \citet{kavanagh15b,2015A&A...583A.121K,kavanagh15}, \citet{2015PKAS...30..149B} and \citet{2016A&A...586A...4K}.

These major contributions, coupled with various additional studies, led to the discovery of \confcount\ confirmed, and an additional \candcount\ candidate {SNRs}. Therefore, this is the first opportunity to perform a complete statistical study on a type of object that is crucial in galaxy evolution in one of the best laboratories available -- the {LMC}. Here, we report on a radio-continuum study of the most up-to-date sample of the LMC SNRs and SNR candidates, consisting of \totalcount\ of these objects.

\section{Observations}

Common analysis methodologies were undertaken and shown for all remnants, such as emission images from across the electromagnetic spectrum including radio-continuum 36~cm \citep[{MOST;}][]{1984IAUS..108..283M}, 20~cm \citep{2007MNRAS.382..543H}, 6~cm \citep{2010AJ....140.1511D}, infra-red 24~$\mu$m, 70~$\mu$m, and 160~$\mu$m \citep[{SAGE};][]{2006AJ....132.2268M}, optical \citep[{Magellanic Cloud Emission Line Survey, MCELS;}][]{2000ASPC..221...83S}, and X-ray \citep[LMC \xmm\ Large Project;][]{2014xru..confE...4H,maggi16}. The aforementioned radio-continuum mosaic images were also the default images used for flux density measurements at these wavelengths. Our own spectroscopic surveys of the MCs \citep{2008MNRAS.383.1175P,2005MNRAS.364..217F,2007MNRAS.376.1793P} and its {SNR} sample were mainly taken with the SAAO 1.9-m and MSSSO 2.3-m telescopes.

\section{Source List}

The number of confirmed SNRs in the LMC is currently at \confcount\ (as shown in \citet{maggi16}; also see Table~\ref{tab:lmcsnrs}). In this paper, we used the same sample with the addition of a further \candcount\ candidates, of which 7 are presented here for the first time (Table~\ref{tbl:candsnrs}). These 7 new SNR candidates are: MCSNR\,J0447-6919, MCSNR\,J0456-6950, MCSNR\,J0457-6739, MCSNR\,J0507-7110, MCSNR\,J0510-6708, MCSNR\,J0512-6716 and MCSNR\,J0527-7134 (for more details see Table~\ref{tbl:candsnrs} and Section~\ref{notes}).


\begin{longrotatetable}
\begin{deluxetable*}{llcccccccccc}
\tabletypesize{\scriptsize}
\tablecaption{\confcount\ confirmed SNRs in the LMC. This is a compilation of our own measurements and measurements taken from the literature. The types listed are as follows: TN -- thermonuclear (type~Ia) SNR, CC -- core-collapse SNR, CCpwn -- core-collapse SNR with associated PWN and X -- unknown type. Types listed with a ``q'' (e.g., CCq, TNq, etc.) are questionable and/or candidates for that type \label{tab:lmcsnrs}}
\tablecolumns{12}
\tablehead{
\colhead{Name} & \colhead{Other} & \colhead{RA} & \colhead{DEC} & \colhead{D$_{maj} \times D_{min}$} &
\colhead{PA} & \colhead{D$_{av}$} & \colhead{$\alpha \pm \Delta \alpha$} & \colhead{S$_{1 \rm{GHz}}$} &
\colhead{$\Sigma_{1 \rm{GHz}}$ ($\times 10^{-20}$)} & \colhead{Type} & \colhead{Age$\pm\Delta$Age\tablenotemark{a}}\\
\colhead{MCSNR\,J}  & \colhead{Name}   & \colhead{(J2000)} & \colhead{(J2000)} & \colhead{(\arcsec)} & \colhead{(\degr)} & \colhead{(pc)}   &    & \colhead{(Jy)}  & \colhead{(W\,m$^{-2}$Hz$^{-1}$sr$^{-1}$)}           &       & \colhead{(yr)}
}
\startdata
0448-6700 & HP\,460  & 04$^{h}$48$^{m}$26.3$^{s}$ & --67\degr00\arcmin24\arcsec & 290$\times$196 & 135   & 57.9 & --0.11$\pm$0.05 & 0.0334 & 0.0317 & X     & \nodata          \\
0449-6920 &          & 04$^{h}$49$^{m}$22.3$^{s}$ & --69\degr20\arcmin25\arcsec & 115$\times$115 & 0     & 27.9 & --0.40$\pm$0.08 & 0.0869 & 0.3560 & X     & \nodata          \\
0450-7050 &          & 04$^{h}$50$^{m}$23.5$^{s}$ & --70\degr50\arcmin23\arcsec & 535$\times$340 & 30    & 103.4& --0.41$\pm$0.03 & 0.6893 & 0.2050 & CCq   & 70000$\pm$25000$^{1}$  \\
0453-6655 & N\,4D    & 04$^{h}$53$^{m}$14.0$^{s}$ & --66\degr55\arcmin10\arcsec & 222$\times$263 & 15    & 58.6 & --0.58$\pm$0.06 & 0.1116 & 0.1040 & X     & \nodata          \\
\smallskip
0453-6829 & LHG\,1   & 04$^{h}$53$^{m}$37.2$^{s}$ & --68\degr29\arcmin28\arcsec & 120$\times$123 & 20    & 29.4 & --0.34$\pm$0.01 & 0.2100 & 0.7730 & CCpwn & 13500$\pm$1500$^{2}$   \\
0454-6712 & N\,9     & 04$^{h}$54$^{m}$33.0$^{s}$ & --67\degr12\arcmin50\arcsec & 140$\times$120 & 0     & 31.4 & --0.51$\pm$0.03 & 0.0767 & 0.2470 & TN    & 29500$\pm$7500$^{3}$   \\
0454-6625 & N11 L    & 04$^{h}$54$^{m}$49.9$^{s}$ & --66\degr25\arcmin36\arcsec & 95$\times$68   & 45    & 19.4 & --0.50$\pm$0.03 & 0.1539 & 1.3000 & CCq   & 11000$\pm$4000$^{4}$   \\
0455-6839 & N\,86    & 04$^{h}$55$^{m}$43.7$^{s}$ & --68\degr39\arcmin02\arcsec & 279$\times$213 & 170   & 59.1 & --0.51$\pm$0.04 & 0.3345 & 0.3050 & X     & 53000$\pm$33000$^{4}$  \\
0459-7008 & N\,186D  & 04$^{h}$59$^{m}$57.3$^{s}$ & --70\degr08\arcmin07\arcsec & 116$\times$116 & 0     & 28.1 & \nodata         & \nodata&\nodata & CCq   & 11000$^{5}$            \\
\smallskip
0505-6752 & DEM\,L71 & 05$^{h}$05$^{m}$41.9$^{s}$ & --67\degr52\arcmin39\arcsec & 88$\times$61   & 170   & 17.8 & --0.60$\pm$0.02 & 0.0087 & 0.0882 & TN    & 4360$\pm$290$^{6}$     \\
0505-6801 & N\,23    & 05$^{h}$05$^{m}$54.1$^{s}$ & --68\degr01\arcmin42\arcsec & 97$\times$92   & 20    & 23.0 & --0.60$\pm$0.04 & 0.3926 & 2.3700 & CC    & 4600$\pm$1200$^{7}$    \\
0506-6542 & DEM\,L72 & 05$^{h}$06$^{m}$08.2$^{s}$ & --65\degr42\arcmin10\arcsec & 410$\times$360 & 50    & 93.1 & \nodata         & \nodata& \nodata& X     & 115000$\pm$35000$^{8}$ \\
0506-7025 & DEM\,L80 & 05$^{h}$06$^{m}$47.9$^{s}$ & --70\degr25\arcmin38\arcsec & 183$\times$157 & 17    & 41.2 & \nodata         & \nodata& \nodata& TN    & 19000$\pm$2000$^{9}$   \\
0508-6902 & HP\,791  & 05$^{h}$08$^{m}$33.9$^{s}$ & --69\degr02\arcmin40\arcsec & 302$\times$234 & 30    & 64.5 & \nodata         & \nodata& \nodata& TN    & 22500$\pm$2500$^{10}$   \\
\smallskip
0508-6830 &          & 05$^{h}$08$^{m}$49.5$^{s}$ & --68\degr30\arcmin41\arcsec & 138$\times$108 & 45    & 29.8 & \nodata         & \nodata& \nodata& TN    & 20000$^{11}$            \\
0508-6843 & N\,103B  & 05$^{h}$08$^{m}$59.4$^{s}$ & --68\degr43\arcmin35\arcsec & 27$\times$29   & 0     & 6.8  & --0.65$\pm$0.03 & 0.5780 & 39.700 & TN    & 860$^{12}$              \\
0509-6731 & LHG\,14  & 05$^{h}$09$^{m}$31.1$^{s}$ & --67\degr31\arcmin17\arcsec & 32$\times$29   & 0     & 7.4  & --0.73$\pm$0.02 & 0.0974 & 5.7200 & TN    & 310$\pm$120$^{13}$      \\
0511-6759 &          & 05$^{h}$11$^{m}$10.7$^{s}$ & --67\degr59\arcmin07\arcsec & 228$\times$216 & 0     & 53.8 & \nodata         & \nodata& \nodata& TN    & 20000$^{11}$            \\
0512-6707 & HP\,483  & 05$^{h}$12$^{m}$28.8$^{s}$ & --67\degr07\arcmin15\arcsec & 55$\times$45   & 0     & 12.1 & --0.49$\pm$0.01 & 0.1046 & 2.2900 & TN    & 3150$\pm$1250$^{14}$  \\
\smallskip
0513-6912 & N\,112   & 05$^{h}$13$^{m}$14.4$^{s}$ & --69\degr12\arcmin15\arcsec & 245$\times$200 & 135   & 53.7 & --0.52$\pm$0.09 & 0.2423 & 0.2680 & X     & 3500$\pm$1500$^{15}$  \\
0514-6840 & HP\,700  & 05$^{h}$14$^{m}$15.5$^{s}$ & --68\degr40\arcmin14\arcsec & 220$\times$220 & 0     & 53.2 & \nodata         & \nodata& \nodata& X     & \nodata          \\
0517-6759 & HP\,607  & 05$^{h}$17$^{m}$10.2$^{s}$ & --67\degr59\arcmin03\arcsec & 324$\times$210 & 40    & 63.2 & \nodata         & \nodata& \nodata& X     & \nodata          \\
0518-6939 & N\,120A  & 05$^{h}$18$^{m}$43.5$^{s}$ & --69\degr39\arcmin11\arcsec & 85$\times$102  & 0     & 22.7 & --0.61$\pm$0.03 & 0.4504 & 2.7900 & CCq   & 7300$^{16}$          \\
0519-6902 & LHG\,26  & 05$^{h}$19$^{m}$34.8$^{s}$ & --69\degr02\arcmin06\arcsec & 36$\times$33   & 0     & 8.3  & --0.64$\pm$0.02 & 0.1316 & 6.0900 & TN    & 600$\pm$200$^{17}$      \\
\smallskip
0519-6926 & LHG\,27  & 05$^{h}$19$^{m}$45.3$^{s}$ & --69\degr26\arcmin01\arcsec & 140$\times$110 & 30    & 30.1 & --0.53$\pm$0.03 & 0.1606 & 0.5650 & X     & \nodata          \\
0521-6542 & DEM\,L142& 05$^{h}$21$^{m}$38.8$^{s}$ & --65\degr42\arcmin58\arcsec & 135$\times$141 & 0     & 33.4 & \nodata         & \nodata& \nodata& X     & \nodata          \\
0523-6753 & N\,44I   & 05$^{h}$23$^{m}$06.5$^{s}$ & --67\degr53\arcmin09\arcsec & 230$\times$230 & 0     & 55.8 & \nodata         & \nodata& \nodata& CCq   & 18000$^{18}$            \\
0524-6623 & N\,48E   & 05$^{h}$24$^{m}$18.9$^{s}$ & --66\degr23\arcmin33\arcsec & 145$\times$145 & 0     & 35.1 & --0.41$\pm$0.02 & 0.0725 & 0.1870 & CCq   & \nodata          \\
0525-6938 & N\,132D  & 05$^{h}$25$^{m}$02.7$^{s}$ & --69\degr38\arcmin33\arcsec & 114$\times$90  & 30    & 24.5 & --0.65$\pm$0.04 & 5.2642 & 27.900 & CC    & 3150$\pm$200$^{19}$     \\
\smallskip
0525-6559 & N\,49B   & 05$^{h}$25$^{m}$24.9$^{s}$ & --65\degr59\arcmin18\arcsec & 155$\times$155 & 0     & 37.6 & --0.56$\pm$0.03 & 0.6344 & 1.4300 & CC    & 10000$^{20}$            \\
0526-6605 & N\,49A   & 05$^{h}$26$^{m}$00.1$^{s}$ & --66\degr05\arcmin00\arcsec & 75$\times$75   & 0     & 18.2 & --0.59$\pm$0.03 & 1.6618 & 16.000 & CCq   & 4800$^{21}$             \\
0527-6912 & LHG\,40  & 05$^{h}$27$^{m}$39.3$^{s}$ & --69\degr12\arcmin07\arcsec & 157$\times$123 & 80    & 33.8 & \nodata         & \nodata& \nodata& CCq   & \nodata          \\
0527-6549 & DEM\,L204& 05$^{h}$27$^{m}$54.9$^{s}$ & --65\degr49\arcmin49\arcsec & 335$\times$275 & 45    & 73.6 & --0.51$\pm$0.04 & 0.1365 & 0.0803 & X     & \nodata          \\
0528-7104 & HP\,1234 & 05$^{h}$28$^{m}$04.3$^{s}$ & --71\degr04\arcmin40\arcsec & 328$\times$234 & 155   & 67.1 & \nodata         & \nodata& \nodata& X     & 25000$^{22}$          \\
\smallskip
0528-6726 & DEM\,L205& 05$^{h}$28$^{m}$11.1$^{s}$ & --67\degr26\arcmin49\arcsec & 260$\times$180 & 30    & 52.4 & \nodata         & \nodata& \nodata& CCq   & 33500$\pm$3500$^{23}$   \\
0528-6713 & HP\,498  & 05$^{h}$28$^{m}$18.5$^{s}$ & --67\degr13\arcmin49\arcsec & 216$\times$216 & 0     & 52.4 & --0.28$\pm$0.09 & 0.1113 & 0.1290 & X     & \nodata          \\
0529-6653 & DEM\,L214& 05$^{h}$29$^{m}$51.0$^{s}$ & --66\degr53\arcmin27\arcsec & 137$\times$128 & 0     & 29.1 & --0.68$\pm$0.03 & 0.0863 & 0.2670 & X     & \nodata          \\
0530-7007 & DEM\,L218& 05$^{h}$30$^{m}$40.4$^{s}$ & --70\degr07\arcmin27\arcsec & 215$\times$180 & 45    & 47.7 & --0.27$\pm$0.01 & 0.0718 & 0.1010 & TNq   & \nodata          \\
0531-7100 & N\,206   & 05$^{h}$31$^{m}$57.9$^{s}$ & --71\degr00\arcmin16\arcsec & 190$\times$170 & 90    & 43.6 & --0.66$\pm$0.03 & 0.4086 & 0.6850 & CCpwn & 25000$\pm$2000$^{24}$   \\
\smallskip
0532-6731 & N\,56    & 05$^{h}$32$^{m}$19.9$^{s}$ & --67\degr31\arcmin37\arcsec & 180$\times$180 & 0     & 43.6 & --0.63$\pm$0.04 & 0.2152 & 0.3600 & X     & \nodata          \\
0533-7202 & RASS\,236& 05$^{h}$33$^{m}$51.0$^{s}$ & --72\degr02\arcmin50\arcsec & 200$\times$160 & 45    & 43.5 & --0.47$\pm$0.06 & 0.1333 & 0.2250 & TNq   & 23000$\pm$5000$^{25}$   \\
0534-6955 & LHG\,53  & 05$^{h}$34$^{m}$00.8$^{s}$ & --69\degr55\arcmin08\arcsec & 120$\times$110 & 155   & 27.2 & --0.51$\pm$0.01 & 0.0889 & 0.3820 & TN    & 10100$^{26}$            \\
0534-7033 & DEM\,L238& 05$^{h}$34$^{m}$23.0$^{s}$ & --70\degr33\arcmin25\arcsec & 222$\times$158 & 190   & 45.4 & --0.44$\pm$0.09 & 0.0796 & 0.1230 & TN    & 12500$\pm$2500$^{27}$   \\
0535-6916 & SNR1987A & 05$^{h}$35$^{m}$28.0$^{s}$ & --69\degr16\arcmin12\arcsec & 1.8$\times$1.8 & 0     & 0.4  & --0.68$\pm$0.03 & 0.8200 &  13700 & CC    & 30               \\
\smallskip
0535-6602 & N\,63A   & 05$^{h}$35$^{m}$43.8$^{s}$ & --66\degr02\arcmin13\arcsec & 81$\times$67   & 45    & 17.8 & --0.74$\pm$0.02 & 1.8641 & 18.700 & CCq   & 3500$\pm$1500$^{28}$          \\
0535-6918 & Honeycomb& 05$^{h}$35$^{m}$45.5$^{s}$ & --69\degr18\arcmin08\arcsec & 91$\times$59   & 160   & 17.7 & --0.71$\pm$0.05 & 0.1483 & 1.5200 & X     & \nodata          \\
0536-6735 & N\,59B   & 05$^{h}$36$^{m}$04.2$^{s}$ & --67\degr35\arcmin11\arcsec & 147$\times$125 & 35    & 32.9 & \nodata         & \nodata& \nodata& CCpwn & 60000$\pm$10000$^{29}$          \\
0536-7038 & DEM\,L249& 05$^{h}$36$^{m}$06.6$^{s}$ & --70\degr38\arcmin38\arcsec & 187$\times$127 & 25    & 37.4 & --0.52$\pm$0.06 & 0.0699 & 0.1600 & TN    & 12500$\pm$2500$^{27}$   \\
0536-6913 &          & 05$^{h}$36$^{m}$17.0$^{s}$ & --69\degr13\arcmin28\arcsec & 66$\times$66   & 0     & 16.0 & \nodata         & \nodata& \nodata& CC    & 3550$\pm$1350$^{30}$    \\
\smallskip
0537-6627 & DEM\,L256& 05$^{h}$37$^{m}$30.3$^{s}$ & --66\degr27\arcmin45\arcsec & 210$\times$165 & 45    & 45.1 & --0.47$\pm$0.06 & 0.0729 & 0.1140 & X     & 50000$^{8}$            \\
0537-6910 &30\,Dor\,B& 05$^{h}$37$^{m}$45.6$^{s}$ & --69\degr10\arcmin20\arcsec & 136$\times$116 & 155   & 30.4 & --0.38$\pm$0.03 & 2.8817 & 9.9000 & CCpwn & 5000$^{31}$             \\
0540-6944 & N\,159   & 05$^{h}$40$^{m}$00.0$^{s}$ & --69\degr44\arcmin06\arcsec & 120$\times$90  & 90    & 25.2 & \nodata         & \nodata& \nodata& CC    & 18000$^{32}$            \\
0540-6919 & N\,158A  & 05$^{h}$40$^{m}$11.3$^{s}$ & --69\degr19\arcmin54\arcsec & 67$\times$58   & 0     & 15.1 & --0.63$\pm$0.03 & 1.0272 & 14.300 & CCpwn & 1100$\pm$340$^{33}$     \\
0541-6659 & HP\,456  & 05$^{h}$41$^{m}$51.5$^{s}$ & --66\degr59\arcmin03\arcsec & 300$\times$272 & 45    & 69.2 & \nodata         & \nodata& \nodata& X     & 23000$^{34}$            \\
\smallskip
0543-6900 & DEM\,L299& 05$^{h}$43$^{m}$02.2$^{s}$ & --69\degr00\arcmin00\arcsec & 226$\times$226 & 0     & 54.9 & \nodata         & \nodata& \nodata& X     & 10800$\pm$7300$^{35}$   \\
0547-6942 &DEM\,L316B& 05$^{h}$47$^{m}$00.0$^{s}$ & --69\degr42\arcmin50\arcsec & 200$\times$160 & 115   & 43.4 & --0.53$\pm$0.16 & 0.7282 & 1.2300 & CCq   & 40500$\pm$1500$^{36}$   \\
0547-6941 &DEM\,L316A& 05$^{h}$47$^{m}$20.9$^{s}$ & --69\degr41\arcmin27\arcsec & 122$\times$118 & 140   & 29.1 & --0.54$\pm$0.16 & 0.5217 & 1.9600 & TNq   & 33000$\pm$6000$^{37}$   \\
0547-7024 & LHG\,89  & 05$^{h}$47$^{m}$48.8$^{s}$ & --70\degr24\arcmin52\arcsec & 120$\times$105 & 0     & 27.2 & --0.56$\pm$0.03 & 0.0632 & 0.2720 & X     & 7100$^{9}$             \\
0550-6823 & DEM\,L328& 05$^{h}$50$^{m}$30.7$^{s}$ & --68\degr23\arcmin37\arcsec & 373$\times$282 & 95    & 78.6 & --0.41$\pm$0.02 & 0.6495 & 0.3350 & CCq   & \nodata                \\
  \enddata

  \tablenotetext{a}{References for LMC SNRs ages: $^{1}$\citet{2004ApJ...613..948W}, $^{2}$\citet{2004ApJ...613..948W}, $^{3}$\citet{2006ApJ...640..327S}, $^{4}$\citet{1999ApJS..123..467W}, $^{5}$\citet{2011ApJ...729...28J}, $^{6}$\citet{2003ApJ...590..833G}, $^{7}$\citet{2006ApJ...645L.117H}, $^{8}$\citet{2010ApJ...725.2281K}, $^{9}$\citet{maggi16}, $^{10}$\citet{2014MNRAS.439.1110B}, $^{11}$\citet{Maggi14}, $^{12}$\citet{1995ApJ...444L..81H}, $^{13}$\citet{2015ApJ...809..119H}, $^{14}$\citet{2015MNRAS.454..991R}, $^{15}$\citet{2010AJ....140..584D}, $^{16}$\citet{rosado93}, $^{17}$\citet{2006ApJ...642L.141B}, $^{18}$\citet{2006AJ....132.1877W}, $^{19}$\citet{2007ApJ...671L..45B}, $^{20}$\citet{2003ApJ...592L..41P}, $^{21}$\citet{2012ApJ...748..117P}, $^{22}$\citet{kavanagh13}, $^{23}$\citet{maggi12}, $^{24}$\citet{2005ApJ...628..704W}, $^{25}$\citet{kavanagh15}, $^{26}$\citet{2003ApJ...593..370H}, $^{27}$\citet{2006ApJ...652.1259B}, $^{28}$\citet{2003ApJ...583..260W}, $^{29}$\citet{2012ApJ...759..123S}, $^{30}$\citet{kavanagh15b}, $^{31}$\citet{2010AJ....140..177S}, $^{32}$\citet{2010AJ....140..177S}, $^{33}$\citet{2014AJ....148...99C}, $^{34}$\citet{grondin12}, $^{35}$\citet{warth14}, $^{36}$\citet{2001PASJ...53...99N}, $^{37}$\citet{2005ApJ...635.1077W},}
 \end{deluxetable*}

\end{longrotatetable}

\clearpage

\begin{longrotatetable}
  \begin{deluxetable*}{llccccccclcl}
\tabletypesize{\small}
\tablecaption{Details of the \candcount\ candidate SNRs in the LMC. New SNR candidates are marked with $^{\star}$. We assume $\alpha=-0.5$ for MCSNR\,J0457-6739 and MCSNR\,J0510-6708 (marked with $\ddag$). References in column 10 are: 1 -- \cite{1984PASAu...5..537T}, 2 -- HP99, 3  -- \citet{2000AJ....119.2242C} \label{tbl:candsnrs}}
\tablewidth{0pt}
\tablecolumns{10}
\tablehead{
\colhead{Name} & \colhead{RA} & \colhead{DEC} & \colhead{D$_{maj} \times D_{min}$} &
\colhead{PA} & \colhead{D$_{av}$} & \colhead{$\alpha \pm \Delta \alpha$} & \colhead{S$_{1 \rm{GHz}}$} &
\colhead{$\Sigma_{1 \rm{GHz}}$ ($\times 10^{-20}$)} & \colhead{Reference \&} \\
\colhead{MCSNR\,J} & \colhead{(J2000)}          & \colhead{(J2000)}        & \colhead{(\arcsec)} & \colhead{(\degr)} & \colhead{(pc)}   &  & \colhead{(Jy)}    & \colhead{(W\,m$^{-2}$Hz$^{-1}$sr$^{-1}$)} & \colhead{Other Names}
}
\startdata
0447-6918$^{\star}$& 04$^{h}$47$^{m}$09.7$^{s}$ &  --69\degr18\arcmin58\arcsec  & 245$\times$245 & \p00    & \p059.4  & \nodata         & \nodata & \nodata&  \\
0449-6903          & 04$^{h}$49$^{m}$34.0$^{s}$ &  --69\degr03\arcmin34\arcsec  & 135$\times$135 & \p00    & \p032.7  & --0.50$\pm$0.01 & 0.0984  & 0.2926 & 1\\
0456-6950$^{\star}$& 04$^{h}$56$^{m}$30.3$^{s}$ &  --69\degr50\arcmin47\arcsec  & 180$\times$180 & \p00    & \p043.6  & \nodata         & \nodata & \nodata&  \\
0457-6923          & 04$^{h}$57$^{m}$07.8$^{s}$ &  --69\degr23\arcmin58\arcsec  & 180$\times$120 & 90      & \p035.6  & \nodata         & \nodata & \nodata& 2 \\
\smallskip
0457-6739$^{\star}$& 04$^{h}$57$^{m}$33.0$^{s}$ &  --67\degr39\arcmin05\arcsec  & 150$\times$150 & \p00    & \p036.4  & $-0.5^{\ddag}$  & 0.0304$^{\ddag}$ & 0.0732$^{\ddag}$& \\
0506-6815          & 05$^{h}$06$^{m}$05.3$^{s}$ &  --68\degr15\arcmin47\arcsec  & 255$\times$210 & 30      & \p056.1  &\p00.00$\pm$0.41 & 0.0371  & 0.0037 & 2; [HP99] 635 \\
0507-7110$^{\star}$& 05$^{h}$07$^{m}$35.3$^{s}$ &  --71\degr10\arcmin15\arcsec  & 270$\times$270 & \p00    & \p065.4  & \nodata         & \nodata & \nodata& DEM\,L81\\
0507-6847          & 05$^{h}$07$^{m}$36.0$^{s}$ &  --68\degr47\arcmin48\arcsec  & 600$\times$400 & 80      & 118.8    & \nodata         & \nodata & \nodata& 3 \\
0510-6708$^{\star}$& 05$^{h}$10$^{m}$11.4$^{s}$ &  --67\degr08\arcmin04\arcsec  & 120$\times$120 & \p00    & \p029.1  & $-0.5^{\ddag}$  & 0.0029$^{\ddag}$ & 0.0110$^{\ddag}$&  \\
\smallskip
0512-6716$^{\star}$& 05$^{h}$12$^{m}$24.7$^{s}$ &  --67\degr16\arcmin55\arcsec  & 240$\times$210 & 45      & \p054.5  & \nodata         & \nodata & \nodata& \\
0513-6731          & 05$^{h}$13$^{m}$29.6$^{s}$ &  --67\degr31\arcmin52\arcsec  & 150$\times$105 & 60      & \p030.4  & --0.56$\pm$0.41 & 0.0261  & 0.0897 & 2; [HP99] 544 \\
0513-6724          & 05$^{h}$13$^{m}$40.0$^{s}$ &  --67\degr24\arcmin20\arcsec  & 150$\times$150 & \p00    & \p036.4  & --0.61$\pm$0.41 & 0.0279  & 0.0673 & 2; [HP99] 530 \\
0527-7134$^{\star}$& 05$^{h}$27$^{m}$48.5$^{s}$ &  --71\degr34\arcmin06\arcsec  & 180$\times$145 & 45      & \p039.2  & --0.52$\pm$0.41 & 0.0284  & 0.0589 &  \\
0538-6921          & 05$^{h}$38$^{m}$12.9$^{s}$ &  --69\degr21\arcmin41\arcsec  & 169$\times$169 & \p00    & \p041.0  & --0.59$\pm$0.04 & 0.5207  & 0.9878 & 1 \\
0539-7001          & 05$^{h}$39$^{m}$36.2$^{s}$ &  --70\degr01\arcmin44\arcsec  & 210$\times$120 & 45      & \p038.5  & --0.47$\pm$0.41 & 0.0054  & 0.0115 & 2; [HP99] 1063 \\
  \enddata
  \end{deluxetable*}
\end{longrotatetable}

\clearpage

We searched all available optical, radio and X-ray surveys in order to secure the most complete population of the LMC SNRs. Primarily, we classified the \candcount\ LMC SNR candidates based on the well established criteria described in \citet{1998A&AS..130..421F}. We emphasise that all these sources require further study in order to secure a bona-fide classification as SNRs. In Figs.~\ref{snrcandidates1}, \ref{snrcandidates2} and \ref{snrcandidates3} we show images of the 14 LMC SNR candidates at various frequencies. An image of MCSNR\,J0507-6846 is shown in \citet{2000AJ....119.2242C}. SNR extent is primarily measured using MCELS images, with some additional information obtained via \cxo\, \xmm\ or \ROSAT\ surveys when needed. Because of their very low surface brightness we could not measure radio emission from six of these \candcount\ LMC SNRs candidates.

All LMC SNRs and SNR candidates' radio flux density measurements are shown here for the first time and their associated errors are well below 10\%. We determined source diameters from the highest resolution image available including optical and X-ray images. We estimated that the error in diameter is smaller than 2\arcsec\ or $\sim$0.5~pc. { We found that our diameters estimated here are $\sim$10\% smaller compared to \citet{maggi16}. The reason for this small discrepancy is due to a better resolution images that we used here compared to \xmm. Also, we show here for the first time a compilation of estimated LMC SNR ages using various methods.} Therefore, Table~\ref{tab:lmcsnrs} is a compilation of our own measurements as well as those of other papers for this well established sample of the LMC SNRs.

\subsection{Notes on LMC SNR candidates}
 \label{notes}

\begin{figure*}
 \begin{center}
  \resizebox{0.8\columnwidth}{!}{\includegraphics[angle=-90,trim = 0 0 0 0]{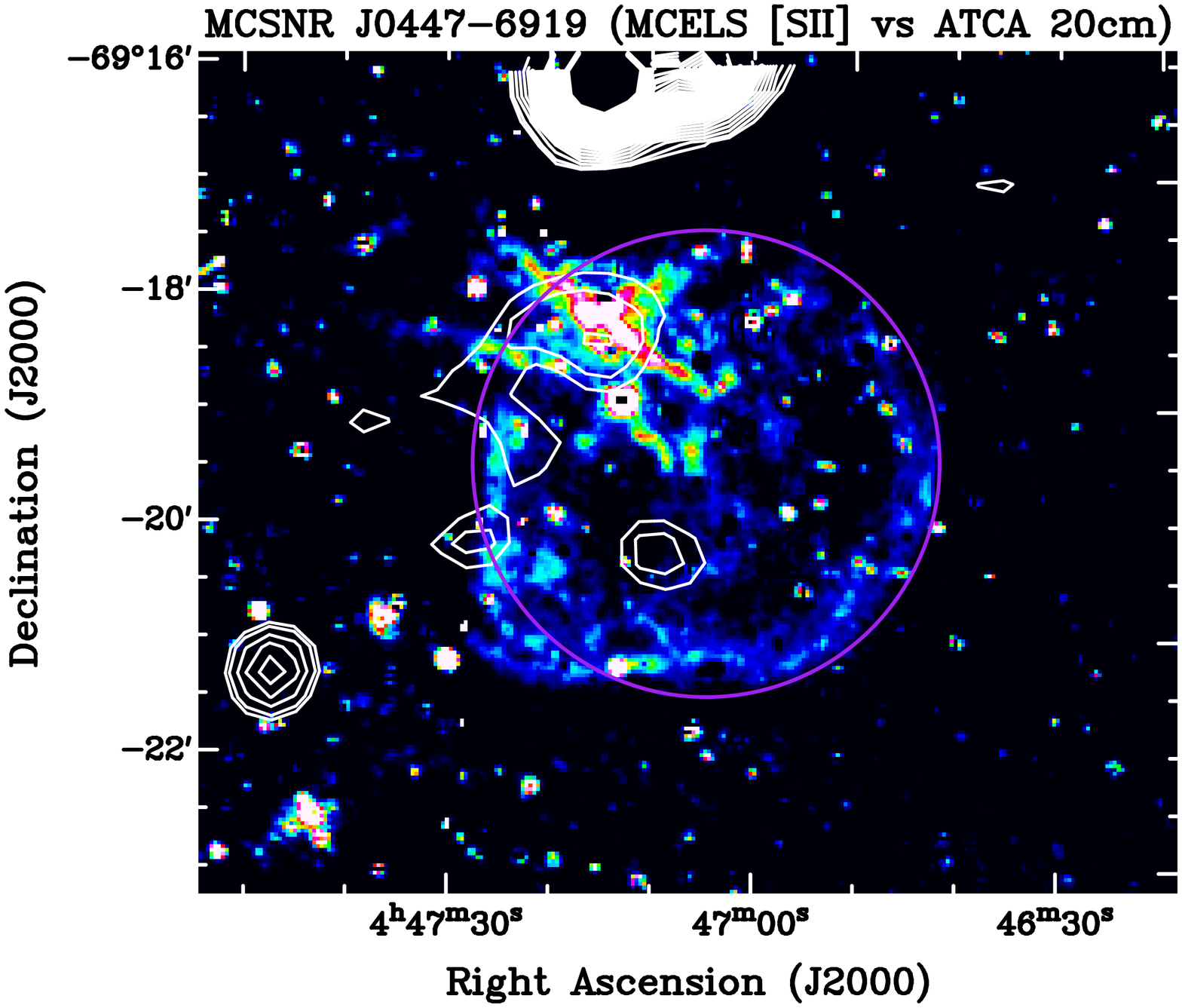}}
  \resizebox{0.8\columnwidth}{!}{\includegraphics[angle=-90,trim = 0 0 0 0]{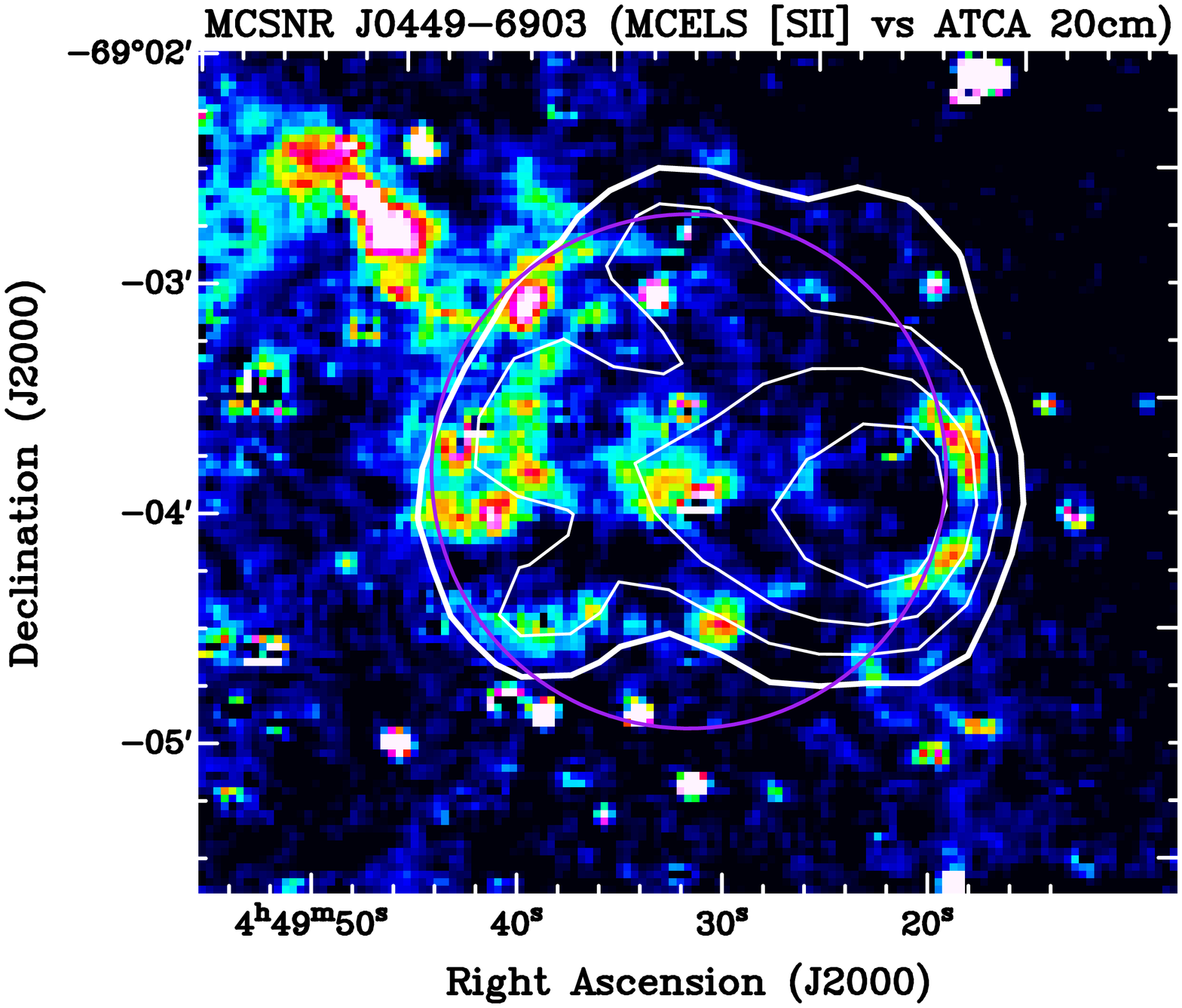}}
  \resizebox{0.8\columnwidth}{!}{\includegraphics[angle=-90,trim = 0 0 0 0]{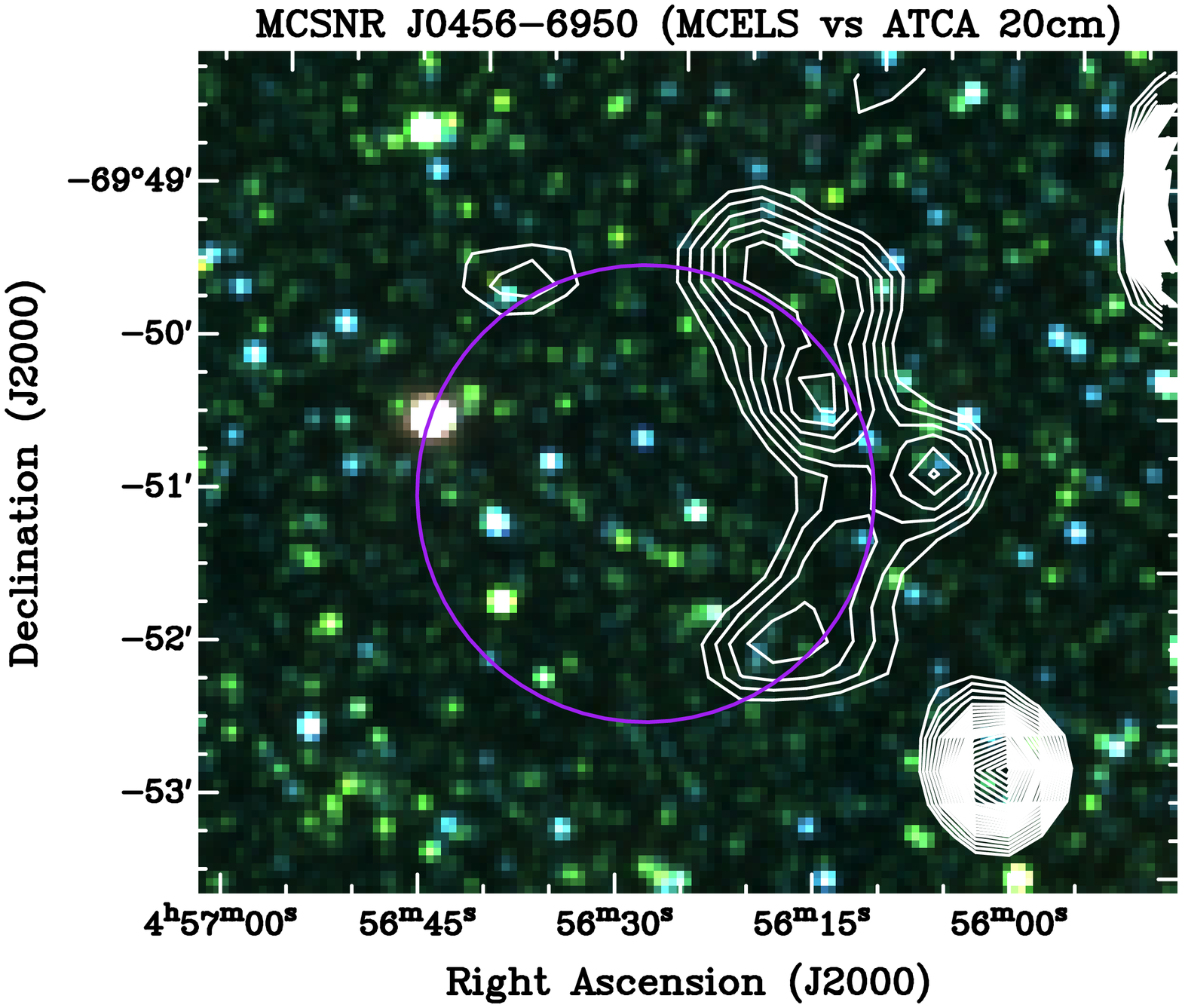}}
  \resizebox{0.8\columnwidth}{!}{\includegraphics[angle=-90,trim = 0 0 0 0]{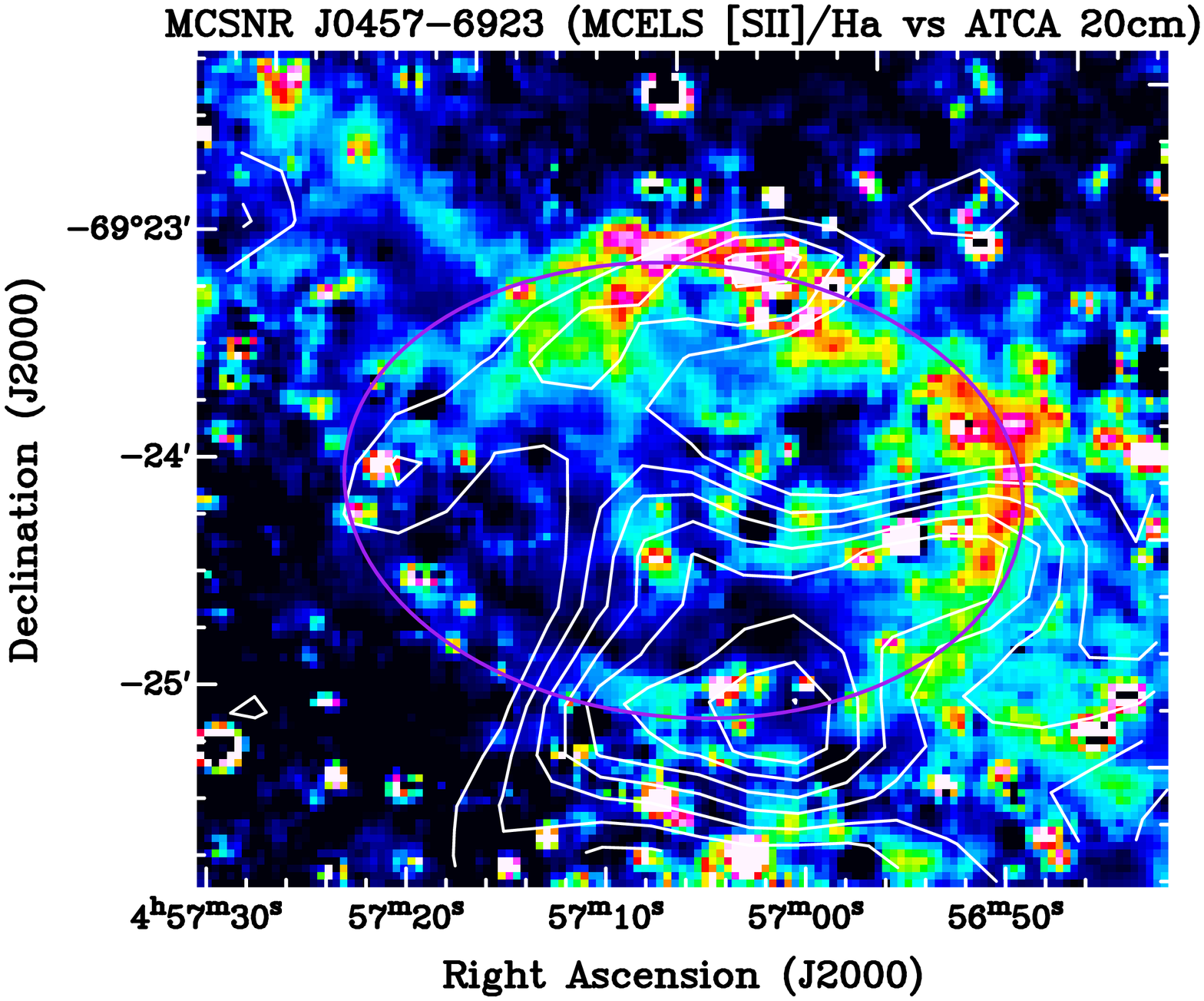}}
  \resizebox{0.8\columnwidth}{!}{\includegraphics[angle=-90,trim = 0 0 0 0]{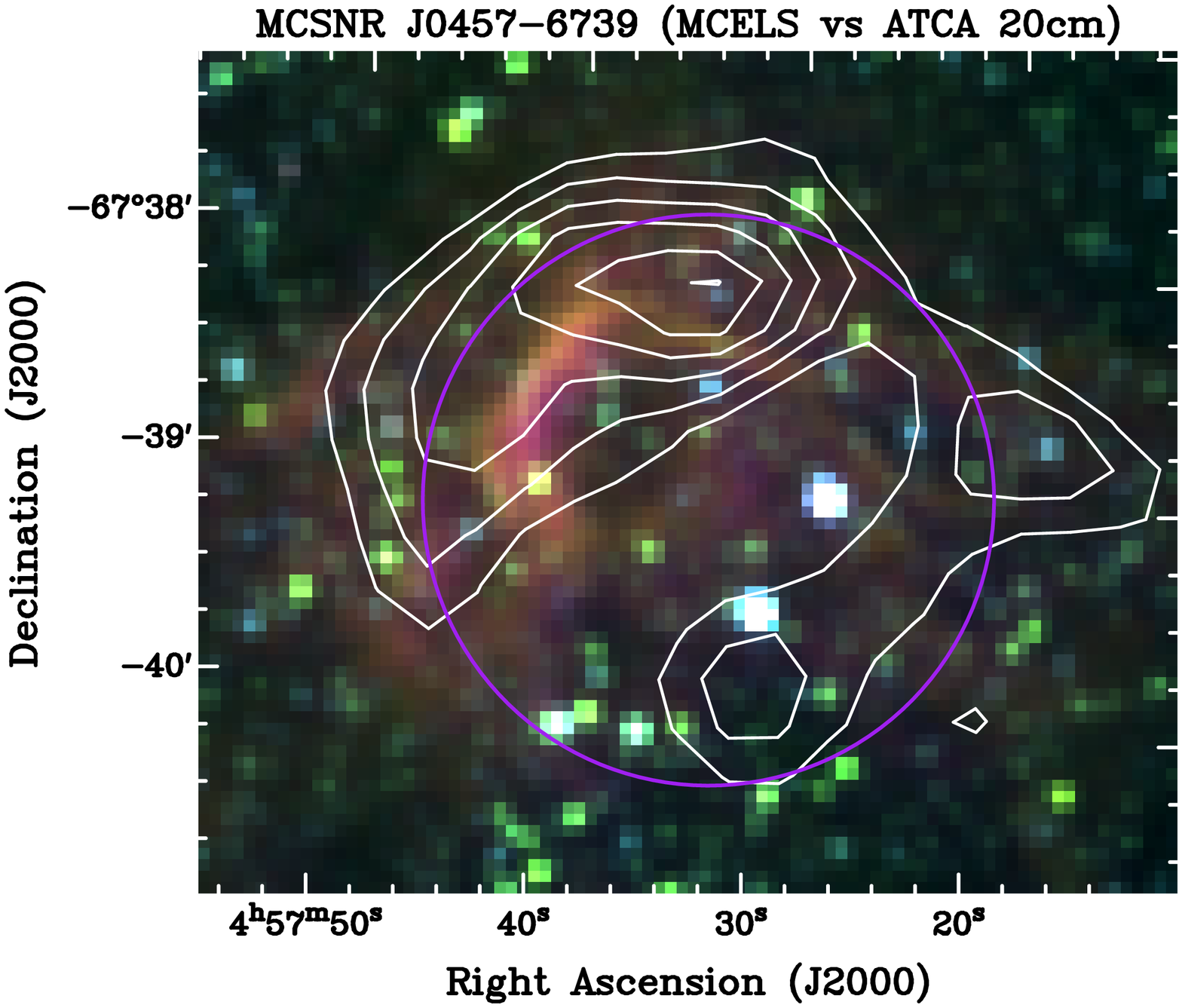}}
  \resizebox{0.8\columnwidth}{!}{\includegraphics[angle=-90,trim = 0 0 0 0]{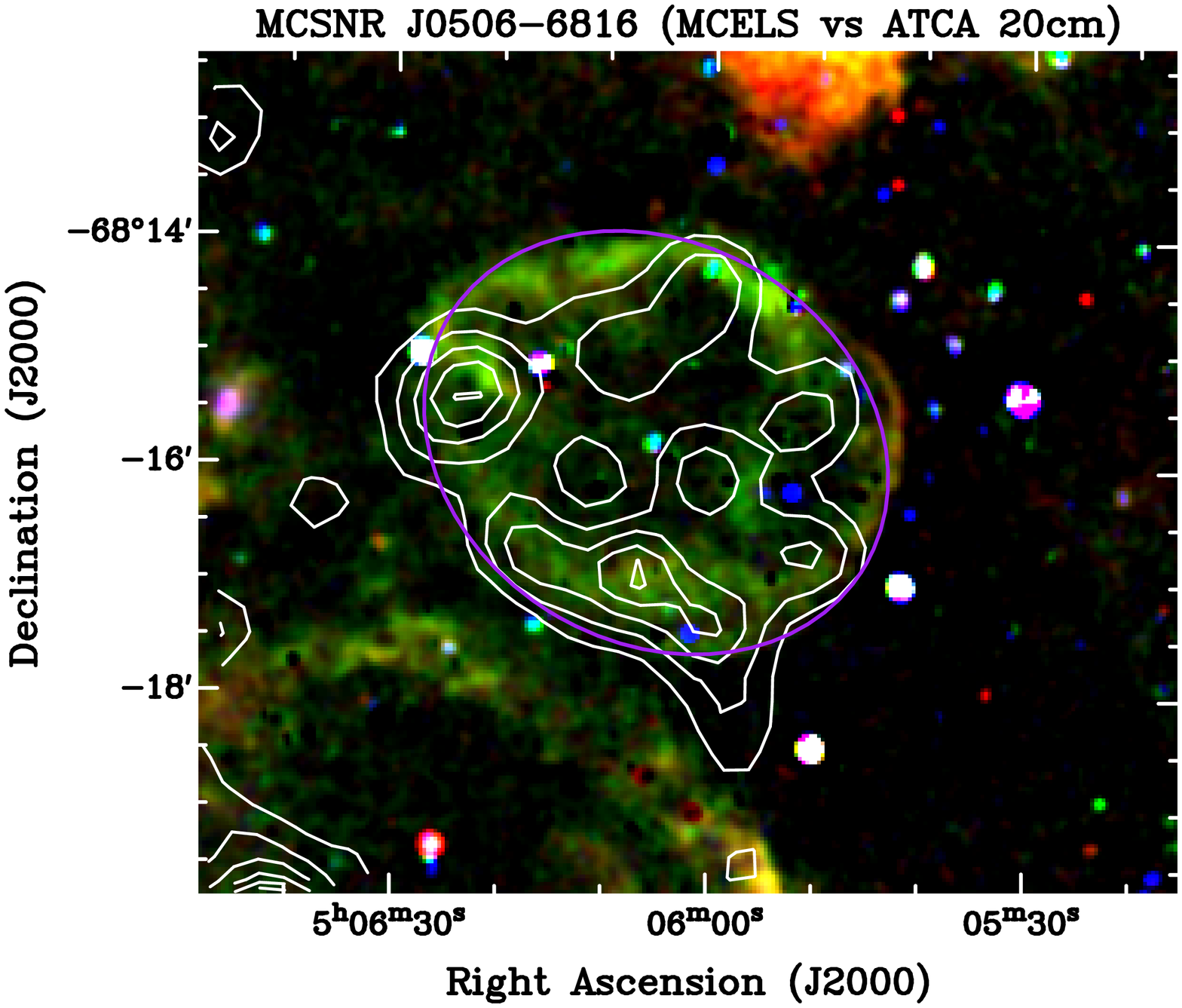}}
  \caption{The LMC SNR candidates 1. Colour images are MCELS where RGB corresponds to H$\alpha$, \SII\ and \OIII. The colour image of MCSNR\,J0457-6923 is ratio map between \SII\ and H$\alpha$. Contours are from the ATCA 20~cm mosaic survey and start at the 3$\sigma$ local noise level with spacing of 1$\sigma$. The circles/ellipses (purple) represent approximate extent of the SNR candidates.}
  \label{snrcandidates1}
 \end{center}
\end{figure*}

\noindent\textbf{MCSNR\,J0447-6918} (Figure~\ref{snrcandidates1}; top-left) -- A large optical shell (245\arcsec$\times$245\arcsec) was present with enhanced \SII /H$\alpha$ ratio of $>$0.4. Also, some weak 20~cm emission was detected in the NE part but no reliable flux density estimate was possible. No sensitive X-ray coverage is available in this field.

\noindent\textbf{MCSNR\,J0449-6903} (Figure~\ref{snrcandidates1}; top-right) -- \citet{1984PASAu...5..537T} originally proposed this source to be an {SNR} exemplifying typical evolved shell type SNR morphology. While there was no obvious optical identification in MCELS, we estimated a radio spectral index of $\alpha=-0.50\pm0.01$ using measured flux densities of 108~mJy, 83~mJy and 45~mJy at 36~cm (843~MHz), 20~cm (1377~MHz) and 6~cm (4800~MHz), respectively. Unfortunately, no sensitive X-ray coverage is available at this point.

\noindent\textbf{MCSNR\,J0456-6950} (Figure~\ref{snrcandidates1}; middle-left) -- This source is a potential radio SNR based on a shell-like radio structure. We have no definite optical confirmation and X-ray surveys have not covered this region.

\noindent\textbf{MCSNR\,J0457-6923} (Figure~\ref{snrcandidates1}; middle-right) -- This source was classified as a potential optical SNR based on a \SII /H$\alpha$ ratio of $>$0.4 as well as an evident radio emission. No sensitive X-ray coverage is available.

\noindent\textbf{MCSNR\,J0457-6739} (Figure~\ref{snrcandidates1}; bottom-left) -- This object exhibits a shell-like optical nebula with a somewhat enhanced \SII /H$\alpha$ ratio of $\sim$0.4 and a shell-like radio-continuum morphology. However, we were only able to measure a flux density at 20~cm of 25.9~mJy. No sensitive X-ray coverage is available.

\noindent\textbf{MCSNR\,J0506-6815 = [HP99] 635} (Figure~\ref{snrcandidates1}; bottom-right) -- HP99 recorded an object at this position, giving it the name [HP99]~635\footnote{[HP99] xxx is SIMBAD nomenclature with source number xxx from HP99}. They listed an extent of 31.3\arcsec\ (very low likelihood), and estimated HR1 and HR2 values of $1.00\pm0.76$ and $-0.05\pm0.19$, respectively. Therefore, the X-ray source could be a point source unrelated to the SNR candidate. While there was no optical (MCELS) signature, we detected extended radio-continuum emission and estimated flux densities of 37~mJy at both frequencies (36~cm and 20~cm) pointing to a flat spectral index which is indicative of a PWN. This object is classified as an SNR candidate primarily based on its radio and optical morphology as the X-ray emission association is not clear at this point.

\noindent\textbf{MCSNR\,J0507-6847} -- \citet{2000AJ....119.2242C} observed this source, consisting of a large ring of diffuse X-ray emission and proposed it as an {SNR} candidate. They found an X-ray luminosity within the range expected for {SNR}s, and predicted an age of $\sim5\times10^{4}$~yr based on the Sedov solution. \citet{2006ApJS..165..480B} did not detect this object in their far-UV survey of the {MCs}. We did not detect associated optical (MCELS) or radio-continuum features in this large object and we propose that this object might represent a superbubble similar to 30~Dor~C \citep{2017arXiv170101962S}.

\begin{figure*}
 \begin{center}
  \resizebox{0.8\columnwidth}{!}{\includegraphics[angle=-90,trim = 0 0 0 0,clip]{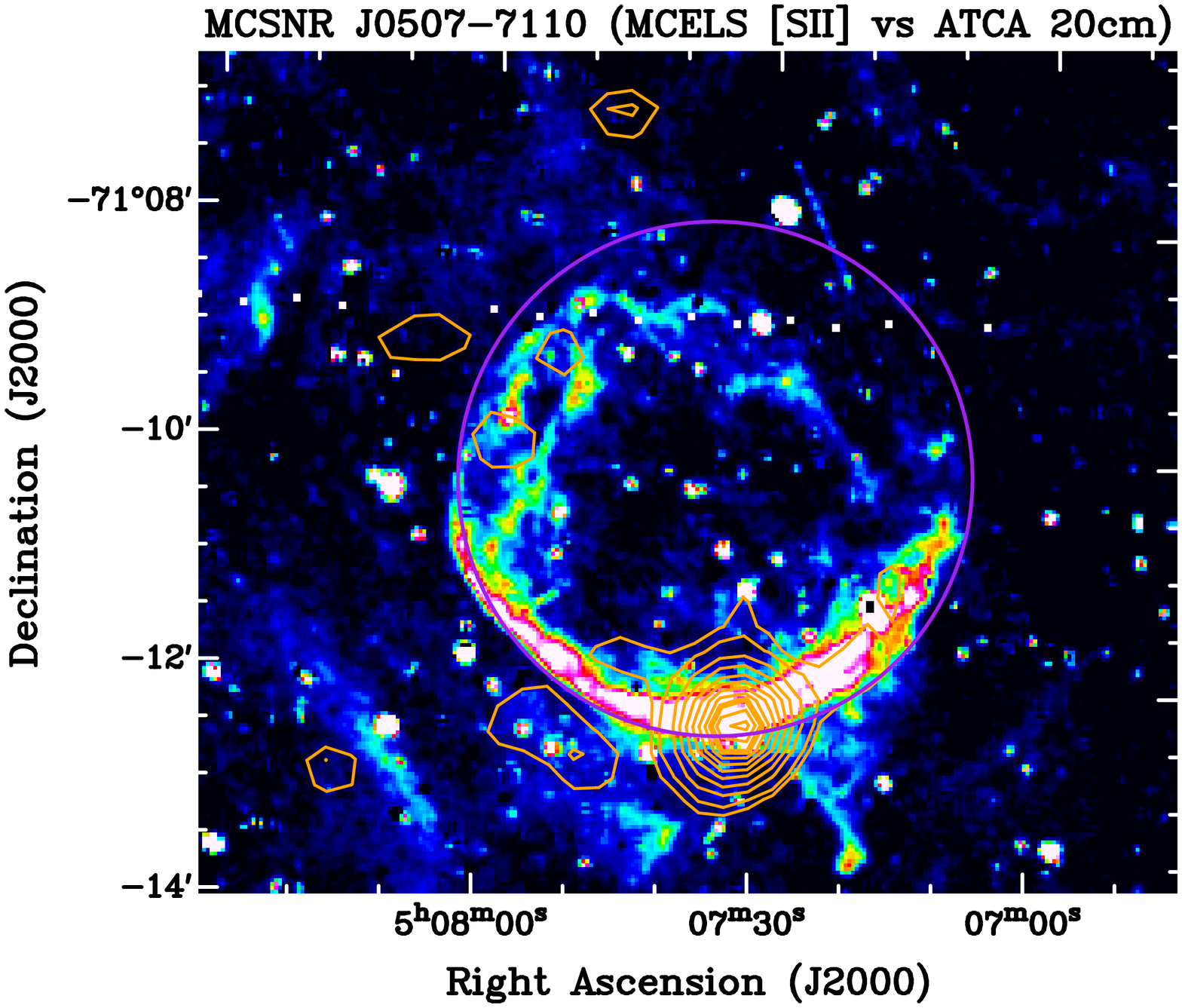}}
  \resizebox{0.8\columnwidth}{!}{\includegraphics[angle=-90,trim = 0 0 0 0,clip]{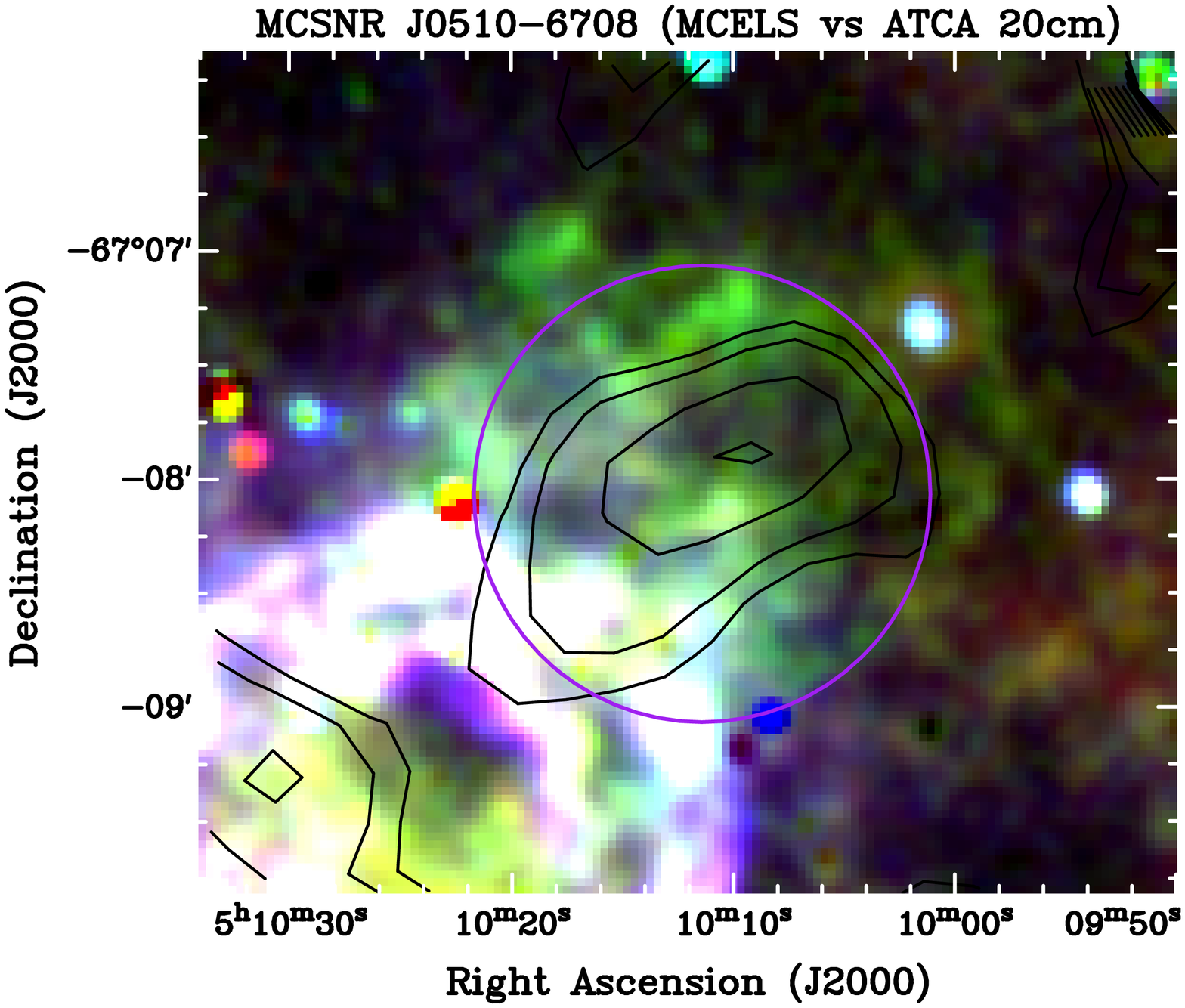}}
  \resizebox{0.8\columnwidth}{!}{\includegraphics[angle=-90,trim = 0 0 0 0,clip]{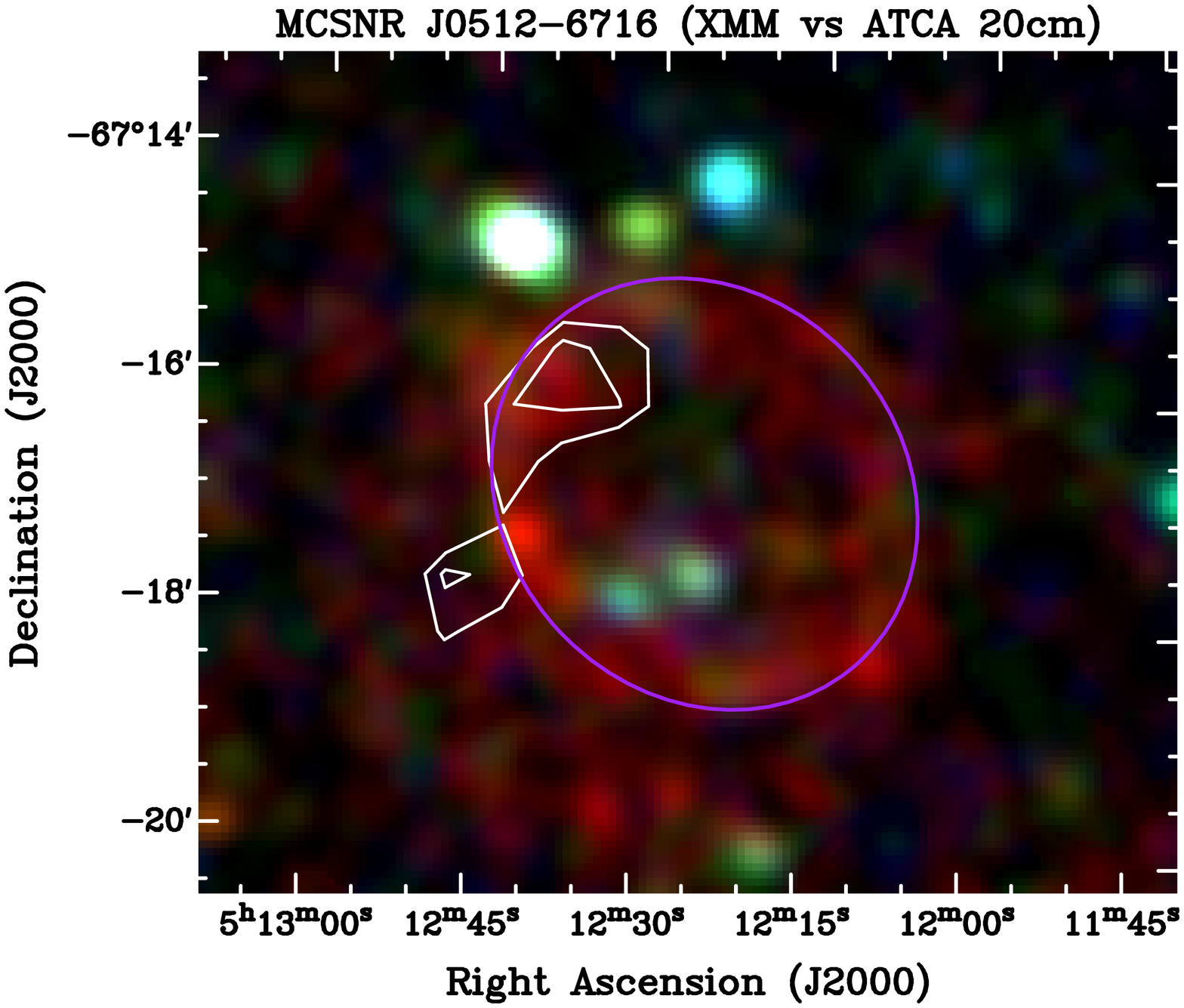}}
  \resizebox{0.8\columnwidth}{!}{\includegraphics[angle=-90,trim = 0 0 0 0,clip]{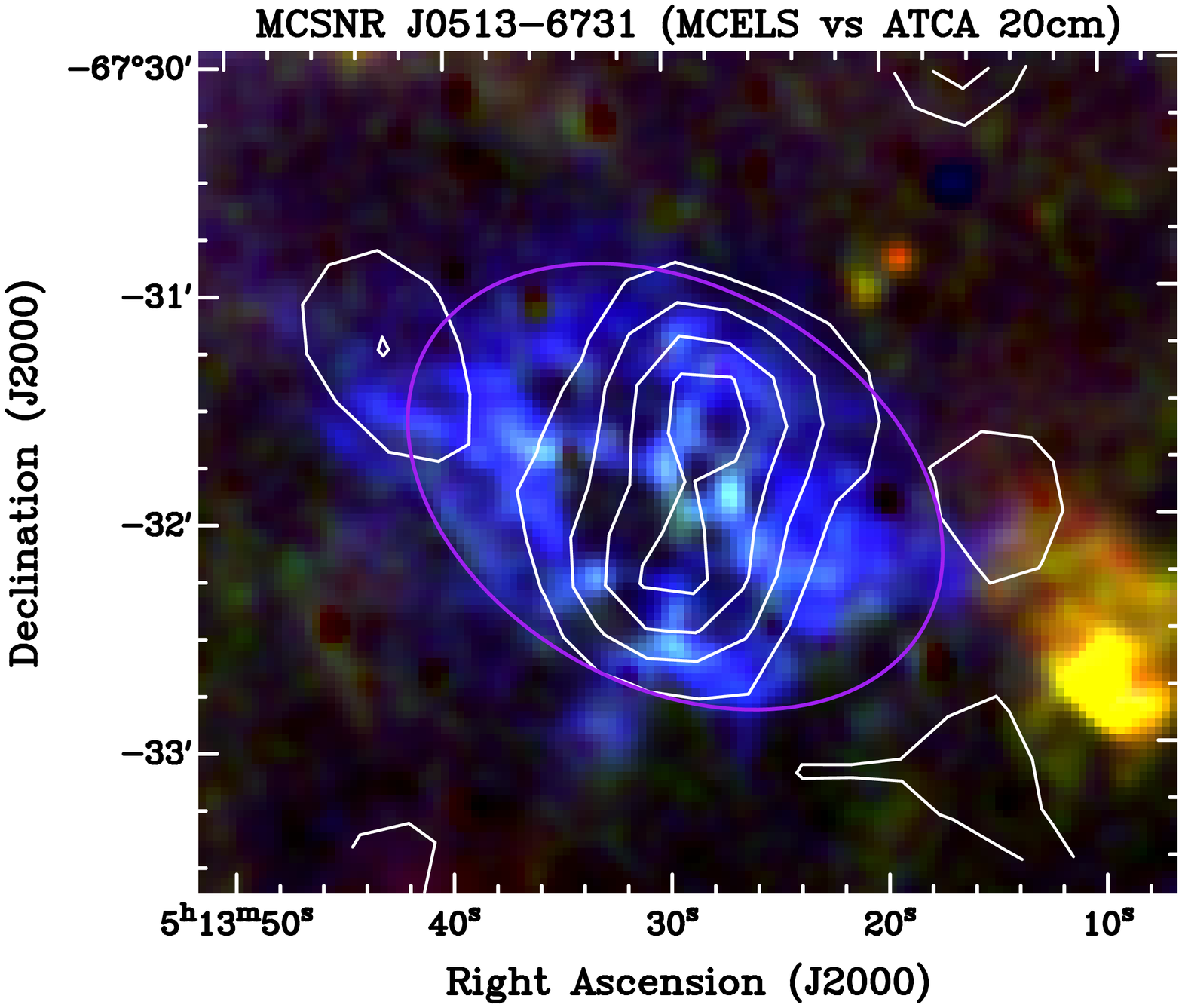}}
  \resizebox{0.8\columnwidth}{!}{\includegraphics[angle=-90,trim = 0 0 0 0,clip]{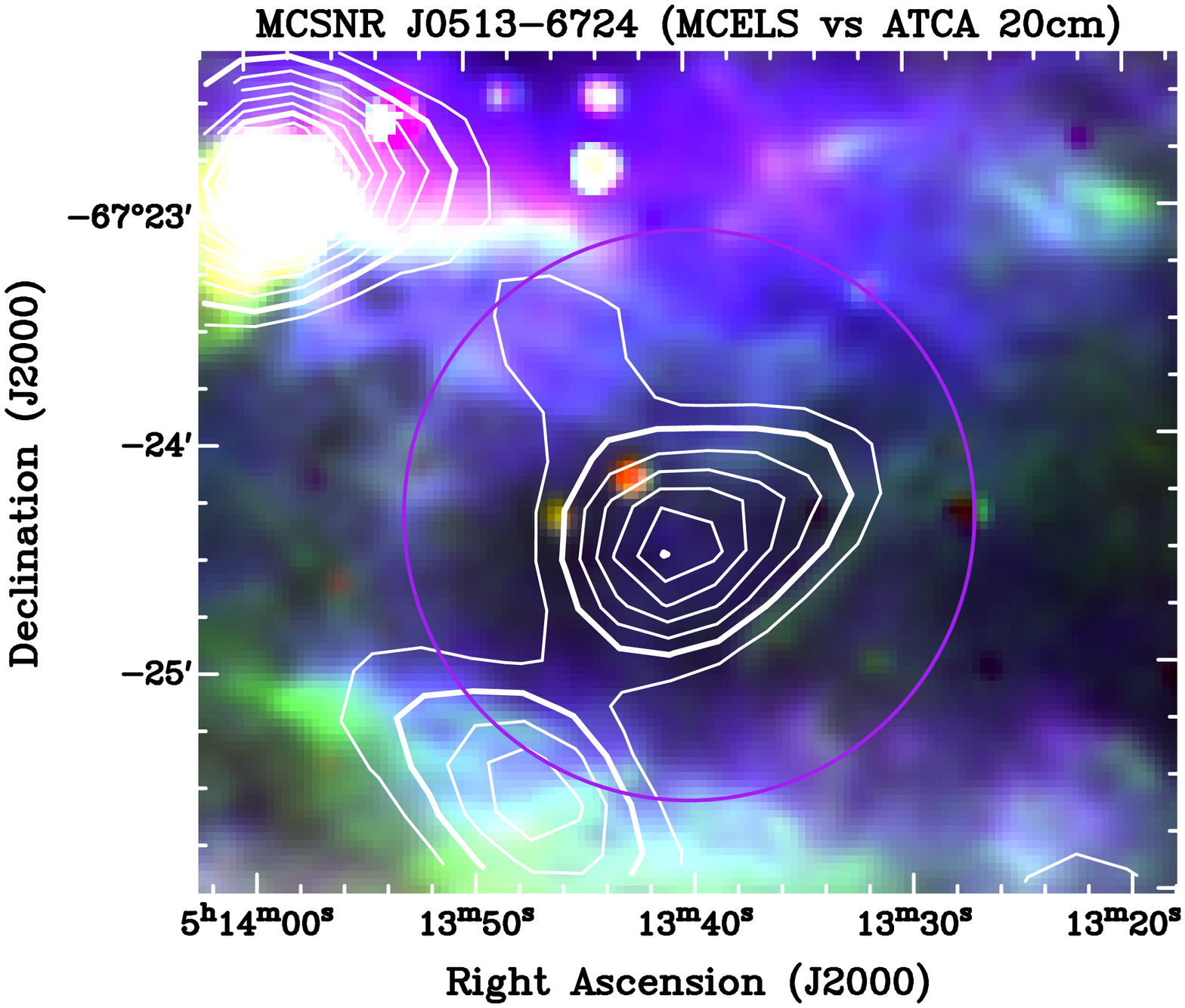}}
  \resizebox{0.8\columnwidth}{!}{\includegraphics[angle=-90,trim = 0 0 0 0,clip]{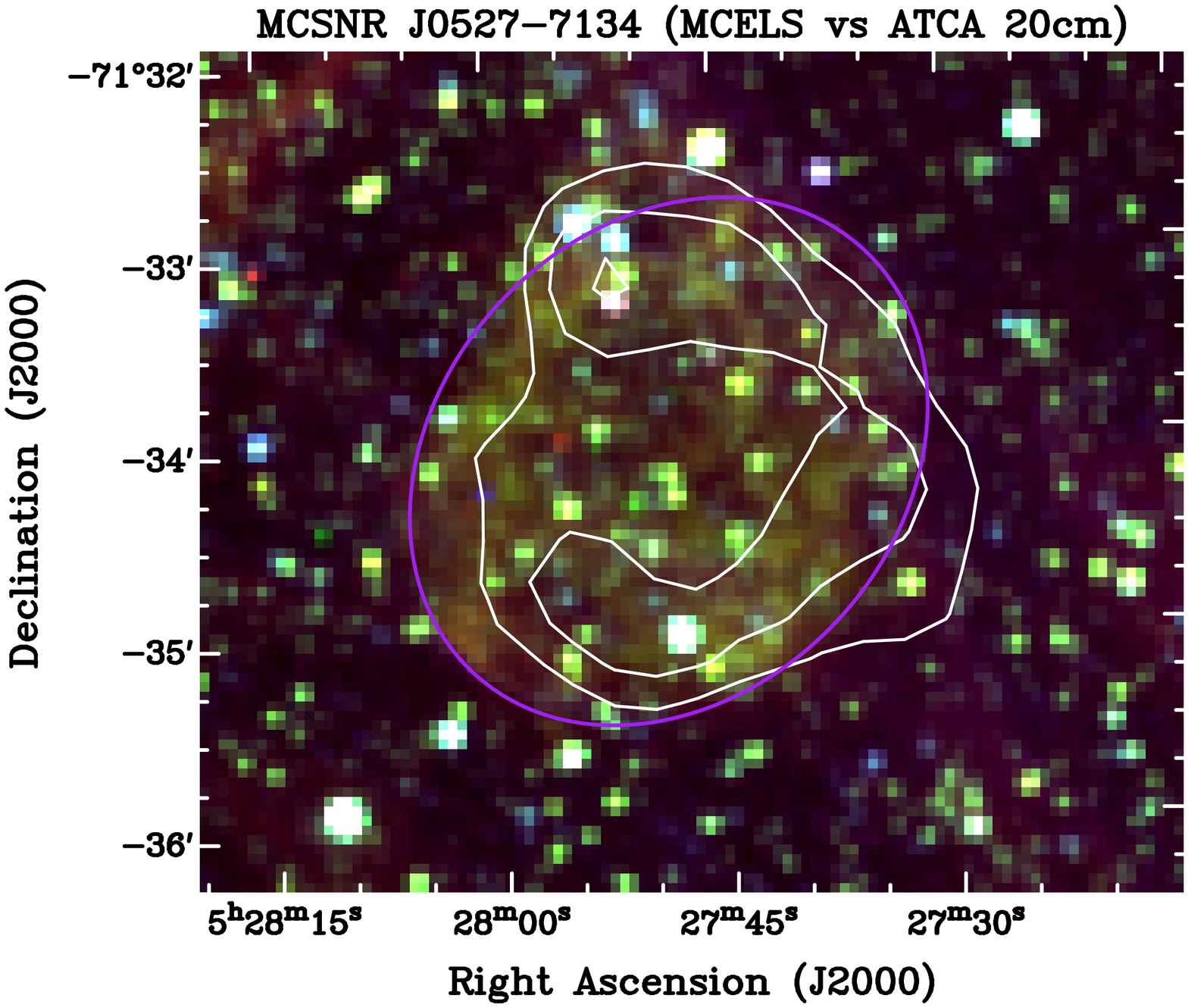}}
  \caption{The LMC SNR candidates 2. Colour images are MCELS where RGB corresponds to H$\alpha$, \SII\ and \OIII. The MCSNR\,J0512-6716 colour image is from the \xmm\ X-ray survey of the LMC SNRs \citep{2015A&A...583A.121K}. Contours are from the ATCA 20~cm mosaic survey and start at the 3$\sigma$ local noise level with spacing of 1$\sigma$. The circles/ellipses (purple) represent approximate extent of the SNR candidates.}
  \label{snrcandidates2}
 \end{center}
\end{figure*}

\noindent\textbf{MCSNR\,J0507-7110 = DEM\,L81} (Figure~\ref{snrcandidates2}; top-left) -- \citet{1976MmRAS..81...89D} listed this object as DEM\,L81, describing the source as a faint semicircular arc extending  $4.5\arcmin$. We identified a southern arc in the MCELS images ahead of the extended radio-continuum emission. As the whole source is very complex, we could not measure any flux density at our radio frequencies. However, the source appears stronger at lower frequencies and is indicative of a steep non-thermal radio-continuum spectrum. Also, there is no sensitive X-ray coverage in this field.

\noindent\textbf{MCSNR\,J0510-6708} (Figure~\ref{snrcandidates2}; top-right) -- We determined an enhanced \SII /H$\alpha$ ratio of $>$0.5 in the shell. The radio-continuum emission is centrally located and very weak (S$_{20\,cm}$=2.5~mJy). At present, there is no sensitive X-ray coverage in the direction of this object.

\noindent\textbf{MCSNR\,J0512-6716} (Figure~\ref{snrcandidates2}; middle-left) -- A prominent X-ray ring from \xmm\ images can be seen with some very weak radio emission overlapping. We could not confirm optical identification of this object.

\noindent\textbf{MCSNR\,J0513-6731 = [HP99]~544} (Figure~\ref{snrcandidates2}; middle-right) -- HP99 named this object [HP99]~544, recording an extent of 27.4\arcsec, in addition to a HR1 measurement of $1.00\pm0.29$. A low likelihood for the extent leaves the possibility of an X-ray point source unrelated to the SNR candidate. We detected weak but distinctive \OIII\ emission surrounding a distinct radio source. The spectral index was determined to be $\alpha$=--0.56 based on measured flux densities of 28.7~mJy at 36~cm and 21.5~mJy at 20~cm.

\noindent\textbf{MCSNR\,J0513-6724 = [HP99]~530} (Figure~\ref{snrcandidates2}; bottom-left) -- HP99 gave this object the name [HP99]~530, recording an extent of 17.5\arcsec, in addition to HR1 and HR2 ratios of $1.00\pm0.21$ and $0.15\pm0.17$, respectively. Because these hardness ratios suggest a ``hard'' source and given also the low source extent likelihood, HP99 suggest the source could be unrelated to the SNR candidate. We found a strong radio point source with flux densities of 31~mJy (at 36~cm) and 23~mJy (at 20~cm) implying non-thermal spectral index of $\alpha$=--0.61. There is also a weak \SII\ ring, though somewhat smaller than the X-ray extent.

\noindent\textbf{MCSNR\,J0527-7134} (Figure~\ref{snrcandidates2}; bottom-right) -- We classified this object as an {SNR} candidate based on an enhanced \SII /H$\alpha$ ratio of $>$0.4 as well as a shell-like radio-continuum morphology. We estimated flux densities of 31~mJy (at 36~cm) and 24~mJy (at 20~cm) implying a non-thermal spectral index of $\alpha$=--0.52. This candidate was observed very recently with $\xmm$ (October 2016, PI: P. Kavanagh). The detection of soft X-ray emission correlated with the optical and radio shells suggests this source as a \textit{bona-fide} SNR. A detailed study of MCSNR\,J0527-7134 will be presented elsewhere.

\begin{figure*}
 \begin{center}
  \resizebox{0.8\columnwidth}{!}{\includegraphics[angle=-90,trim = 0 0 0 0,clip]{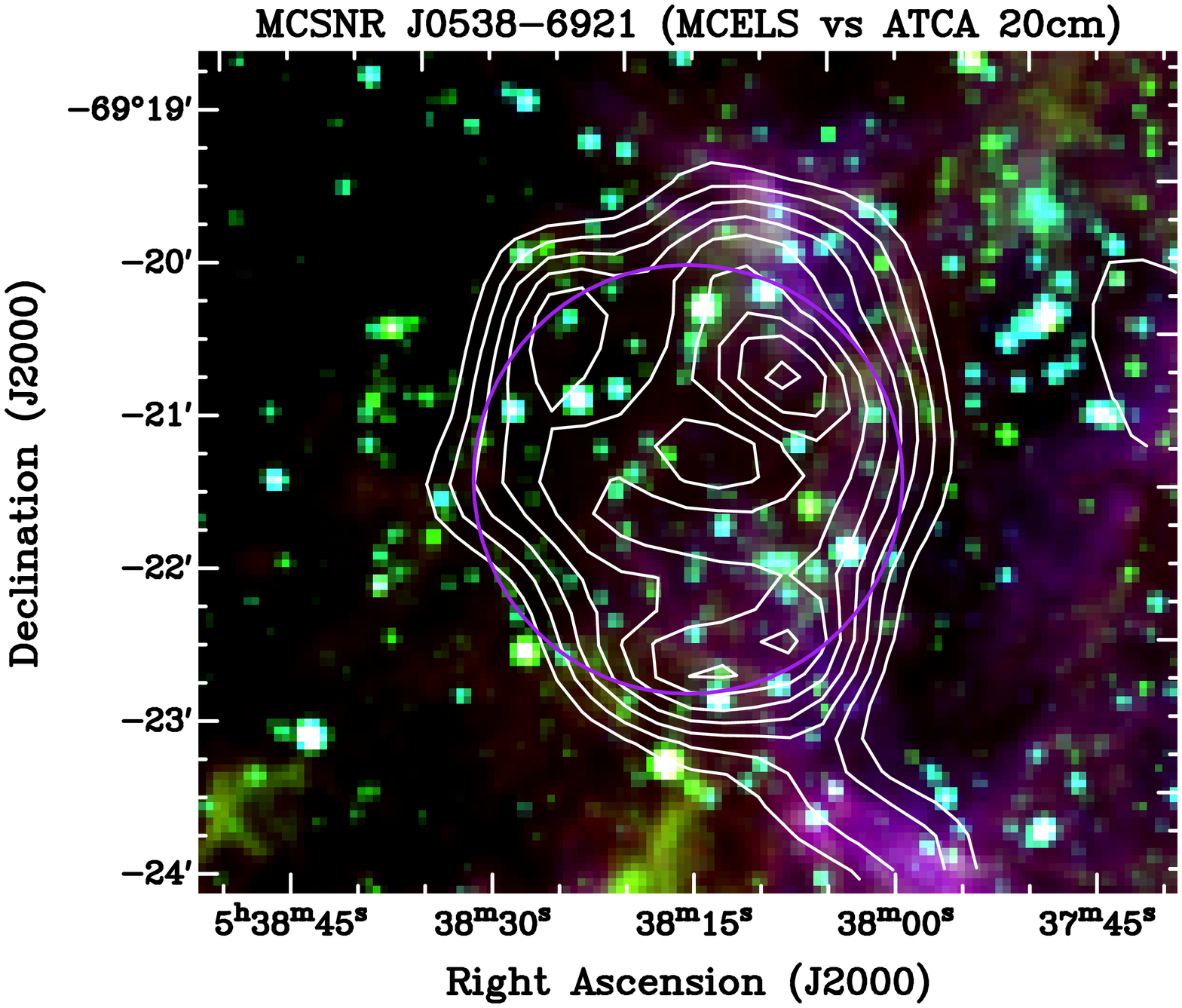}}
  \resizebox{0.8\columnwidth}{!}{\includegraphics[angle=-90,trim = 0 0 0 0,clip]{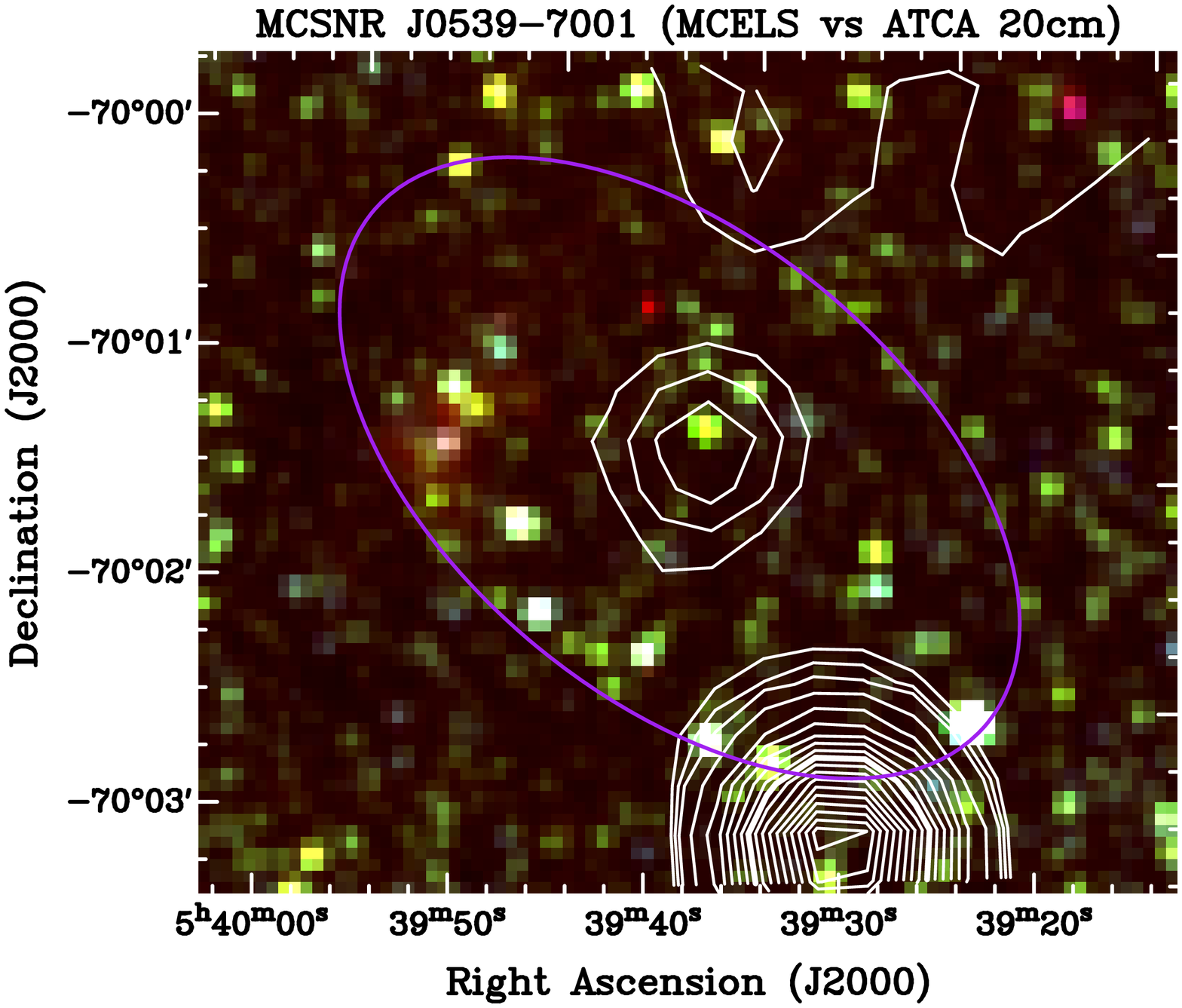}}
  \caption{The LMC SNR candidates 3. Colour images are MCELS where RGB corresponds to H$\alpha$, \SII\ and \OIII. Contours are from the ATCA 20~cm mosaic survey and start at the 3$\sigma$ local noise level with spacing of 1$\sigma$. The ellipses (purple) represent approximate extent of the SNR candidates.}
  \label{snrcandidates3}
 \end{center}
\end{figure*}

\noindent\textbf{MCSNR\,J0538-6921} (Figure~\ref{snrcandidates3}; left) -- \citet{1984PASAu...5..537T} originally proposed this source to be an {SNR}. This is the only strong LMC {SNR} candidate to date that has been detected in radio frequencies alone. The estimated spectral index is $\alpha=-0.59\pm0.04$ based on measured flux densities at various frequencies.

\noindent\textbf{MCSNR\,J0539-7001 = [HP99]~1063} (Figure~\ref{snrcandidates3}; right) -- HP99 assigned this source with the name [HP99]~1063, recording an extent of 18.2\arcsec, in addition to listing HR1 and HR2 values of $1.00\pm0.17$ and $-0.17\pm0.10$, respectively. They classified the X-ray source, which was constant in flux during the \ROSAT\ observations, as an {SNR} candidate. We found a weak radio point source in the centre of this remnant with measured flux densities at 36~cm of 5.8~mJy and at 20~cm of 4.6~mJy, giving a spectral index of $\alpha$=--0.47.

%
%

\section{Physical properties from kernel density estimates}
  \label{kds}

This most complete sample of LMC SNRs allows us to better study their morphology, evolution and physical processes that are responsible for their observed emission. To get better statistical insight into the physical properties of the LMC SNRs, we reconstructed the probability density function (PDF) for diameter, flux density, spectral index and ovality. We note that the errors associated with the diameter ($<$2\arcsec) and flux density ($<$10\%) are small and not displayed in this analysis. For $n$ independent and identically distributed (iid) measurements of a variable $X(x_1, x_2, x_3, ..., x_n)$, the PDF ($f(x)$) is reconstructed using the Gaussian kernel smoothing \citep{Wasserman:2010:SCC:1965575,Feigelsonbabu2012astrostatistics,Sheather04densityestimation}, with the kernel function:
\begin{equation}
K\left(\frac{x-x_i}{h}\right) = \frac{1}{\sqrt{2\pi}}\exp{\frac{(x-x_i)^2}{2h^2}}.
\end{equation}

The common way to select the kernel bandwidth $h$ and obtain $f(x)$ estimate $f_h(x)$ as:
\begin{equation}
f_h(x) = \frac{1}{nh}\sum_{i=1}^{n} K\left(\frac{x-x_i}{h}\right),
\end{equation}
is to minimise the mean integrated square error (MISE):
\begin{equation}
MISE(h) = \int _{-\infty}^{\infty}(f_h(x) - f(x))^2 dx.
\end{equation}
For a Gaussian kernel, under asymptotic conditions, $n \rightarrow \infty$ and $h \rightarrow 0$, such that $nh \rightarrow \infty$, $MISE(h)$ translates to asymptotic $MISE$ ($AMISE$):
\begin{equation}
AMISE(h) = \frac{1}{4}h^4 R(f) + \frac{h}{2\sqrt{\pi} n},
\label{amise}
\end{equation}
where $R(f)$ is the roughness of $f$ calculated as:
\begin{equation}
R(f) = \int _{-\infty}^{\infty} f''(x)^2 dx.
\end{equation}
It can be shown that $AMISE(h)$ has the minimum at:
\begin{equation}
h = (2\sqrt{\pi} n R(f))^{-0.2} .
\label{amise_h}
\end{equation}

To estimate $ f''(x) = d^2f(x)/dx^2, $ a ``plug-in''  bandwidth value $h_0$ is required so $f_{h_0}''(x)$ can be calculated. Usually, a rule-of-thumb value can be used \citep{silverman1986density}. If $\sigma$ is the standard deviation of the $n$ data points sample, the rule-of-thumb bandwidth calculates as $h_0 = \sigma n^{-0.2}$. This value is obtained by minimising $MISE$ as described above under the additional assumption that the data are normally distributed.

\begin{figure*}[]
\includegraphics[width = 0.4\textwidth,clip]{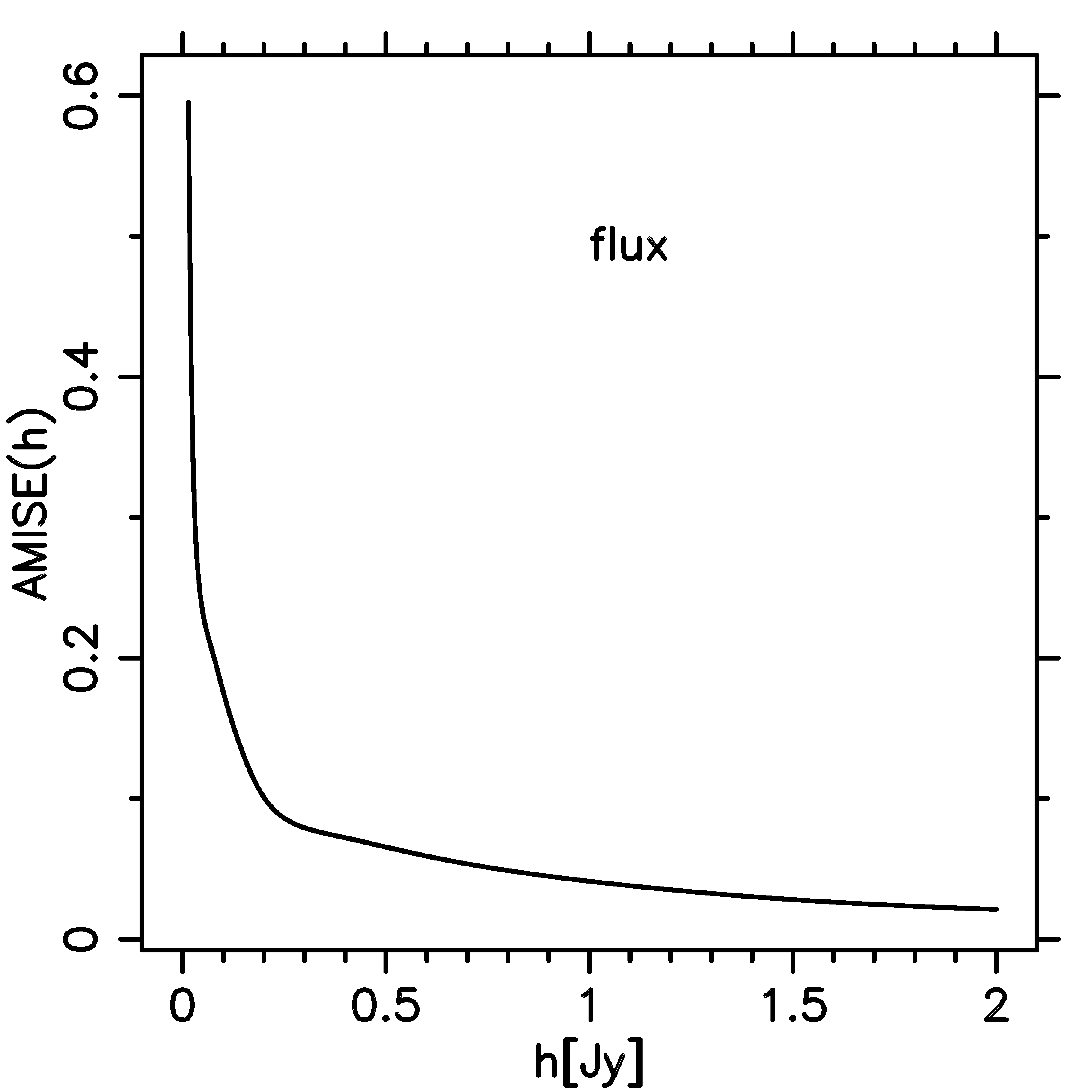}
\includegraphics[width = 0.4\textwidth,clip]{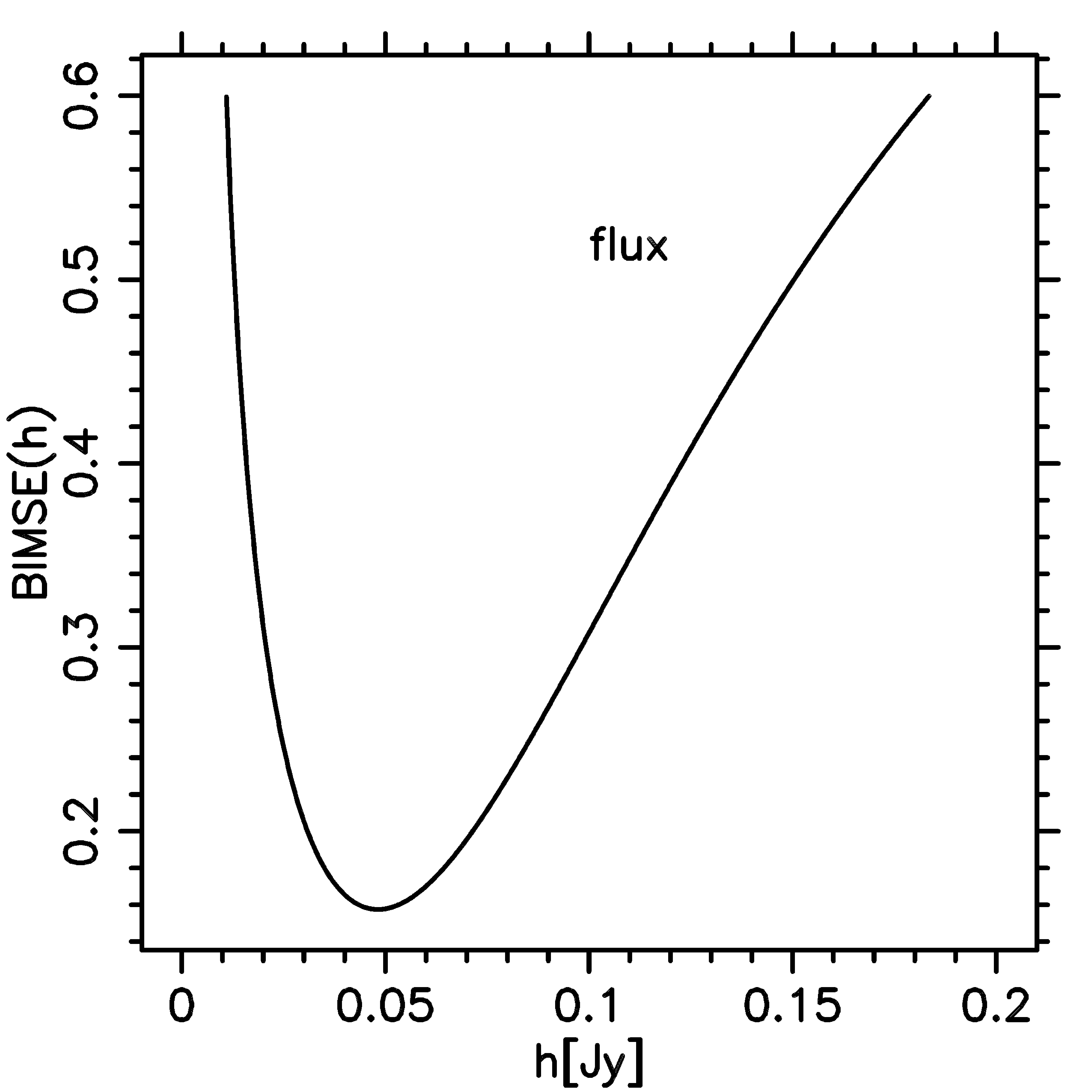}
\caption{Asymptotic mean integrated squared error (AMISE) and bootstrap integrated mean squared error (BIMSE) for the LMC flux density data at 1~GHz, shown in Figure~\ref{fig:fluxhist}. {Unlike AMISE, the BIMSE merit estimator shows a clear minima at the $h$ value which is considered as an optimal smoothing bandwidth.}}
\label{fig_amise_bimse}
\end{figure*}

However, the $h_0$ optimal for calculating $f(x)$ need not be an optimal choice for estimating $f''(x)$. In addition, the asymptotic nature of Eq.~\ref{amise_h} may give incorrect results when there is a lot of fine q in the data. We illustrate this in Figure~\ref{fig_amise_bimse}. The AMISE data {does not
appear to have a noticeable minima value.}

As a more robust way of estimating $h$ we used a procedure similar to that described in \citet{Faraway:1990:BCB}. Instead of minimising $AMISE(h)$, they minimised the bootstrap mean integrated square error:
\begin{equation}
BIMSE(h) = B^{-1}\sum\limits_{i=1}\limits^{B}\int _{-\infty}^{\infty}(f^*_h(x) - f(x))^2 dx.
\end{equation}
where $B$ is the number of smooth bootstrap re-samplings from the original data sample. At a given $h$ each re-sample gives $f^*_h(x)$. Unlike the common bootstrap \citep{tEFR93a}, the smooth bootstrap also requires a ``plug-in'' bandwidth $h_0$ for re-sampling and estimating $f(x)$. To each re-sampled data point $x^*_i$ an offset is added as $x^{*}_{\mathrm{sb}~i} = x^*_i + \theta(h_0,x^*_i)$, where $\theta$ is normally distributed with standard deviation $h_0$ around a mean $x^*_i$. The $x^{*}_{\mathrm{sb}~i}$ values obtained in this manner are then used to calculate $f^*_h(x)$. The $BIMSE(h)$ in Figure~\ref{fig_amise_bimse} is calculated with optimal bandwidth resulting from the procedure described below, the results of which are presented in Section~\ref{sec_flux_density}. Unlike $AMISE$, it shows a very distinctive, unambiguous, minimum value of global character. Although the $BIMSE$ calculation is much more intensive than the calculation of $AMISE$, we find it significantly more robust and reliable for estimating optimal smoothing bandwidths. Since AMISE data do not appear to have a strong minimum value in many of the cases examined in this work, we adopted the BIMSE procedure for estimating optimal smoothing bandwidths {which (as evident from Figure~\ref{fig_amise_bimse}) shows a distinctive BIMSE(h) minima value even on an order of magnitude smaller scale than AMISE(h).}

\subsection{Procedure description}
Using the rule-of-thumb bandwidth we made smooth bootstrap re-samplings and estimated $f(x)$ from the original data points. To minimise $BIMSE(h)$ we used a golden section search algorithm \citep{KieferFibonacciSearch}, that narrows down the interval that contains the minima value by comparing the $BIMSE(h)$ values at four points. To reduce the influence of Monte Carlo error on the minimising procedure we allowed the $h$ values within the algorithm to change only in increments of $h_\mathrm{s} = 10^{\mu(h_0) - 2}$, where $\mu$ presents the order of magnitude of its argument. The search was stopped when the size of the $h$ interval containing the minima fell below $h_\mathrm{s}$.  At each given $h$ value the $f^*_h(x)$ was calculated for all re-samplings and the $(f^*_h(x) - f(x))^2$ term was averaged to obtain $BIMSE(h)$.

In the work of \citet{Faraway:1990:BCB} the authors stated that their procedure can be used in an iterative manner and hence improve the values obtained for an optimal smoothing bandwidth. We iterated the above procedure until the difference between the input and output $h$ value fell below $h_\mathrm{s}$ or until this difference $h_\mathrm{dif}$ changes sign in which case we took the input value $h_\mathrm{in}$ of the final iteration as the outcome of the procedure. Even in such a case, $h$ is calculated with $\approx \mu(h_\mathrm{in})-1$ order of magnitude accuracy. The change of sign in $h_\mathrm{dif}$ can be avoided and the accuracy improved, simply by applying a larger number of bootstrap re-samplings. However, \citet{Faraway:1990:BCB} noted that, at the time, their procedure was very computationally intensive and that they obtained satisfactory results using only $100$ re-samplings. The calculations described in this work took up to $\approx 1$ day of computing per examined data sample on a standard desktop PC using $500$ bootstrap re-samplings. However, this computing time was dominated by repeating the procedure for computing confidence bands (described below) which required much greater computation intensity. To increase the speed of the computations we calculated the Gaussian distribution values only within the five standard deviations from the mean considering the values outside this interval to be zero.

\subsection{Confidence bands and parameters}
\label{kde_conf}
\citet{Faraway:1990:BCB} noted that a by-product of their procedure is the estimation of confidence intervals as desired quantiles of the $f^*_{\hat{h}}(x)$ distributions, where $\hat{h}$ is the output optimal bandwidth of their smoothed bootstrap procedure. However, the confidence band derived in this manner depends on $\hat{h}$ which depends on $h_0$. To avoid this we applied a common bootstrap re-sampling (without the addition of a smoothing term $\theta$) to the original data sample. For each of the re-samples obtained in this way, we applied the described smoothed bootstrap procedure to obtain $f^*_{\hat{h_\bullet}}(x)$, where the bullet sign designates that an optimal smoothing bandwidth was obtained for the common bootstrap re-sample of the original data sample. At each selected point $x_j$ along $x$-axis, we then calculated the desired confidence fraction $f_\mathrm{ci}$ of $f^*_{\hat{h_\bullet}}(x_j)$ values. First we calculated the median of the given $f^*_{\hat{h_\bullet}}(x_j)$ array of values and then upper and lower limits of confidence bands as $f_\mathrm{ci} / 2$ fraction of the total number of array elements from the median value. The median value was also used as the value of the smoothed density distribution $\hat{f}(x_j)$ at a point $x_j$.

If the testing $x_\mathrm{min}$ and $x_\mathrm{max}$ are the lowest and highest values of the data sample, respectively, the density distributions were calculated in the $[x_\mathrm{min} - 5 * h_0, x_\mathrm{max} + 5 * h_0]$ interval at the centres of $10^3$ bins of equal width. The confidence bands are calculated as an $f_\mathrm{ci}=0.75$ fraction of the total number of $500$ common bootstrap re-samplings ($\bullet$) of the original data sample. We also used $500$ smooth bootstrap re-samplings ($*$) for each ($\bullet$) re-sampling. For each ($\bullet$) distribution we calculated the mean, mode and median. The uncertainties of the mean, mode and median were calculated similar to the confidence bands for the smooth density distribution, and we give the confidence band with higher discrepancy from the median as the uncertainty. Confidence bands were obtained also at $f_\mathrm{ci}=0.75$, similar to the $\hat{f}(x_j)$. The smooth density functions, with their parameters obtained as described above, for diameter, radio flux, spectral index and ovality, are presented in Section~\ref{results}.

\subsection{Kernel density smoothing in 2D}
\label{kds_2d}
A similar procedure as described in this section can be generalised to a 2D case. We applied the procedure in 2D to the LMC and SMC radio surface brightness and diameter data to check if there were any significant emergent data features (Figure~\ref{lmc-track}). The generalisation of the procedure can be done in two ways. Either by using a 2D Gaussian kernel (where a kernel is a function of two variables and two smoothing bandwidths), or as a product kernel of separate Gaussian kernels in each dimension. We adopted the latter. As described in \citet{Feigelsonbabu2012astrostatistics}, for a 2D case the density estimate in (x,y) space is calculated as:
\begin{equation}
\small
f_{h^xh^y}(x,y) = \frac{1}{n h^x h^y} \sum_{i=1}^{n}K\left(\frac{x-x_i}{h^x}\right) K\left(\frac{y-y_i}{h^y}\right),
\end{equation}
where $h^x$ and $h^y$ are bandwidths for the kernel components in $x$ and $y$, respectively. For the 2D density estimate procedure, only the smoothed bootstrap was used to minimise over BIMSE. Due to higher computation requirements, we did not use common bootstrap re-samplings to estimate confidence intervals. For the input values to the procedure $h^x_0$ and $h^y_0$ (similar to 1D case), we used rule of thumb bandwidths calculated separately in each dimension. For the range of $h^x$ and $h^y$ we calculated and minimised $BIMSE$. The values of $h^x$ and $h^y$ where $BIMSE$ reached a minimum were used as the next iteration input values. The procedure was repeated until the output values for $h^x$ and $h^y$ equal the input values at the given level of accuracy.

%
%

\section{Results}
 \label{results}

\begin{figure*}
 \begin{center}
\resizebox{2\columnwidth}{!}{\includegraphics[angle=-90, trim = 0 0 0 0,clip]{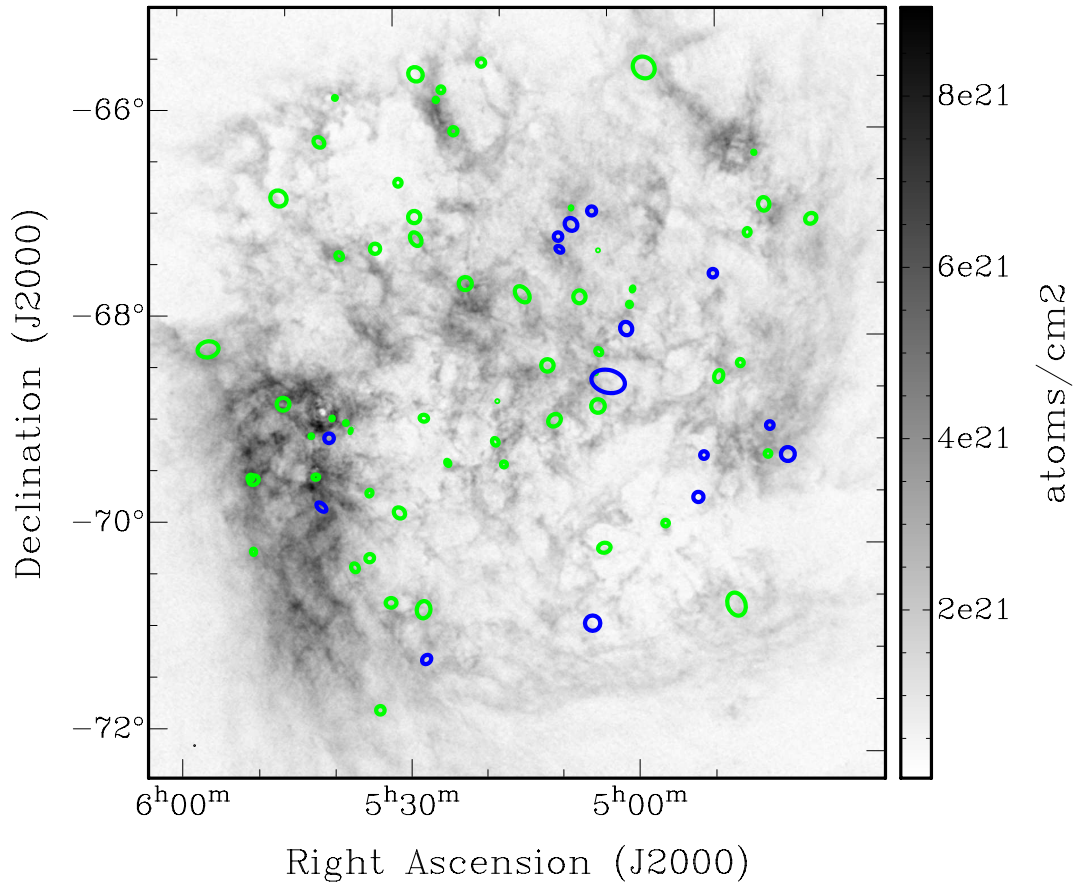}}
  \caption{Spatial distribution of the \confcount\ confirmed and \candcount\ candidate SNRs in the LMC. There appears to be a connection between the location of the remnants and the spiral pattern of the emission from the \HI\ peak temperature map \citep[grey scale]{1998ApJ...503..674K}. Green symbols represent confirmed remnants, while blue symbols show the position of candidate remnants.}
  \label{spatialdist}
 \end{center}
\end{figure*}

Here we present a statistical analysis and discussion of the \confcount\ confirmed (Table~\ref{tab:lmcsnrs}) and \candcount\ candidate {SNRs} (Table~\ref{tbl:candsnrs}) in the {LMC} (Figure~\ref{spatialdist}). This includes \typeiacount\ known SNRs resulting from a thermonuclear SN (type~Ia) explosion and \typecccount\ confirmed to arise from a core-collapse (CC) SN event \citep[and reference therein]{maggi16}.

\subsection{Spatial Distribution}
 \label{Sect_SD}

To investigate the spatial distribution of SNRs in the LMC, annotations containing the size and position angle of the \confcount\ confirmed and \candcount\ candidate remnants were superimposed on the \HI\ peak temperature map from \citet{1998ApJ...503..674K}. There is an indication of a connection between the higher \HI\ density (also including H$\alpha$ and radio-continuum) and the location of SNRs, which seems to follow a spiral structure \citep{1998A&AS..130..421F}. The mean foreground \HI\ column density in the direction to the \confcount\ confirmed and \candcount\ candidate remnants was estimated to be $\sim2\times 10^{21}$~atoms~cm$^{-2}$ (with a Standard Deviation SD=1$\times$10$^{21}$~atoms~cm$^{-2}$), while the ``empty'' LMC\,4 supershell \citep{2012MNRAS.420.2588B} exhibits a mean \HI\ density of 5$\times$10$^{20}$~atoms~cm$^{-2}$ (SD=1$\times$10$^{20}$~atoms~cm$^{-2}$). Curiously, we found only one SNR (MCSNR\,J0529-6653 inside the LMC4 supershell) to be located outside the apparent spiral structure of the \HI\ distribution. Also, MCSNR\,J0527-6549 expands in a very rarified environment with the mean \HI\ column density of 6$\times$10$^{20}$~atoms~cm$^{-2}$. We note that our \typeiacount\ type~Ia SNR sample might be expanding in a somewhat lower density environment (mean=1.9$\times$10$^{21}$~atoms~cm$^{-2}$) while the CC sample of \typecccount\ LMC SNRs exhibits a mean \HI\ column density of 2.4$\times$10$^{21}$~atoms~cm$^{-2}$. However, SDs of both samples are quite large (SD=1.1$\times$10$^{21}$~atoms~cm$^{-2}$). Therefore, this is indicative of a different molecular environment in which type~Ia and CC are expanding, although it should be taken with caution.

\subsection{Multi-frequency Emission Comparison}
 \label{Sect_MFE}

\begin{figure}
\centering
\def\firstcircle{(0,0) circle (1.5cm)}
\def\secondcircle{(1,1.5) circle (1.5cm)}
\def\thirdcircle{(0:2cm) circle (1.5cm)}
\resizebox{1\columnwidth}{!}{\begin{tikzpicture}
    \begin{scope}[shift={(3cm,-4cm)}, fill opacity=0.25]
        \fill[white] \firstcircle;
        \fill[white] \secondcircle;
        \fill[white] \thirdcircle;
        \draw \firstcircle node[below] {\textit{}};
        \draw \secondcircle node [above] {$$};
        \draw \thirdcircle node [below] {$$};
        \node[color=black, opacity=1] at (-0.3,-1) {\textit{Optical}};
        \node[color=black, opacity=1] at (2.3,-1) {\textit{Radio}};
        \node[color=black, opacity=1] at (1,2.5) {\textit{X-ray}};
        \node[color=black, opacity=1] at (1,.7) {{47}};
        \node[color=black, opacity=1] at (1,.4) {{\tiny{3 (X)}}};
        \node[color=black, opacity=1] at (1,.2) {{\tiny{2 (O)}}};
        \node[color=black, opacity=1] at (1,2) {{3}};	
        \node[color=black, opacity=1] at (1.8,0.9) {{3}}; 
        \node[color=black, opacity=1] at (0.2,0.9) {{1}};
        \node[color=black, opacity=1] at (-0.3,-0.3) {{0}};
        \node[color=black, opacity=1] at (2.3,-0.3) {{0}};
        \node[color=black, opacity=1] at (1,-0.4) {{0}};
    \end{scope}
\end{tikzpicture}
}
\caption{A Venn diagram showing the \confcount\ confirmed LMC SNRs in different electromagnetic domains. In the centre pane, the 3 (X) and 2 (O), where X = X-ray and O = optical, shows those remnants which either lack observations, or are entangled in unassociated emission, though if present, would reside in this group.}
 \label{venndiagram}
\end{figure}
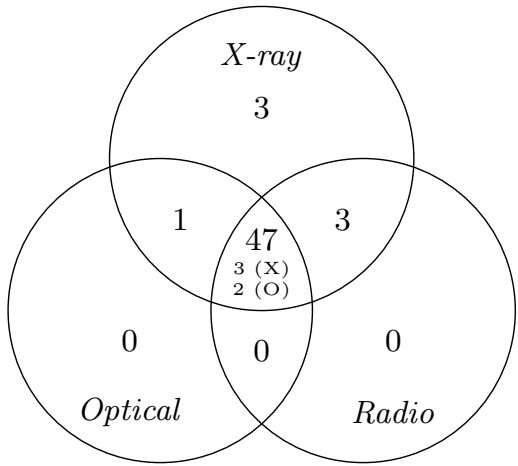

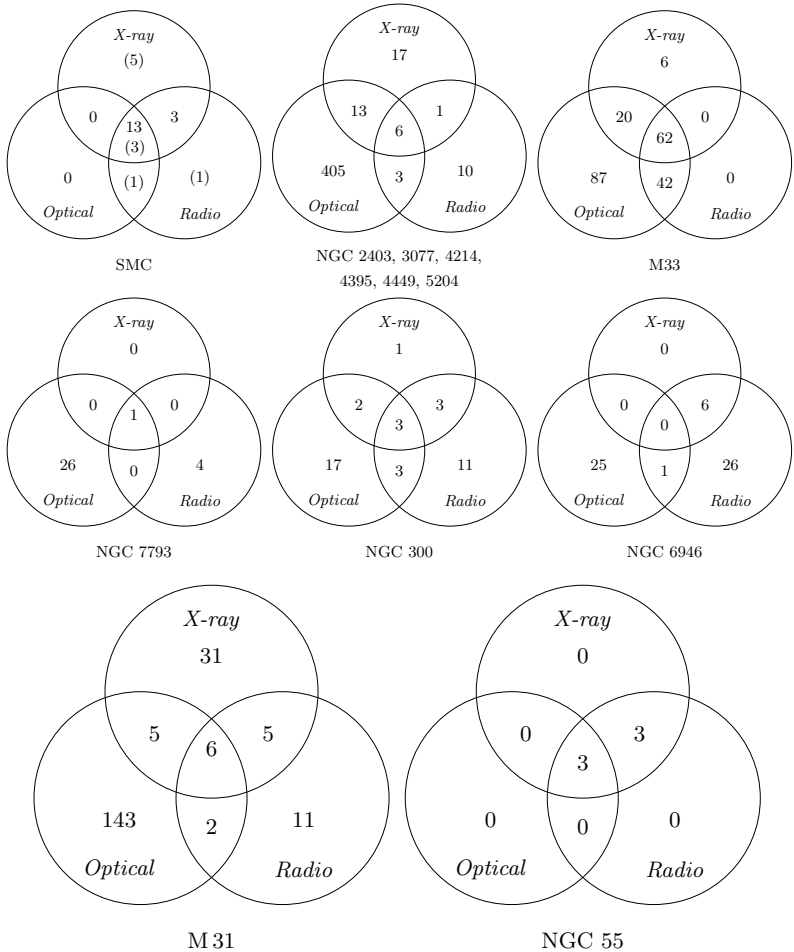
\begin{figure*}[]
\centering
\def\firstcircle{(0,0) circle (1.5cm)}
\def\secondcircle{(1,1.5) circle (1.5cm)}
\def\thirdcircle{(0:2cm) circle (1.5cm)}
\resizebox{1.5\columnwidth}{!}{\begin{tikzpicture}
    \begin{scope}[shift={(3cm,-4cm)}, fill opacity=0.25]
        \fill[white] \firstcircle;
        \fill[white] \secondcircle;
        \fill[white] \thirdcircle;
        \draw \firstcircle node[below] {\textit{}};
        \draw \secondcircle node [above] {$$};
        \draw \thirdcircle node [below] {$$};
        \node[color=black, opacity=1] at (-0.3,-1) {\textit{Optical}};
        \node[color=black, opacity=1] at (2.3,-1) {\textit{Radio}};
        \node[color=black, opacity=1] at (1,2.5) {\textit{X-ray}};
        \node[color=black, opacity=1] at (1,.7) {{13}};		
        \node[color=black, opacity=1] at (1.02,.3) {{{(3)}}};	
        \node[color=black, opacity=1] at (1,2) {{(5)}};		
        \node[color=black, opacity=1] at (1.8,0.9) {{3}}; 		
        \node[color=black, opacity=1] at (0.2,0.9) {{0}};		
        \node[color=black, opacity=1] at (-0.3,-0.3) {{0}};	
        \node[color=black, opacity=1] at (2.3,-0.3) {{(1)}};	
        \node[color=black, opacity=1] at (1,-0.4) {{(1)}};		
        \node[color=black, opacity=1] at (1,-2) {{SMC}};		%
        \node[color=black, opacity=1] at (1,-2.5) {{}};		%
    \end{scope}
\end{tikzpicture}
\begin{tikzpicture}
    \begin{scope}[shift={(3cm,-4cm)}, fill opacity=0.25]
        \fill[white] \firstcircle;
        \fill[white] \secondcircle;
        \fill[white] \thirdcircle;
        \draw \firstcircle node[below] {\textit{}};
        \draw \secondcircle node [above] {$$};
        \draw \thirdcircle node [below] {$$};
        \node[color=black, opacity=1] at (-0.3,-1) {\textit{Optical}};
        \node[color=black, opacity=1] at (2.3,-1) {\textit{Radio}};
        \node[color=black, opacity=1] at (1,2.5) {\textit{X-ray}};
        \node[color=black, opacity=1] at (1,.5) {{6}};		
        \node[color=black, opacity=1] at (1,2) {{17}};		
        \node[color=black, opacity=1] at (1.8,0.9) {{1}}; 		
        \node[color=black, opacity=1] at (0.2,0.9) {{13}};		
        \node[color=black, opacity=1] at (-0.3,-0.3) {{405}};	
        \node[color=black, opacity=1] at (2.3,-0.3) {{10}};	
        \node[color=black, opacity=1] at (1,-0.4) {{3}};		
        \node[color=black, opacity=1] at (1,-2) {{NGC 2403, 3077, 4214,}};		%
        \node[color=black, opacity=1] at (1,-2.5) {{4395, 4449, 5204}};		%
    \end{scope}
\end{tikzpicture}
\begin{tikzpicture}
    \begin{scope}[shift={(3cm,-4cm)}, fill opacity=0.25]
        \fill[white] \firstcircle;
        \fill[white] \secondcircle;
        \fill[white] \thirdcircle;
        \draw \firstcircle node[below] {\textit{}};
        \draw \secondcircle node [above] {$$};
        \draw \thirdcircle node [below] {$$};
        \node[color=black, opacity=1] at (-0.3,-1) {\textit{Optical}};
        \node[color=black, opacity=1] at (2.3,-1) {\textit{Radio}};
        \node[color=black, opacity=1] at (1,2.5) {\textit{X-ray}};
        \node[color=black, opacity=1] at (1,.5) {{62}};		
        \node[color=black, opacity=1] at (1,2) {{6}};		
        \node[color=black, opacity=1] at (1.8,0.9) {{0}}; 		
        \node[color=black, opacity=1] at (0.2,0.9) {{20}};		
        \node[color=black, opacity=1] at (-0.3,-0.3) {{87}};	
        \node[color=black, opacity=1] at (2.3,-0.3) {{0}};	
        \node[color=black, opacity=1] at (1,-0.4) {{42}};		
        \node[color=black, opacity=1] at (1,-2) {{M33}};		%
        \node[color=black, opacity=1] at (1,-2.5) {{}};		%
    \end{scope}
\end{tikzpicture}
}
\resizebox{1.5\columnwidth}{!}{\begin{tikzpicture}
    \begin{scope}[shift={(3cm,-4cm)}, fill opacity=0.25]
        \fill[white] \firstcircle;
        \fill[white] \secondcircle;
        \fill[white] \thirdcircle;
        \draw \firstcircle node[below] {\textit{}};
        \draw \secondcircle node [above] {$$};
        \draw \thirdcircle node [below] {$$};
        \node[color=black, opacity=1] at (-0.3,-1) {\textit{Optical}};
        \node[color=black, opacity=1] at (2.3,-1) {\textit{Radio}};
        \node[color=black, opacity=1] at (1,2.5) {\textit{X-ray}};
        \node[color=black, opacity=1] at (1,.7) {{1}};		
        \node[color=black, opacity=1] at (1,2) {{0}};		
        \node[color=black, opacity=1] at (1.8,0.9) {{0}}; 		
        \node[color=black, opacity=1] at (0.2,0.9) {{0}};		
        \node[color=black, opacity=1] at (-0.3,-0.3) {{26}};	
        \node[color=black, opacity=1] at (2.3,-0.3) {{4}};	
        \node[color=black, opacity=1] at (1,-0.4) {{0}};		
        \node[color=black, opacity=1] at (1,-2) {{NGC~7793}};		%
        \node[color=black, opacity=1] at (1,-2.5) {{}};		%
    \end{scope}
\end{tikzpicture}
\begin{tikzpicture}
    \begin{scope}[shift={(3cm,-4cm)}, fill opacity=0.25]
        \fill[white] \firstcircle;
        \fill[white] \secondcircle;
        \fill[white] \thirdcircle;
        \draw \firstcircle node[below] {\textit{}};
        \draw \secondcircle node [above] {$$};
        \draw \thirdcircle node [below] {$$};
        \node[color=black, opacity=1] at (-0.3,-1) {\textit{Optical}};
        \node[color=black, opacity=1] at (2.3,-1) {\textit{Radio}};
        \node[color=black, opacity=1] at (1,2.5) {\textit{X-ray}};
        \node[color=black, opacity=1] at (1,.5) {{3}};		
        \node[color=black, opacity=1] at (1,2) {{1}};		
        \node[color=black, opacity=1] at (1.8,0.9) {{3}}; 		
        \node[color=black, opacity=1] at (0.2,0.9) {{2}};		
        \node[color=black, opacity=1] at (-0.3,-0.3) {{17}};	
        \node[color=black, opacity=1] at (2.3,-0.3) {{11}};	
        \node[color=black, opacity=1] at (1,-0.4) {{3}};		
        \node[color=black, opacity=1] at (1,-2) {{NGC~300}};		%
        \node[color=black, opacity=1] at (1,-2.5) {{}};		%
    \end{scope}
\end{tikzpicture}
\begin{tikzpicture}
    \begin{scope}[shift={(3cm,-4cm)}, fill opacity=0.25]
        \fill[white] \firstcircle;
        \fill[white] \secondcircle;
        \fill[white] \thirdcircle;
        \draw \firstcircle node[below] {\textit{}};
        \draw \secondcircle node [above] {$$};
        \draw \thirdcircle node [below] {$$};
        \node[color=black, opacity=1] at (-0.3,-1) {\textit{Optical}};
        \node[color=black, opacity=1] at (2.3,-1) {\textit{Radio}};
        \node[color=black, opacity=1] at (1,2.5) {\textit{X-ray}};
        \node[color=black, opacity=1] at (1,.5) {{0}};		
        \node[color=black, opacity=1] at (1,2) {{0}};		
        \node[color=black, opacity=1] at (1.8,0.9) {{6}}; 		
        \node[color=black, opacity=1] at (0.2,0.9) {{0}};		
        \node[color=black, opacity=1] at (-0.3,-0.3) {{25}};	
        \node[color=black, opacity=1] at (2.3,-0.3) {{26}};	
        \node[color=black, opacity=1] at (1,-0.4) {{1}};		
        \node[color=black, opacity=1] at (1,-2) {{NGC 6946}};		%
        \node[color=black, opacity=1] at (1,-2.5) {{}};		%
    \end{scope}
\end{tikzpicture}
}
\resizebox{1.4\columnwidth}{!}{\begin{tikzpicture}
    \begin{scope}[shift={(3cm,-4cm)}, fill opacity=0.25]
        \fill[white] \firstcircle;
        \fill[white] \secondcircle;
        \fill[white] \thirdcircle;
        \draw \firstcircle node[below] {\textit{}};
        \draw \secondcircle node [above] {$$};
        \draw \thirdcircle node [below] {$$};
        \node[color=black, opacity=1] at (-0.3,-1) {\textit{Optical}};
        \node[color=black, opacity=1] at (2.3,-1) {\textit{Radio}};
        \node[color=black, opacity=1] at (1,2.5) {\textit{X-ray}};
        \node[color=black, opacity=1] at (1,.7) {{6}};		
        \node[color=black, opacity=1] at (1,2) {{31}};		
        \node[color=black, opacity=1] at (1.8,0.9) {{5}}; 		
        \node[color=black, opacity=1] at (0.2,0.9) {{5}};		
        \node[color=black, opacity=1] at (-0.3,-0.3) {{143}};	
        \node[color=black, opacity=1] at (2.3,-0.3) {{11}};	
        \node[color=black, opacity=1] at (1,-0.4) {{2}};		
        \node[color=black, opacity=1] at (1,-2) {{M\,31}};		%
        \node[color=black, opacity=1] at (1,-2.5) {{}};		%
    \end{scope}
\end{tikzpicture}
\begin{tikzpicture}
    \begin{scope}[shift={(3cm,-4cm)}, fill opacity=0.25]
        \fill[white] \firstcircle;
        \fill[white] \secondcircle;
        \fill[white] \thirdcircle;
        \draw \firstcircle node[below] {\textit{}};
        \draw \secondcircle node [above] {$$};
        \draw \thirdcircle node [below] {$$};
        \node[color=black, opacity=1] at (-0.3,-1) {\textit{Optical}};
        \node[color=black, opacity=1] at (2.3,-1) {\textit{Radio}};
        \node[color=black, opacity=1] at (1,2.5) {\textit{X-ray}};
        \node[color=black, opacity=1] at (1,.5) {{3}};		
        \node[color=black, opacity=1] at (1,2) {{0}};		
        \node[color=black, opacity=1] at (1.8,0.9) {{3}}; 		
        \node[color=black, opacity=1] at (0.2,0.9) {{0}};		
        \node[color=black, opacity=1] at (-0.3,-0.3) {{0}};	
        \node[color=black, opacity=1] at (2.3,-0.3) {{0}};	
        \node[color=black, opacity=1] at (1,-0.4) {{0}};		
        \node[color=black, opacity=1] at (1,-2) {{NGC~55}};		%
        \node[color=black, opacity=1] at (1,-2.5) {{}};		%
    \end{scope}
\end{tikzpicture}
}
\caption{A series of Venn diagrams showing the detection of extragalactic SNRs in their host galaxies. The numbers in brackets denote candidate SNRs. \textit{Top Row:} Left: Data from the SMC \citep{2005MNRAS.364..217F,2012A&A...545A.128H} Middle: Data from six galaxies (NGC\,2403, 3077, 4214, 4395, 4449, and 5204 as described in \citealt{2013MNRAS.429..189L}). Right: Data from M\,33 (Long~et~al.~2016; priv. comm.). \textit{Middle row:} Left: Data from NGC\,7793 \citep{2011AJ....142...20P,2014Ap&SS.353..603G}. Middle: Data from NGC\,300 \citep{2011Ap&SS.332..221M,2012Ap&SS.340..133G}. Right: Data from NGC\,6946 \citep{2007AJ....133.1361P}. \textit{Bottom row}: Data from M\,31 \citep{2014SerAJ.189...15G} and data from NGC\,55 \citep{2013Ap&SS.347..159O}. { We note that NGC\,55 is an edge-on spiral galaxy which shows only a fraction of its SNRs due to obscuration. }}
\label{othervenndiagram}
\end{figure*}

To compare the multi-frequency emission from the \confcount\ known SNRs in the LMC, we plotted a Venn diagram (Figure~\ref{venndiagram}) that summarises the number of SNRs exhibiting emission in the different electromagnetic domains. It is important to note that the lack of detected emission does not mean that the remnant does not emit such radiation. However, it may indicate that the emission is under the sensitivity level of current surveys. As for the candidate remnants, many of them were entangled in unrelated emission or not part of current surveys, making it difficult to construct a worthwhile Venn diagram for these sources. We also note that there are examples of SNRs such as the SMC SNR HFPK\,334 \citep{2014AJ....148...99C} or the Galactic Vela~Jr SNR \citep{2001AIPC..565..267F,2005AdSpR..35.1047S} that could not be identified in optical frequencies despite extensive searches.

For comparison to our LMC results, Venn diagrams were constructed for various other nearby galaxies, shown in Figure~\ref{othervenndiagram}. We note that some of these galaxies do not have deep X-ray and/or radio coverage. Still, our results are closely comparable to those found for the SMC \citep{2005MNRAS.364..217F}, which is to be expected as it is the most similar to the LMC. The obvious common trait between the LMC and SMC SNR populations is that they are ubiquitous in X-rays because of low foreground absorption towards the MCs. We concluded from Figure~\ref{venndiagram} that this sample is not under severe influence from observational biases. While the LMC and SMC \xmm\ surveys reach similar depth, the LMC X-ray field coverage is somewhat incomplete and therefore our present LMC X-ray SNR sample is likely incomplete as well.

Comparing Figures~\ref{venndiagram} and \ref{othervenndiagram}, we note that all other galaxies have high numbers of detected SNRs only in optical frequencies, with small numbers of SNRs in cross sections of the Venn diagrams. This is expected as \citet{2000ApJ...544..780P} argued that X-ray and radio SNRs are mixed/embedded and therefore confused with \HII\ regions in distant galaxies. Also, we argue that the detection of optical SNRs in distant galaxies is biased toward lower densities. Therefore, we detect only smaller numbers of X-ray/radio SNRs in the more distant samples -- ones which are brighter and in denser, star-forming regions.

Interestingly, in a revised catalogue of 294 Galactic SNRs by \citet{2014BASI...42...47G}, 93\% were detected in radio, $\sim$40\% in X-ray and only $\sim$30\% at optical wavelengths. This points to a clear selection bias which limits the optical and X-ray detection. This is most likely due to obscuration from dust and clouds, as well as the lack of deeper observations at various frequencies. Also, we note that NGC~55 is an edge-on spiral galaxy and, therefore, only a small number of SNRs can be detected due to obscuration.

%
%

\subsection{Differential Size Distribution}
 \label{Sect_DSD}

\begin{figure*}[]
 \begin{center}
 \resizebox{1.6\columnwidth}{!}{\includegraphics[trim = 0 0 0 0, clip]{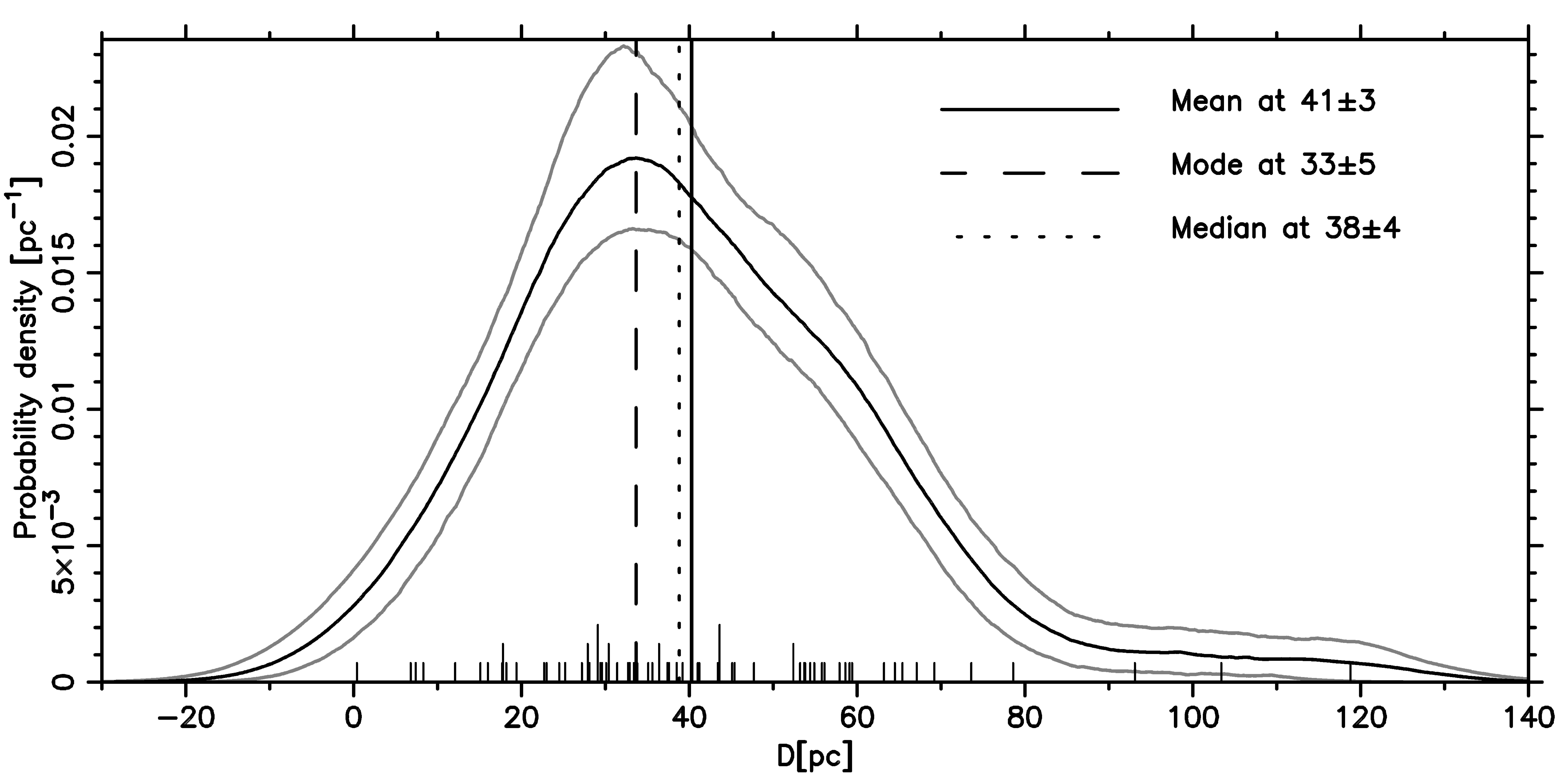}}\\
 \vspace{-7mm}
 \resizebox{1.6\columnwidth}{!}{\includegraphics[trim = 0 0 0 130, clip]{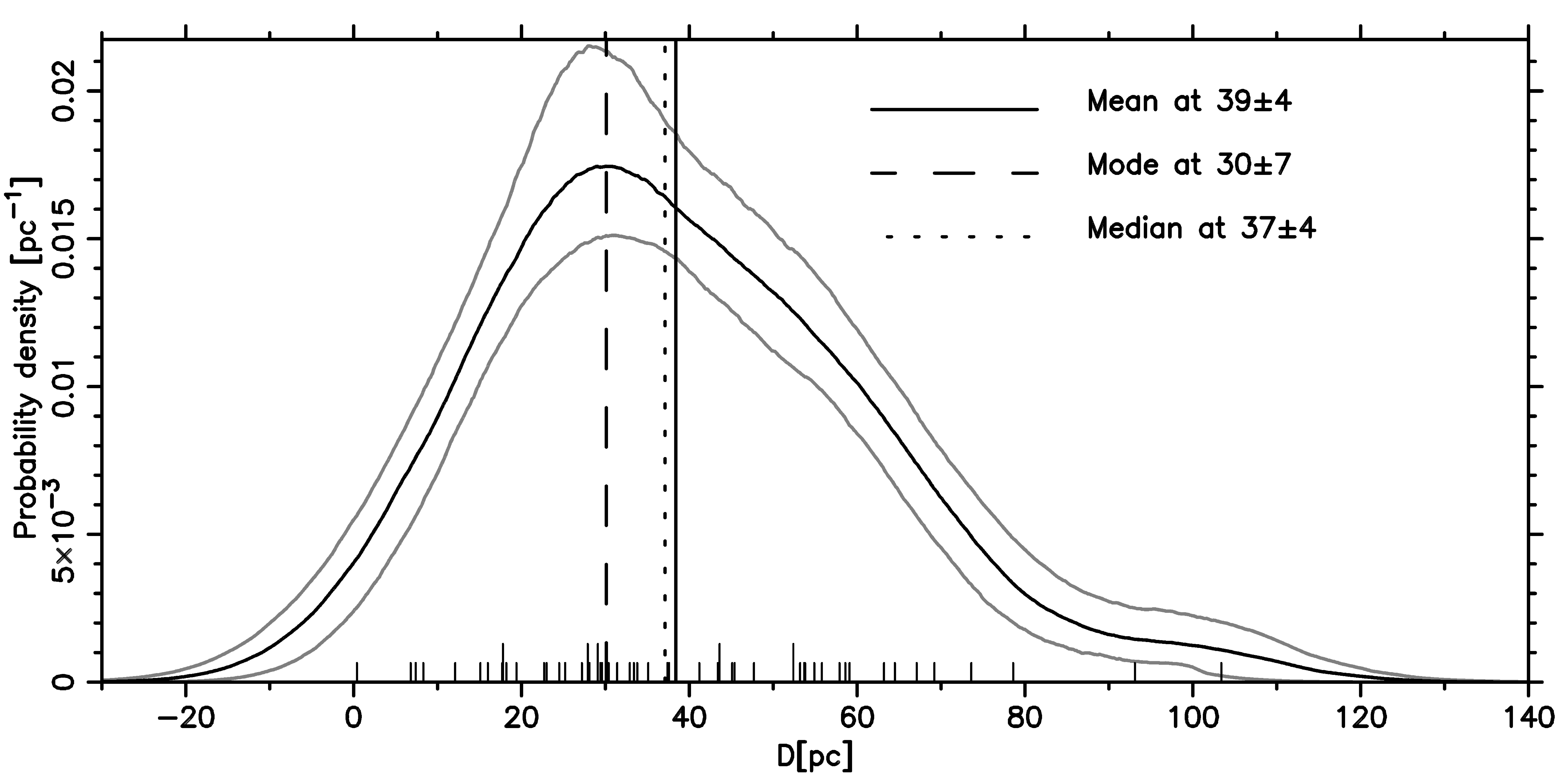}}\\
  \caption{The SNR diameter smoothed density distributions, obtained as described in Section~\ref{kds}, with round-up values for mean, mode and median. The data points are marked with vertical dashes on the horizontal axis{, with dash length proportional to the number of SNRs in the sample with the corresponding diameter}. Distribution parts with $D<0$ have no physical meaning and are plotted for the sake of completeness. The grey lines  represent estimated uncertainties at 75\% level (as explained in Section~\ref{kde_conf}). Top: The sample of {\totalcount} confirmed and candidate LMC SNRs. Bottom: The sample of {\confcount} confirmed LMC SNRs. The optimal smoothing bandwidths for the examined data samples were found to be ${11.4}$~pc for confirmed and ${9.89}$~pc for the bulk sample of confirmed and candidate remnants.
  \label{diameterHistogram}}
 \end{center}
\end{figure*}

To measure the extent of the {SNRs} in the LMC, an ellipse was fitted to delineate the bounds of emission for all confirmed and candidate {SNRs} in this study \citep[see Figure~7]{2014MNRAS.439.1110B}. A multi-wavelength approach was used and the given size takes into account the optical, radio and X-ray emission. Such an approach was taken as some shells may appear incomplete at optical wavelengths, though, completed at radio or X-ray wavelengths and vice versa. A somewhat typical example of this is where emission in one band (e.g., X-rays) is located in the centre and encased by an optical/radio shell, e.g., MCSNR\,J0508$-$6902 presented in \citet[ Figure~2]{2014MNRAS.439.1110B}, where the radio (5500~MHz) and optical (H$\alpha$) emission form the `ring' of emission inside which the X-ray (0.7$-$1.1 keV) emission resides. The major and minor axes, in addition to the position angle of these measurements, are listed in Tables~\ref{tab:lmcsnrs} and \ref{tbl:candsnrs}. The resulting PDF showcasing the distribution of these data is displayed in Figure~\ref{diameterHistogram}, where the diameter was taken as the geometric average of the major and minor axes (Tables~\ref{tab:lmcsnrs} and \ref{tbl:candsnrs}).

The mean value of the reconstructed distribution for the {\confcount} confirmed and {\candcount} candidate remnants was found to be $39 \pm 4$~pc for confirmed SNRs, and $41 \pm 3$~pc for the entire sample (Figure~\ref{diameterHistogram}). This increased mean value for the entire sample is most likely due to many of the candidate remnants being larger and weaker, only being recently detected by newer and more sensitive instruments. The majority ({36} out of {\confcount} or {61}\%) of the remnants were found to exhibit diameters in a range of 15--50~pc. These values are {moderately larger} than the value found in the study of M\,83 SNRs by \citet{2010ApJ...710..964D}, where a mean diameter of 22.7~pc (with standard deviation of {{SD} = 10.3~pc}) was found for a sample of 47 remnants. Also, in a study of the {SMC}, \citet{2005MNRAS.364..217F} found a mean diameter of 30~pc, and argued that such a value indicates that most of the remnants are in the adiabatic (Sedov) evolutionary stage. The results of this study are more in line with those found by \citet{2010ApJS..187..495L} in their study of M\,33, finding a median of 44~pc, and by \citet{2014ApJ...786..130L} of M\,31, which showed a strong peak at $D=48$~pc.

%
%

\subsection{Spherical Symmetry}
 \label{Sect_SS}

The spherical symmetry of the {LMC} {SNR} population was measured to investigate whether or not a trend exists between the type of {SNR} and how circular their morphology appeared. \citet{2011ApJ...732..114L} suggested a link between the spherical thermal X-ray morphology and the remnant type, where those {SNRs} resulting from a type~Ia {SN} explosion were more spherical than those from a CC {SN}. In this study, we define SNR spherical symmetry via:
\begin{equation}
 \text{Ovality}~(\%) = \frac{2(D_{maj} - D_{min})}{D_{maj} + D_{min}}
\end{equation}
\noindent where $D_{maj}$ and $D_{min}$ are the major and minor axes, respectively. This was done for all \confcount\ confirmed remnants using the diameter measurements shown in Table~\ref{tab:lmcsnrs} (Col.~5).

\begin{figure*}[]
 \begin{center}
  \resizebox{2\columnwidth}{!}{\includegraphics{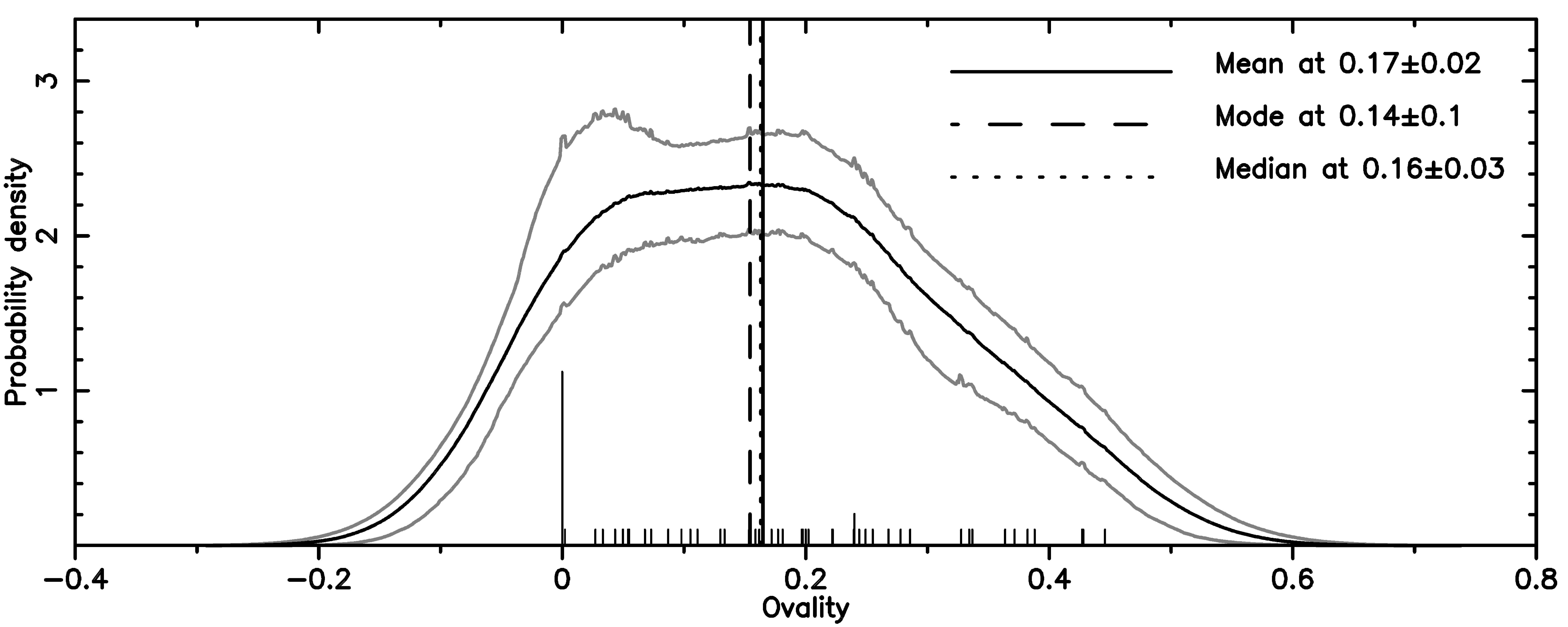}}\\
  \caption{Smoothed density distribution for ovality, obtained as described in Section~\ref{kds}, with round-up values for mean, mode and median. The data points are marked with vertical dashes on the horizontal axis, with dash length proportional to the number of SNRs in the sample with the corresponding ovality. Distribution parts with { ovality smaller than} $0$ have no meaning by definition and are plotted for completeness. The grey lines represent estimated uncertainties at 75\% level (as explained in Section~\ref{kde_conf}) The optimal smoothing bandwidth for the analysed data sample was calculated to be ${0.079}$.
  \label{figure_ovality}}
 \end{center}
\end{figure*}

We plot the smoothed density distribution for the ovality data (as explained in Section~\ref{kds}) in Figure~\ref{figure_ovality}. The large uncertainty for the mode parameter and shape of the confidence bands strongly suggest the existence of statistically significant bi-modality. To test whether this bi-modality is related to a progenitor type we plotted the smoothed density distributions for the sub-samples selected by progenitor type: type~Ia, CC and unknown (with progenitor type being undetermined).

\begin{figure*}[]
 \begin{center}
  \resizebox{1.6\columnwidth}{!}{\includegraphics[trim = 0 38 0 2, clip]{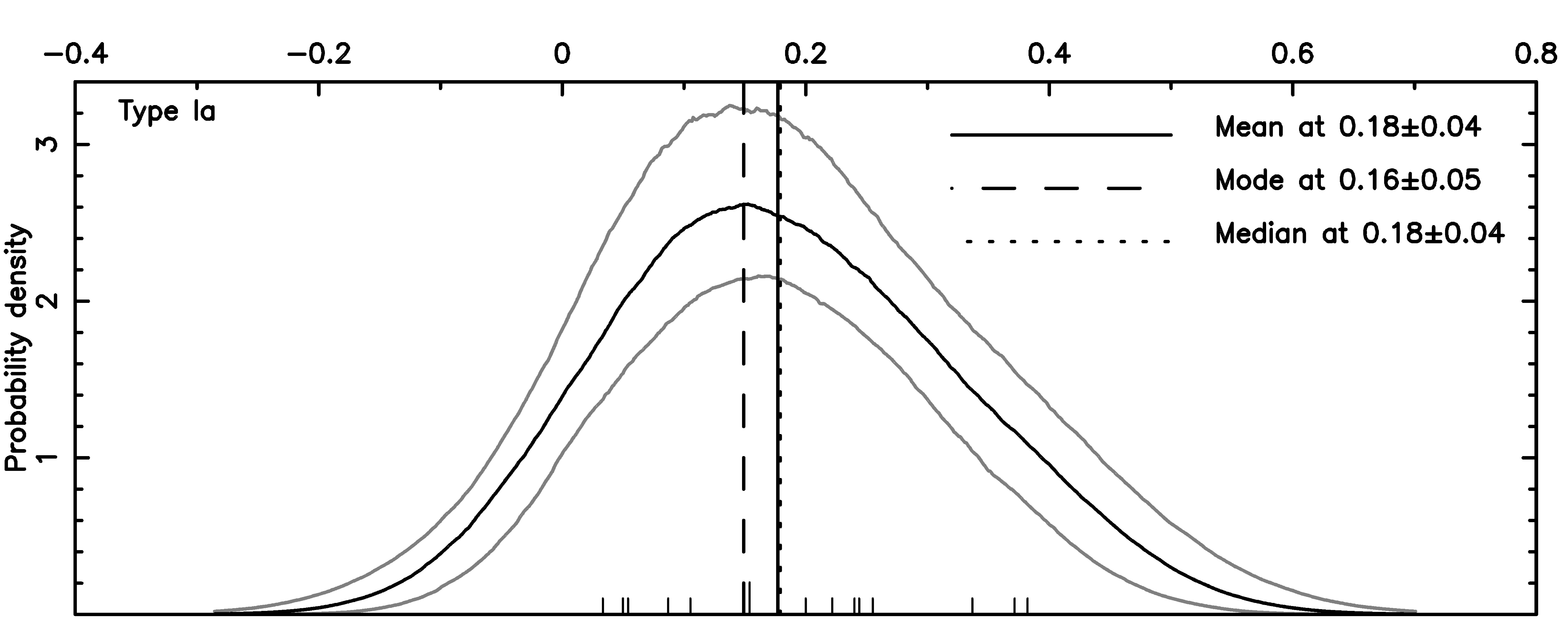}}\\
  \resizebox{1.6\columnwidth}{!}{\includegraphics[trim = 0 20 0 20, clip]{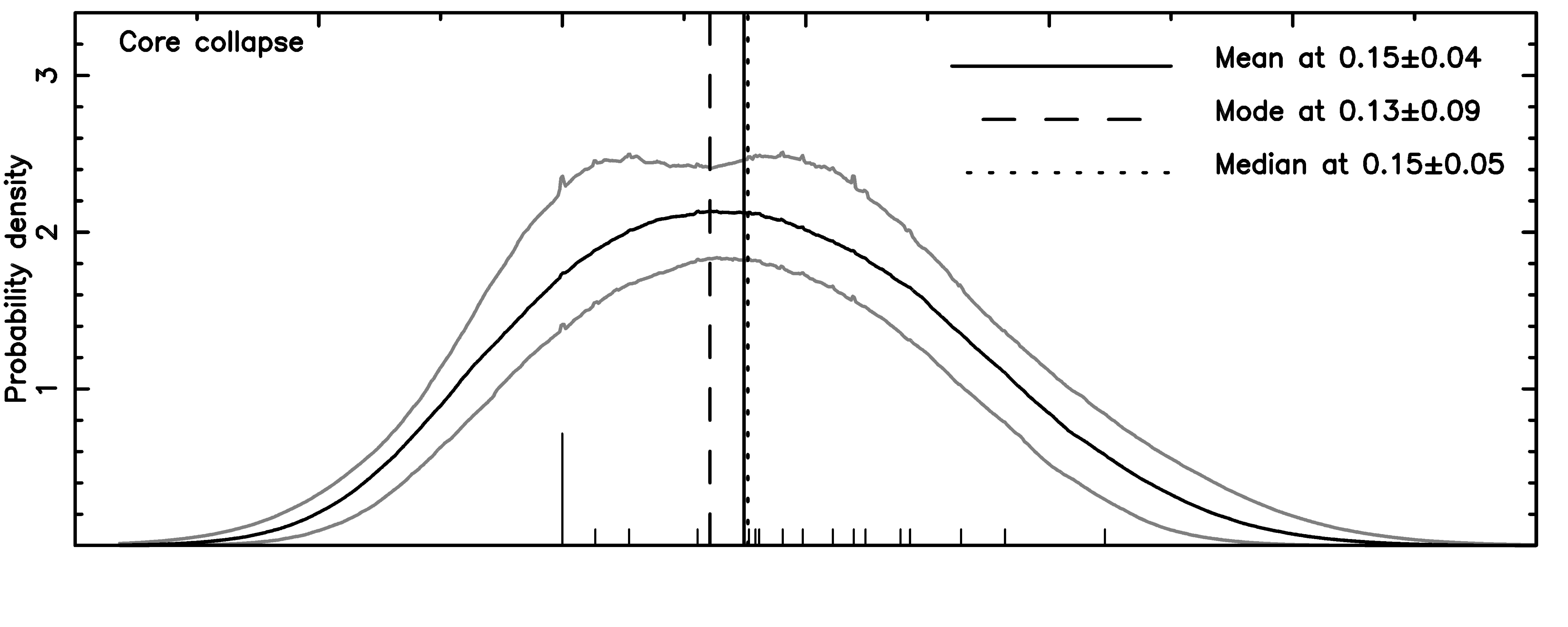}}\\
  \vskip -5mm
  \resizebox{1.6\columnwidth}{!}{\includegraphics[trim = 0 0 0 40, clip]{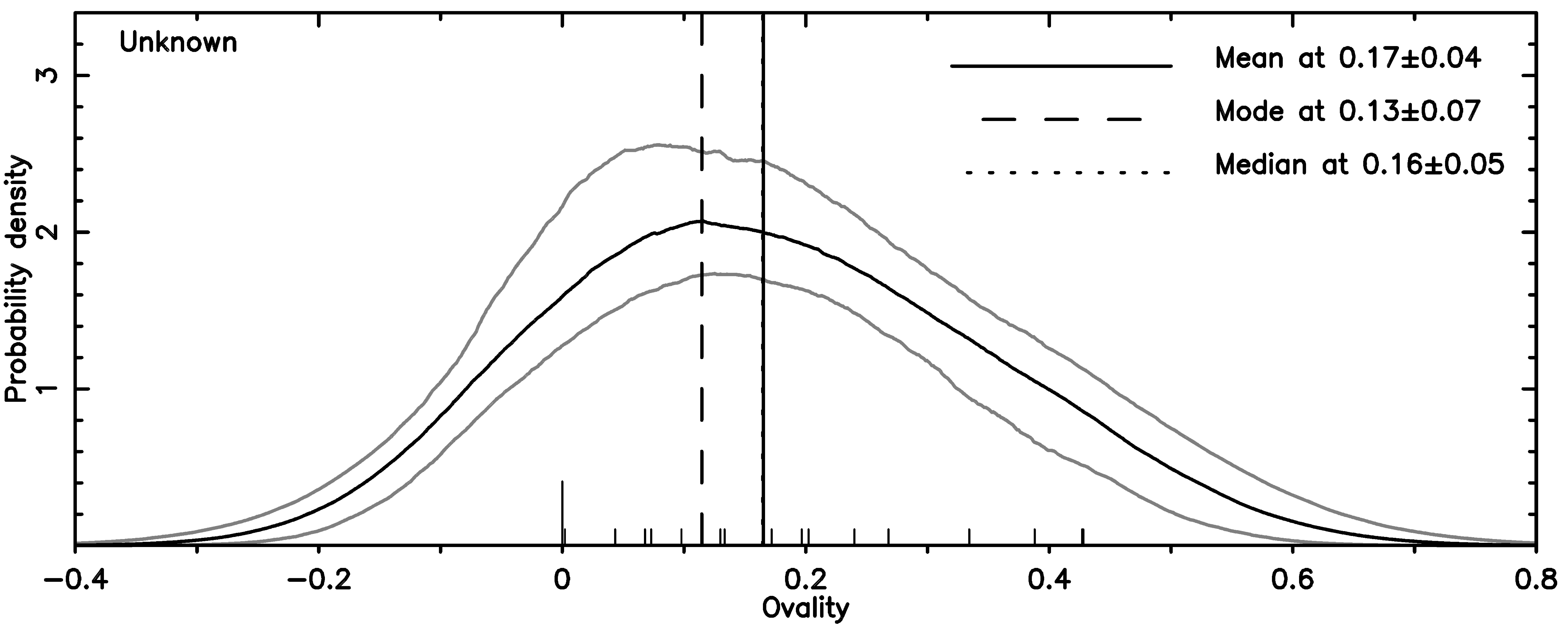}}\\
  \caption{Smoothed density distribution for ovality (as in Figure~\ref{figure_ovality}), for sub-samples based on progenitor type (designated on each panel). The optimal smoothing bandwidths for the analysed data samples were found to be ${0.107}$, ${0.112}$ and ${0.124}$ for Type~Ia, CC and unknown sub-samples, respectively.
  \label{figure_ovality_sub}}
 \end{center}
\end{figure*}

In Figure~\ref{figure_ovality_sub} the mode values for all examined sub-samples overlap within the estimated uncertainties at the $75\%$ confidence level, so it is unlikely that the examined sub-samples are actually coming from statistically distinct populations of SNRs. Also, from the shapes and parameters of the distributions for type~Ia and CC progenitor types, it appears that no distinction based on the progenitor type can be made and that the expected ovality of $\sim$0.16 should be even smaller since the progenitors from the unknown group are more spherical in shape. In total, $\sim${27}\% (16 out of 59) of the LMC SNRs exhibit ovality up to 5\% (0 to 0.05) and only 2 out of 16 ($\sim${13}\%) type~Ia SNRs' ovality are within this range, which does not support the hypothesis that type~Ia SNRs are more spherical in shape.

To further test if there is a statistically significant independence for ovality sub-samples based on SNR progenitor type we conducted an Anderson-Darling {\bf (AD)} test \citep{Anderson54}, studied in more detail for a two-sample case by \citet{10.2307/2335097}. The test calculates the parameter:
\begin{equation}
A^2_{nm} = \frac{nm}{N}\int_{-\infty}^{\infty}\frac{(F_n(x) - G_m(x))^2}{H_N(x)(1.0-H_N(x))}dH_N(x),
\end{equation}
where $F_n(x)$ and $G_n(x)$ are empirical cumulative distribution functions (ECDF) for sample 1 with $n$ points and sample 2 with $m$ points, while $H_N(x)$ is the ECDF of a joint sample with $N=m+n$ points. The integral of the squared deviations weighted with the joint ECDF factor gives a robust measure of the difference, even at the tails of the distribution where, by definition, all ECDFs converge towards $0$ and $1$. The resulting value of $A^2_{nm}$ is then compared against the {\bf critical value ($A^2_{nm}{'}$) at a specific $\alpha$ level} to test the null hypothesis that the two samples represented with $F_n(x)$ and $G_n(x)$ are sampled from the same underlying distribution. The results of the test are shown in Table \ref{table_ad_test}. It is apparent that even for the  $\mathbf{\alpha = 0.1}$, $A_{nm}^2<A_{nm}^2{'}$. {\bf This indicates that the null hypothesis cannot be rejected at $\mathbf{\alpha = 0.1}$ level} and that the tested sub-samples could plausibly be sampled from the same distribution. This argues against the earlier stated existence of a correlation between the type of {SNR} and their spherical symmetry.

\begin{figure}[]
 \begin{center}
  \resizebox{1\columnwidth}{!}{\includegraphics{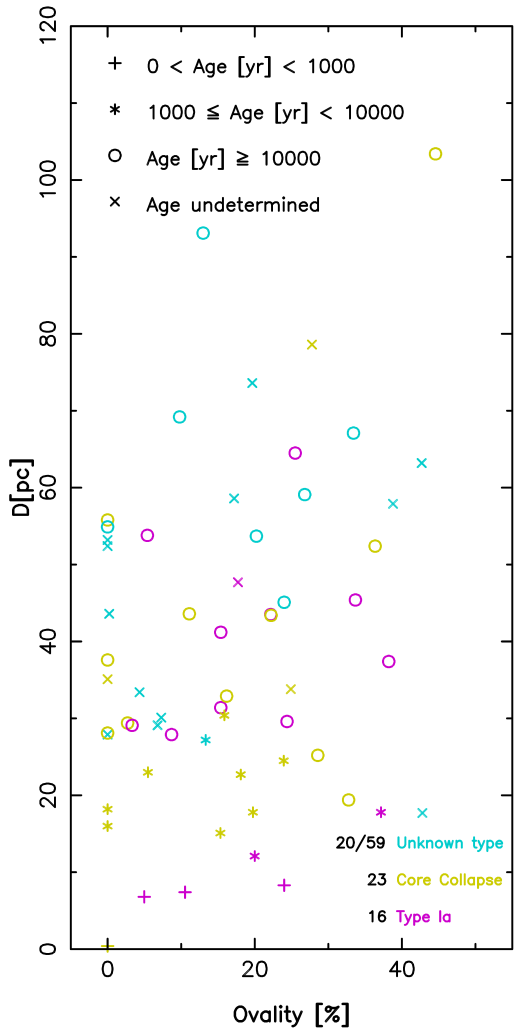}}
  \caption{The { geometric mean} diameter versus Ovality (\%) graph. Also, we plot the SNR age and the SN explosion type where known. No correlation was found between any of these LMC SNR parameters. The measurement uncertainties are significantly smaller than the data scatter and the main reason for the data scatter is likely to come from over-simplifying the model since it is likely that plotted variables are dependent much more on other parameters (e.g., ambient medium properties, explosion energy, ejecta mass, etc). }
  \label{fig11}
 \end{center}
\end{figure}

In addition, we make a scatter plot in Figure~\ref{fig11} of {\bf geometric mean diameter} vs. ovality, color-coded by the ages of the remnants. For this population, there does not appear to be any conclusive evidence that the type of {SN} explosion (Table~\ref{tab:lmcsnrs}; Col.~11) correlates with ovality (as defined here) of the resulting {SNR} or its known age, estimated from the various multi-frequency measurements (Table~\ref{tab:lmcsnrs}; Col.~12).

For the {\bf 28} LMC SNRs older than 10\,000~yrs (Figure~\ref{fig11}) we found a suggestive progenitor type. This is somewhat surprising as for older remnants, signatures of progenitor type are likely to be faint and hence harder to determine. Also, in Figure~\ref{figure_ovality_sub} the distribution of objects with undetermined progenitor type appear to be skewed towards smaller ovality values. This implies that older remnants might be more spherical in shape.

We expect that future surveys with high accuracy ovality measurements, better age determinations and more complete samples are of crucial importance for a better assessment of this subject.

\begin{table}
 \begin{center}
\caption{Results of the Anderson-Darling two-sample test. The comparison samples (S1 and S2) are given in the first column while rest of the columns present the number of objects in S1, number of objects in S2, calculated value of the Anderson-Darling variable for the two-sample test ($A_{nm}^2$) and value for the specific confidence level $\alpha$ ($A_{nm}^2{'}$) at the $\alpha=0.1$ ($10\%$) taken from \citet{10.2307/2335097}, respectively}
\begin{tabular}{lcccc}
\hline\hline
S1 vs. S2           & $n$ & $m$ & $A_{nm}^2$   & $A_{nm}^2{'}$ \\
                    &     &     &              & $\alpha=0.1$ \\
\hline
Unknown\,vs.\,type~Ia & 20  & 16  & { 1.150}  & 1.933  \\
Unknown vs. CC      & 20  & 23  & { 0.331}  & 1.933  \\
type~Ia vs. CC      & 16  & 23  & {1.556}  & 1.933  \\
\hline
 \end{tabular}
 \label{table_ad_test}
\end{center}
\end{table}

%
%
%
%

\subsection{Cumulative Number-Diameter Relation}
 \label{sect:ND}

The cumulative number-diameter relation, also known as \ND$ - D$, shows the number of {SNRs} smaller than a given diameter. Assuming a rough uniformity in SN explosion and environment, it possible to estimate the evolution of {SNRs} and the rate of {SN} occurrence via:
\begin{equation}
 N(<D)=\frac{t(D)}{\tau}
\end{equation}
\noindent where $t(D)$ is the age of the remnant and $\tau$ is the average time between the SN events, i.e. $\tau ^{-1}$ is the average supernova rate. As a remnant evolves, it is expected to pass through different phases of hydrodynamical evolution.  The turnover from the free-expansion phase to the Sedov phase, i.e., when the mass of the swept up material is higher than the ejecta, was approximated using the equation from \citet{1978ppim.book.....S}:
\begin{equation}
 R_{pc} \approx 2.13\left(\frac{M_E}{n_o}\right)^{1/3} \label{eq:fetost}
\end{equation}
\noindent where $R_{pc}$ is the transition radius in parsecs,  $n_0$ is the number density of the {ISM} per cubic centimetres, and $M_E$ is the ejecta mass in solar units. If we assume progenitor masses between the Chandrasekhar limit (1.4~\msun) and that of a large star (40~\msun), expanding in ambient densities between 0.1~--~1.0~cm$^{-3}$, the expected turnover from free-expansion to Sedov expansion would fall between $\sim$4.8~pc and $\sim$31.4~pc. It should be noted however, that a progenitor star's mass may exceed the given 40~\msun\ \citep{2009Natur.462..624G}, and that densities can vary significantly (e.g. cold atomic regions of the ISM have densities of $20-50$~cm$^{-3}$, while molecular regions may be as high as $10^2- 10^6$~cm$^{-3}$ \citep[][and references therein]{2001RvMP...73.1031F}. Therefore, such cut-off levels are far from being well established. The upper limit of Sedov expansion (i.e., the turnover to snowplough evolution) is estimated by \citet{1972ARA&A..10..129W} to be 50~kyr, or $D~\sim$~48~pc. These estimated turnover diameters are in moderate agreement with \citet{1987A&A...181..398B}, who found evidence that {SNRs} are radiative between $D_R(n_0=1) \simeq$~20~pc to 40~pc.

If we assume that the expansion is in the form of a power-law $D \propto t^m$, then \ND$\propto D^{a} $, where $ a = 1/m$. This means that for the free expansion $a\approx 1$ is expected, while in the Sedov phase, the slope in the log-log space would be $a = 2.5$.

Early studies of the  \ND\ relation for {SNRs} in the Galaxy and the {LMC} found exponents of $\approx$1 \citep{1983ApJS...51..345M,1983IAUS..101..551M,1984IAUS..108..283M}, implying that the majority of the remnants are in free expansion. However, \citet{1987A&A...181..398B} made the point that such an exponent is readily mimicked by any influence tending to randomise diameters. Later studies by \citet{1990ApJS...72...61L,1993ApJ...407..564S,1998ApJS..117...89G} found steeper slopes, common to later phases such as Sedov or snowplough expansion. For example, in their study of {SNRs} in M\,33, \citet{1998ApJS..117...89G} found an \ND\ relation consistent with Sedov expansion and inconsistent with free-expansion, inferring that if the {ISM} is similar to our own Galaxy and the {LMC}, that these galaxy surveys are seriously incomplete. In their study of {SNRs} in M\,83 \citet{2010ApJ...710..964D}, found a slope consistent with free-expansion for nuclear remnants, whereas remnants residing in the disk generally followed the expected value of the radiative phase ($N(<D) \propto D^{7/2}$). In M\,31, \citet{2014ApJ...786..130L} found two breaking points in the data: one at 17~pc and another at 50~pc. The first component ({SNRs} with $D <$ 17~pc) showed a power-law slope of $a=1.65\pm0.02$, while the second component (17~pc $ < D <$ 50~pc) was in line with Sedov expansion, with a slope of $a=2.53\pm0.04$.

Here, we make distinctions between the types of {SNR} based on their morphologies. \citet{2014ApJ...786..130L} defined so-called A-type remnants as those with well-defined shells. The slope for this sub-class of M\,31 {SNRs} was found to be $a=2.15\pm0.09$ for 25~pc~$<D<$~45~pc. In their study of M\,33, \citet{1998ApJS..117...89G}, using a maximum likelihood estimate fit to the \ND\ relation, found slopes for $D_{max}$ = 30~pc and $D_{min}$= 8~pc and 10~pc of $a=2.2\pm0.05$ and $a=2.3\pm0.06$, respectively. While $D_{max}$ = 35~pc and $D_{min}$= 8~pc, 10~pc and 15~pc showed slopes of $a=2.0\pm0.04$, $a=2.0\pm0.05$ and  $a=1.1\pm0.06$, respectively. For the SMC, \citet{2005MNRAS.364..217F} found an overall slope for the galaxy of $a=1.7\pm0.2$.

\begin{figure*}[]
 \begin{center}
 \resizebox{2\columnwidth}{!}{\includegraphics{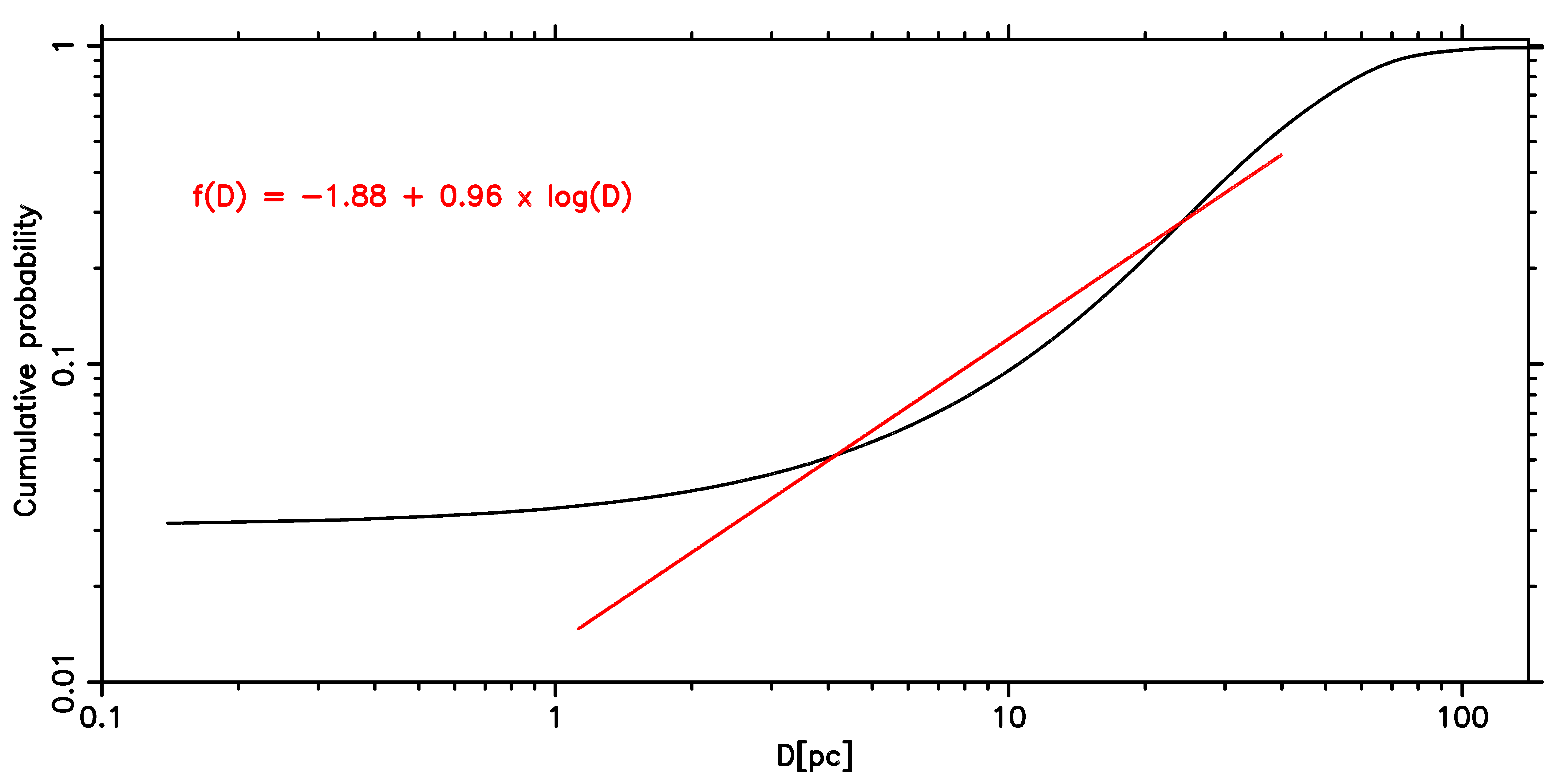}}\\
  \resizebox{2\columnwidth}{!}{\includegraphics{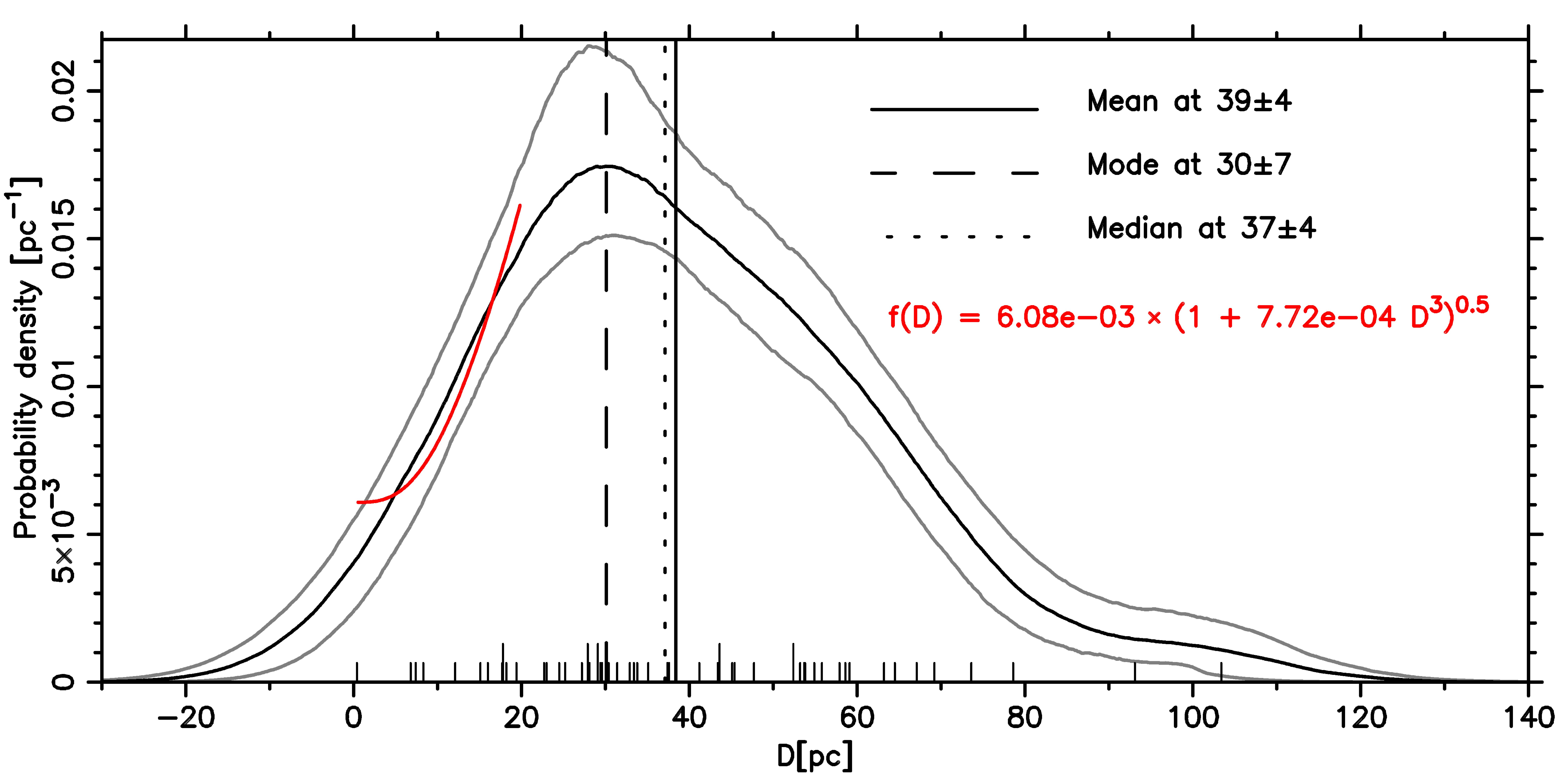}}\\
  \caption{Top: Cumulative (integrated) probability distribution from the bottom panel presented on a log-log scale. The red coloured lines are the best fit lines to the plotted distributions, from approximately 1 to 40~pc (top), i.e., from the data point with the smallest diameter to the diameter value at the estimated inflexion point at $\approx~{20}$~pc (bottom). Bottom: The distribution from the bottom panel of Figure~\ref{diameterHistogram}.
  }
  \label{NDgraph}
 \end{center}
\end{figure*}

In Figure~\ref{NDgraph}, we present our results for the LMC. The top of Figure~\ref{NDgraph} shows cumulative counts versus diameter for all {\confcount} confirmed LMC SNRs in log-log scale. The red line shows the best fit with a slope $a = 1/m = {0.96}$. Since the sample above D$>$40~pc seems to be somewhat incomplete, only the first part of the curve was fitted \citep[for an alternative view, see e.g.][]{2010MNRAS.407.1301B}. The population exponent $a = {0.96}$ is close to 1, even given the larger population and more complete sample size of this study. Therefore, the earlier suggestion by \citet{1987A&A...181..398B} regarding randomised diameters readily mimicking such an exponent is probably the case in our LMC sample, and not that the relation is indicative of the SNR population in the galaxy to be in free-expansion. The exponent $1< a < 2.5$ { may } indicate that { a} population is somewhere between free expansion and the Sedov phase. { Although this is unlikely in the case of LMC sample}, we have used a simple model for the SNRs' expansion velocity \citep{2005Arb,2012ApJ...751...65F}:
\begin{equation}
 \frac{1}{2} v^2= \frac{k_1E_o}{k_2{M}_E+4\pi R^3 \rho _o /3} , \ \ \ v =\frac{dR}{dt}
 \label{eq:ndlimits}
\end{equation}
\noindent and combined it with $dN/dD = 1/(2\tau v)$ in order to perform a non-linear fit to the data (again, only the first part of the curve was fitted). In Eq.~\ref{eq:ndlimits}, $E_o$ is the explosion energy, $M_E$ mass of the ejecta and $\rho _o$ is the ISM density. The constants $k_1$ and $k_2$ are determined in such a way that when $E_o$ = 1 foe\footnote{Energy unit; 1~foe=$10^{51}$~erg} and $M_E$=1.4~\msun, $v \approx 20\,000$~km/s, and when $R \gg R_{pc}$ the velocity tends toward the Sedov solution. In Figure~\ref{NDgraph} (bottom) we show the differential distribution for the number of the LMC SNRs versus diameter. The red curve represents the best {fit} obtained by applying Eq.~\ref{eq:ndlimits}, {which is still far from good.} For a fixed energy of 1~foe and $M_E \sim 10$~\msun\ (although the sample could contain few type~Ia SNe) the fit gives a supernova rate of {0.55}/cy\footnote{cy is centi year or century} \citep[twice as high as rates found in the literature, e.g., 0.23/cy,][]{1991ARA&A..29..363V}, while the density is quite low, $\sim${0.03}~cm$^{-3}$, which {would not be} surprising for SNRs still close to the free expansion phase, {the latter, however, being unrealistic, as already said above.}

%
%

\section{Spectral Indices and Evolution in Radio-continuum}

Following \citet{1998ApJS..117...89G} the radio spectral index ($\alpha$) of an SNR is a measure of the energy distribution of the relativistic electrons producing synchrotron radiation. Assuming a power-law for the injection spectrum of relativistic particles of the form:
\begin{equation} N(E)
 \propto E^{-\gamma}
\end{equation}
\noindent where $E$ is the energy of the relativistic particle and $\gamma$ is the power-law index of the energy spectrum, the radio spectral index is related via:
\begin{equation}
 \alpha = -\frac{\gamma-1}{2} .
\end{equation}
\noindent From \citet{1978MNRAS.182..443B} it is seen that diffusive shock acceleration ({DSA}) can accelerate relativistic particles in a remnant, such that:
\begin{equation}
 \gamma=\frac{\chi+2}{\chi-1}
\end{equation}
\noindent where $\chi$ is the compression ratio in the shock front. In the ideal monatomic gas, the limiting compression ratio is 4, resulting in a power-law index of 2, and thus, a spectral index of $-0.5$.

\subsection{Radio-Continuum Spectral Index Distribution}
 \label{Sect_RCSID}

\begin{figure*}[]
 \begin{center}
  \resizebox{2\columnwidth}{!}{\includegraphics{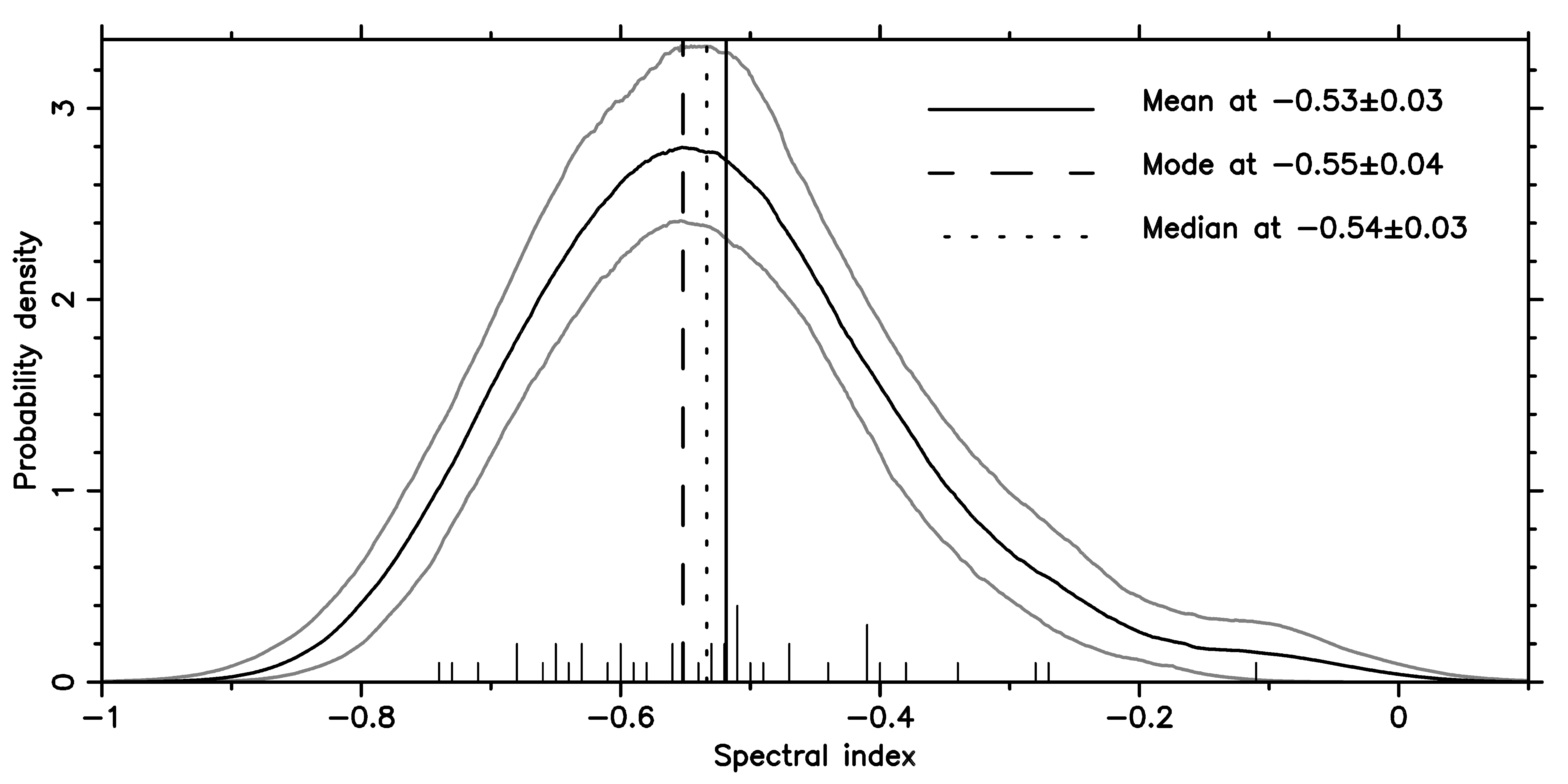}}
  \caption{Spectral index smoothed density distribution (as explained in Section~\ref{kds}) for the sample of \spccount\ LMC SNRs with round-up values for mean, mode and median. We used a $75\%$ confidence interval for estimating the uncertainties (grey lines). We note that a number of SNRs have an overlapping spectral index value, even though they were estimated to accuracy of two decimal digits (see also Table~\ref{tab:lmcsnrs}). The data points are marked with vertical dashes on the horizontal axis, with dash length proportional to the number of SNRs in the sample with the corresponding spectral index value. Younger SNRs are generally found toward the steeper indices, while older remnants tend toward the flatter end of the distribution (physical explanations for why this is believed to occur are discussed in Section~\ref{sect:spcidxevo}) The optimal smoothing bandwidth for the analysed data sample was found to be $0.077$.
  \label{spchist}}
 \end{center}
\end{figure*}

The mean spectral index of the reconstructed distribution is $\alpha=-0.52\pm0.03$ (with sample SD=0.13), calculated from the \spccount\ LMC SNRs where a reliable {spectral energy distribution (SED)} could be estimated. Individual remnant indices as well as associated properties can be found in Table~\ref{tab:lmcsnrs}, while a reconstructed differential distribution (similar to the reconstructed distributions from Figure~\ref{diameterHistogram}) of these values can be seen in Figure~\ref{spchist}. The mean value of $-0.52$ is in line with the theoretically expected spectral index of $\alpha=-0.5$ (as discussed above). The left side of the distribution is predominately composed of young SNRs, while older remnants or those harbouring a pulsar are found towards the right end of the PDF.

Comparing to other galaxies, \citet{2005MNRAS.364..217F} found a mean spectral index of $-0.63$ ({SD} = 0.43) for confirmed and candidate SNRs in the {SMC}. In the MW, \citet{1976MNRAS.174..267C} found a mean spectral index of $-0.45$ with a {SD} of $\sim0.15$.

\subsection{Radio-Continuum Spectral Index Evolution}
\label{sect:spcidxevo}

\begin{figure*}[]
 \begin{center}
  \resizebox{1.5\columnwidth}{!}{\includegraphics[trim = 0 40 0 0, clip]{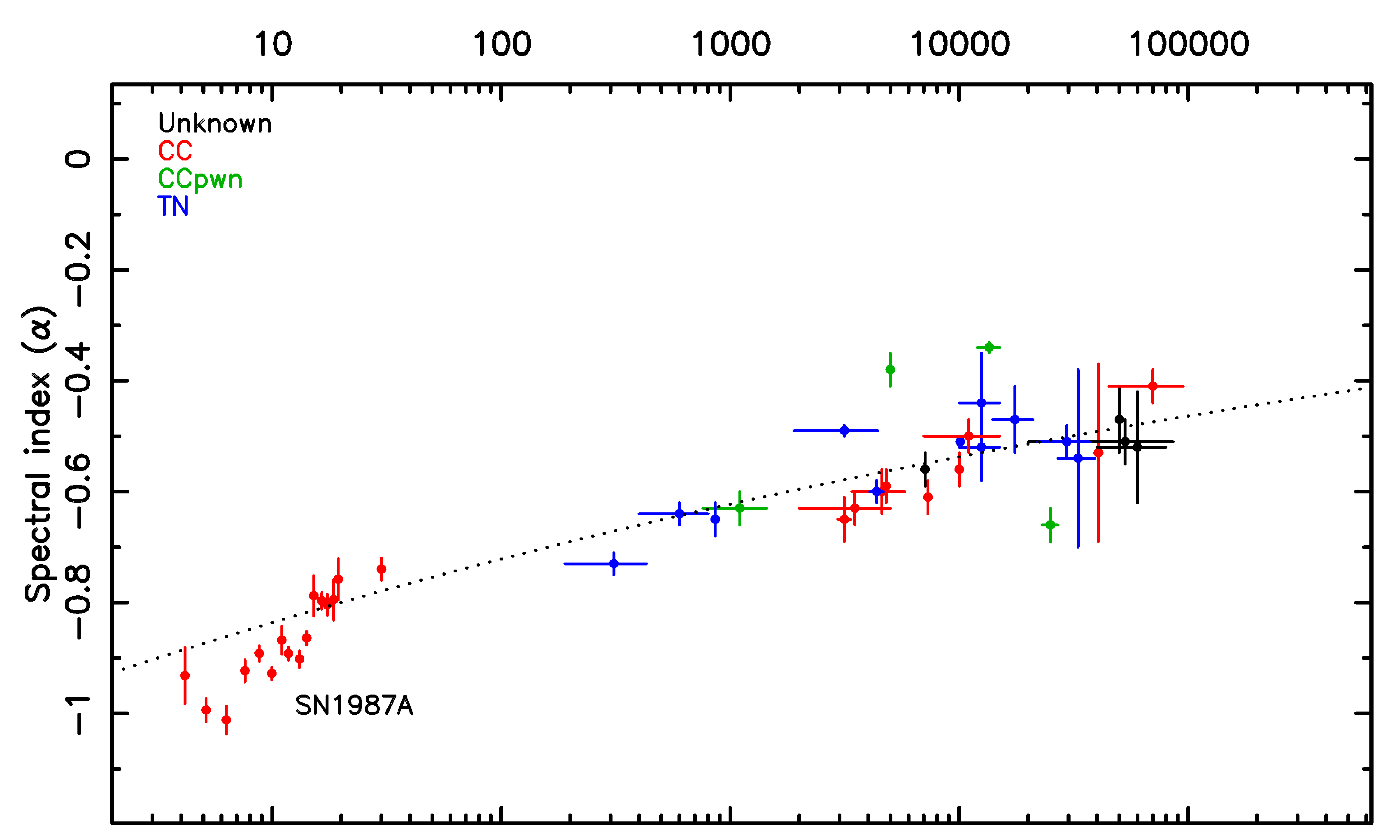}}
  \resizebox{1.5\columnwidth}{!}{\includegraphics[trim = 0 0 0 40, clip]{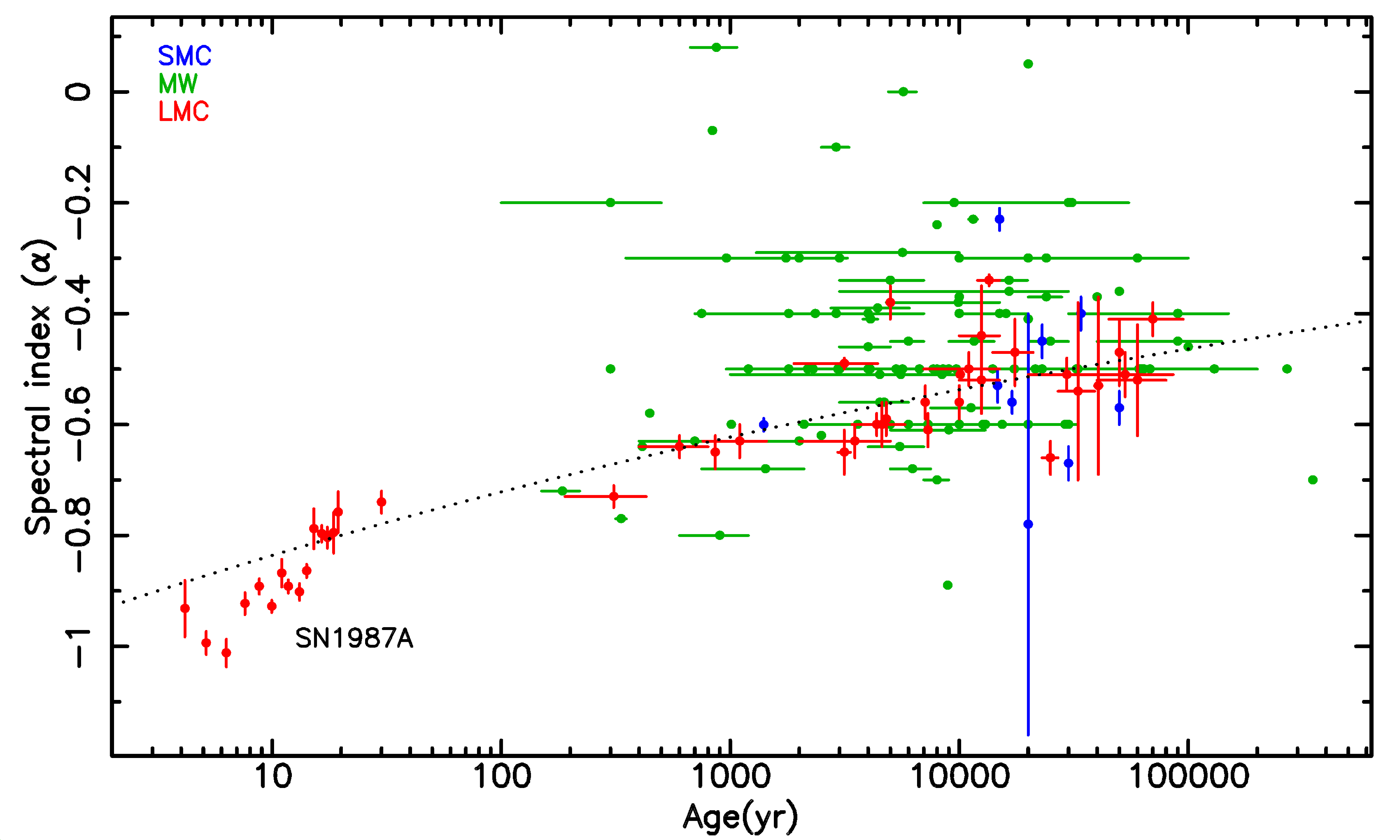}}
  \caption{\textit{Top:} Radio spectral index versus age for the LMC SNR sample of {29} remnants with known ages and spectral indices. The data points located toward the left of the plot are all from SN\,1987A \citep[data from][]{2010ApJ...710.1515Z,2013ApJ...777..131N,2016MNRAS.462..290C}, representing the data from earlier times, however, only the most recent (2016) is included in the population fit. The black dashed line shows a power-law fit to the {29} SNRs, resulting in ${{\alpha=(-0.97\pm0.09) \times t^{-0.06\pm0.02}}}$. \textit{Bottom:} Same as \textit{Top} with the addition of the {SMC} and {MW} samples.}
  \label{avsid}
 \end{center}
\end{figure*}

To investigate the relationship between a remnant's age and its radio spectral index, we plotted these two properties against each other (Figure~\ref{avsid}; top), resulting in a power-law fit for the {29} SNRs in the LMC with established age:
\begin{equation}
{{\alpha=(-0.97\pm0.09) \times t_{yr}^{-0.06\pm0.02}}}
\label{alpha}
\end{equation}
\noindent where $t_{yr}$ represents the remnant's age in years. {The fit parameter values are calculated as mean values from an arrays of $10^3$ values for each parameter obtained by fitting (non weighted fit) the  re-samples of the original data sample \citep[for more details on bootstrap procedure see][]{tEFR93a}. The parameters uncertainties are calculated as standard deviations of the corresponding arrays.} Some of the earlier studies (e.g., \citealt{1976MNRAS.174..267C}) indicated that a remnant's spectral index did not appear to be correlated with any other parameters. However, this conclusion was generally formed when smaller samples of remnants were available or used. In our study, we found the trend that younger remnants exhibit significantly steeper spectral indices, while mid-to-older remnants show flatter indices. Also, we found that the averaged spectral index for this sample of {29} LMC SNRs with known age (and spectral index) is $\alpha=-0.55$, which is fractionally steeper then the whole sample ($\alpha=-{0.52}$). It is likely that this is because it is easier to obtain an age for younger, and therefore brighter SNRs which have steeper spectral indices.

The notion that older {SNRs} exhibit flatter indices was first recognised by \citet{1962ApJ...135..661H} based on observational evidence. \citet{2013Ap&SS.346....3O} and \citet{2014Ap&SS.354..541U} suggested that {SNR} diminishing may be explained by the contribution of the second order Fermi mechanism, higher shock compressions or/and thermal bremsstrahlung. Conversely, the steeper spectra found for younger {SNRs} may be explained by optically thin synchrotron emission produced from the accelerated electrons and compressed magnetic field produced at the shock front \citep{2005coex.conf...89S}. Further, \citet{2011MNRAS.418.1208B} suggested that expansion into a Parker spiral may produce a geometry favouring quasi-perpendicular shocks and spectral steepening.

One also should not forget that there is evidence that the SNR radio spectral index significantly flattens when the SNR shell interacts with surrounding molecular clouds \citep{2014MNRAS.445.4507I}. The linear fits in Figure~\ref{avsid} are not likely to capture all of the features of the data as we do not know precisely what causes the spread. It could be a wide variety of factors spanning from the excess in N$_H$, differing supernova energies and types to differing number densities.

To compare our results with those from other galaxies, we re-plotted all values in Figure~\ref{avsid} (bottom) alongside spectral index and age data from SNRs in the MW \citep{2014BASI...42...47G} and SMC \citep{2005MNRAS.364..217F,2008A&A...485...63F}. As it can be seen, there is a good alignment between the galaxies. SNR ages, except for the few historical SNe, are derived assuming a hydrodynamical model, usually Sedov, linking age $t~\propto~R/v_{\rm s}$. The shock speed $v_{\rm s}$ is derived from X-ray fitting ($\propto \sqrt{kT}$) and is independent of distance ($D$), so $t~\propto~R~(kT)^{-1/2}$~$\sim \theta D (kT)^{-1/2}$, where $\theta$ is the observed angular diameter. Thus, distance uncertainties contribute most of the uncertainty in age measurements. As noted in Section~1, $D$ is most poorly known for Galactic objects, leading to large age error bars. However, for the LMC objects, a common distance of 50~kpc is assumed so the age measurement do not suffer from such large uncertainties.

\begin{figure*}[]
 \begin{center}
  \resizebox{1.6\columnwidth}{!}{\includegraphics[trim = 0 40 0 0, clip]{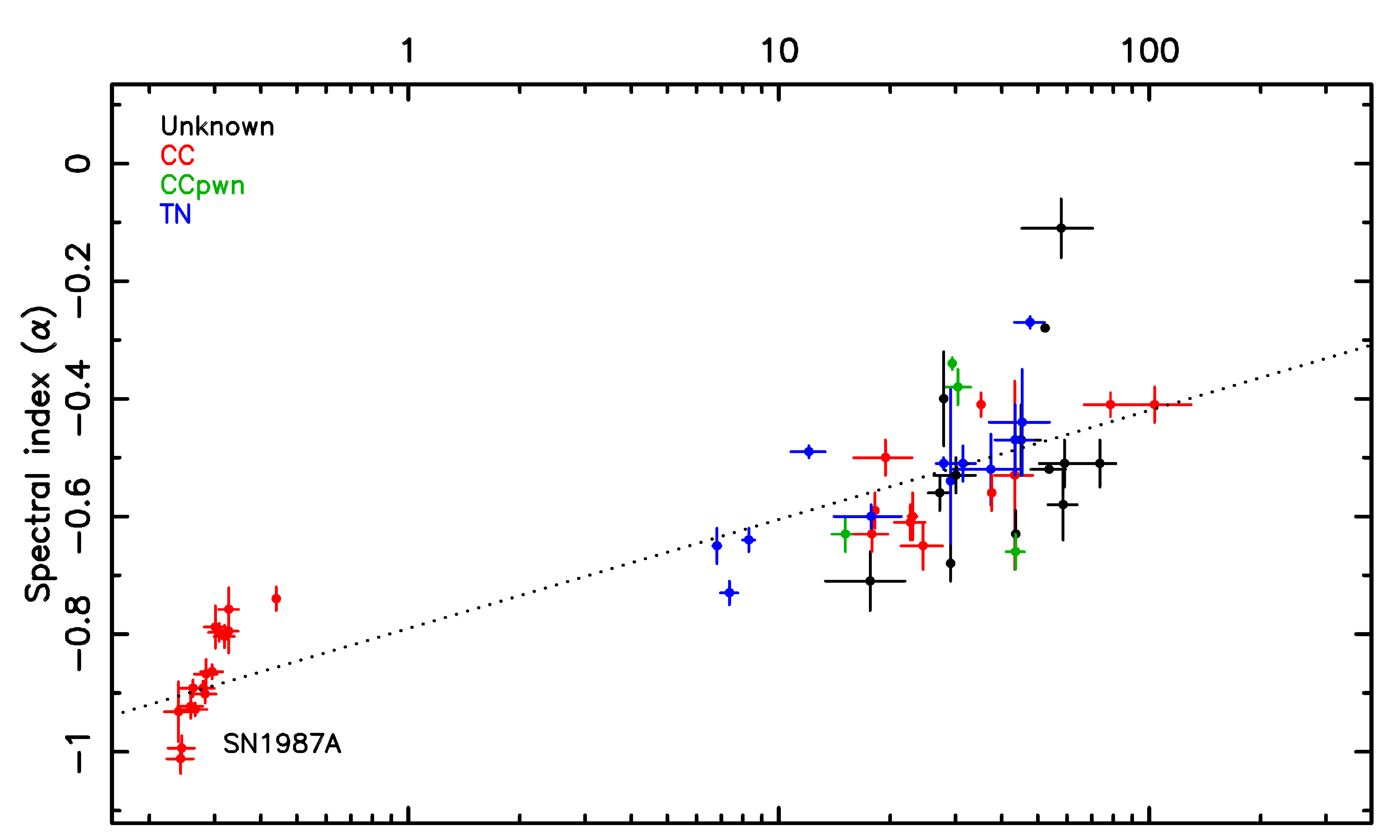}}
  \resizebox{1.6\columnwidth}{!}{\includegraphics[trim = 0 0 0 40, clip]{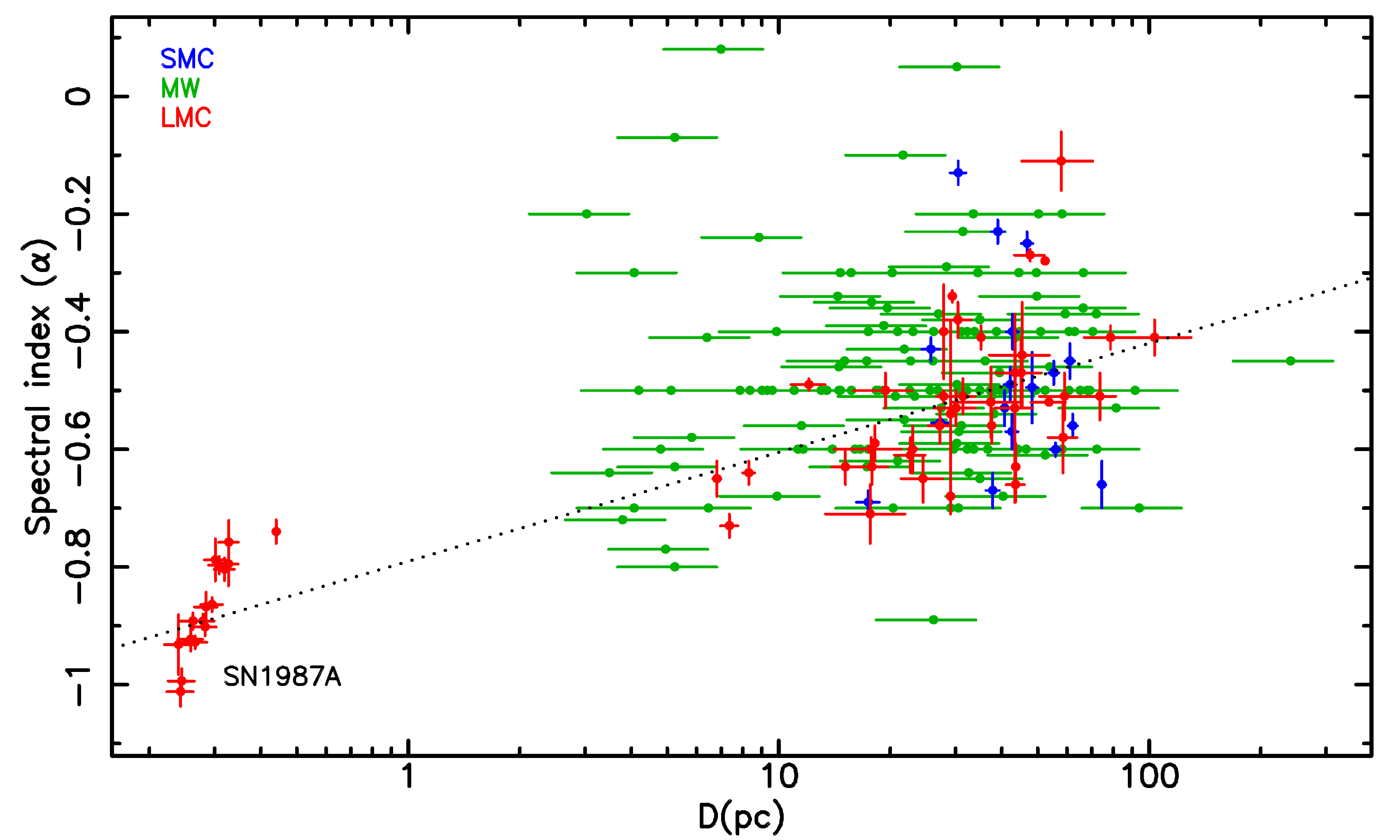}}
  \caption{\textit{Top:} Radio spectral index versus diameter for the sample of \lumcount\ LMC SNRs with measured spectral indices. The data points located toward the lower left corner of the plot are all from SN\,1987A, representing the data from earlier times. However, they are not included in the fit. A fit to the measurements (using only the latest data point for SN\,1987A \citep[from][]{2016MNRAS.462..290C}) results in ${P=0.18\pm0.04, \alpha_0=-0.79\pm0.07}$ from Eq.~\ref{eq:logdalpha}. \textit{Bottom:} Same as \textit{Top}, with the addition of data from remnants in the SMC and MW.}
  \label{dvsid}
 \end{center}
\end{figure*}

In a similar fashion to the age-spectral index relation, a diameter-spectral index relation was created, which gives a better view of the entire sample of {remnants, resulting in:}
\begin{equation}
{ \alpha=(-0.8\pm0.1) \times D_{pc}^{-0.12\pm0.03}}.
 \label{alpha2}
\end{equation}
\noindent As expected from {Figure~\ref{avsid}}, smaller (which would generally imply younger) remnants tend to exhibit steeper spectral indices, while their larger counterparts show flatter indices. Similarly to the age --- spectral index relation, we compare these results with those from the SMC and MW sample (see Figure~\ref{dvsid}; bottom). There is loose alignment in the results, though remnants in the MW generally appear to exhibit flatter indices at the same diameter compared to the LMC sample.

\citet{1985SvAL...11..350G} studied the evolution of young shell {SNRs} using eight Galactic remnants, in addition to one located in the LMC. He found that the remnants' spectral indices flattened as they got larger and older, in line with the results in this study. \citet{1996A&AT...11..317G} used positive $\alpha$-values expressed in the form:
\begin{equation}
  \alpha=P\text{log}\left (\frac{D_{pc}}{pc}\right) + \alpha_0
 \label{eq:logdalpha}
\end{equation}
\noindent to find values of $P=-0.58\pm0.07, \alpha_0=0.62\pm0.03$ for M\,82 and NGC\,253 ($0.3 < D < 4 $pc), and $P=-0.54\pm0.03, \alpha_0=1.03\pm0.02$ for the Galaxy and M\,31 ($3 < D \leq 21$~pc). He concluded that young shell SNRs evolve differently in galaxies with and without star bursts. Using the same method for the sample of LMC SNRs listed in Table~\ref{tab:lmcsnrs}, we found {${P=0.18\pm0.04, \alpha_0=-0.79\pm0.07}$} {(see Figure~\ref{dvsid}; bottom)} which shows that the spectral index flattening is much less severe as the remnant evolves, in line with the age--spectral index relation.

%
%

\subsection{Flux Density Distribution}
\label{sec_flux_density}

\begin{figure*}[]
 \begin{center}
  \resizebox{1.6\columnwidth}{!}{\includegraphics[trim = 0 40 0 0, clip]{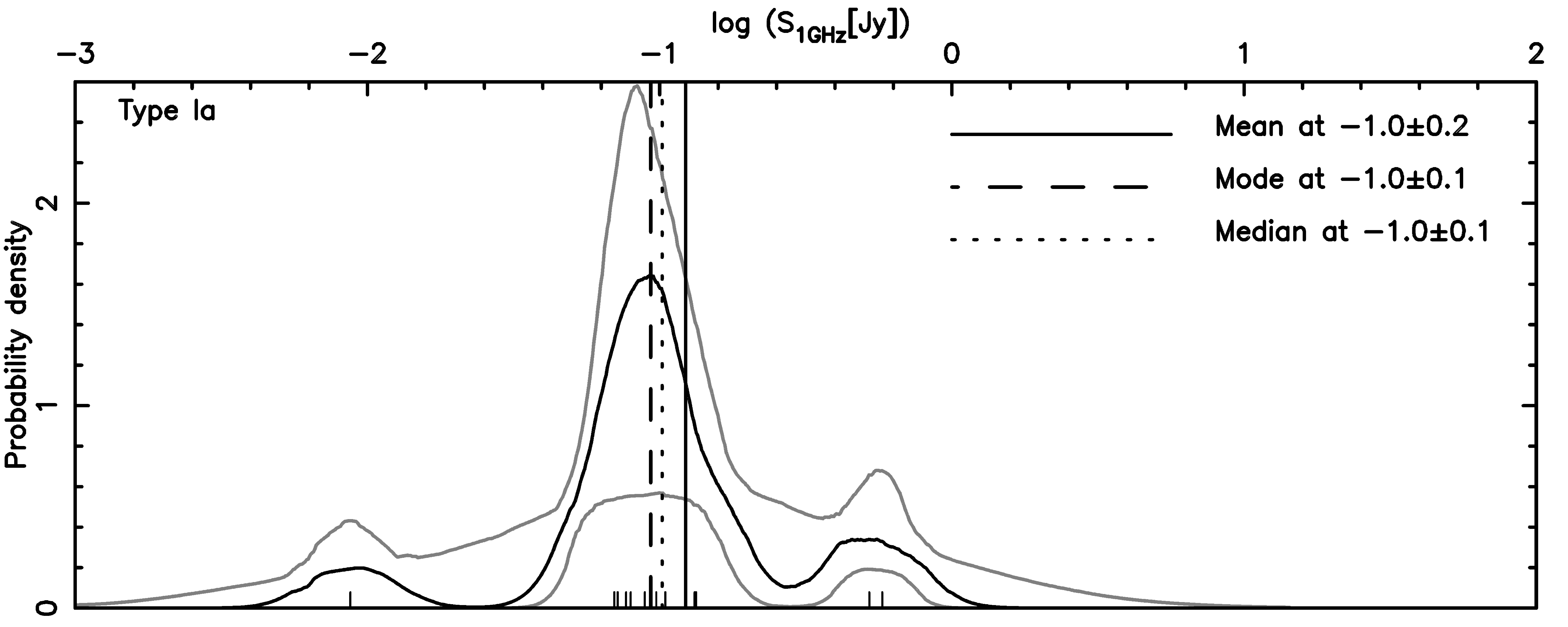}}\\
  \resizebox{1.6\columnwidth}{!}{\includegraphics[trim = 0 140 0 160, clip]{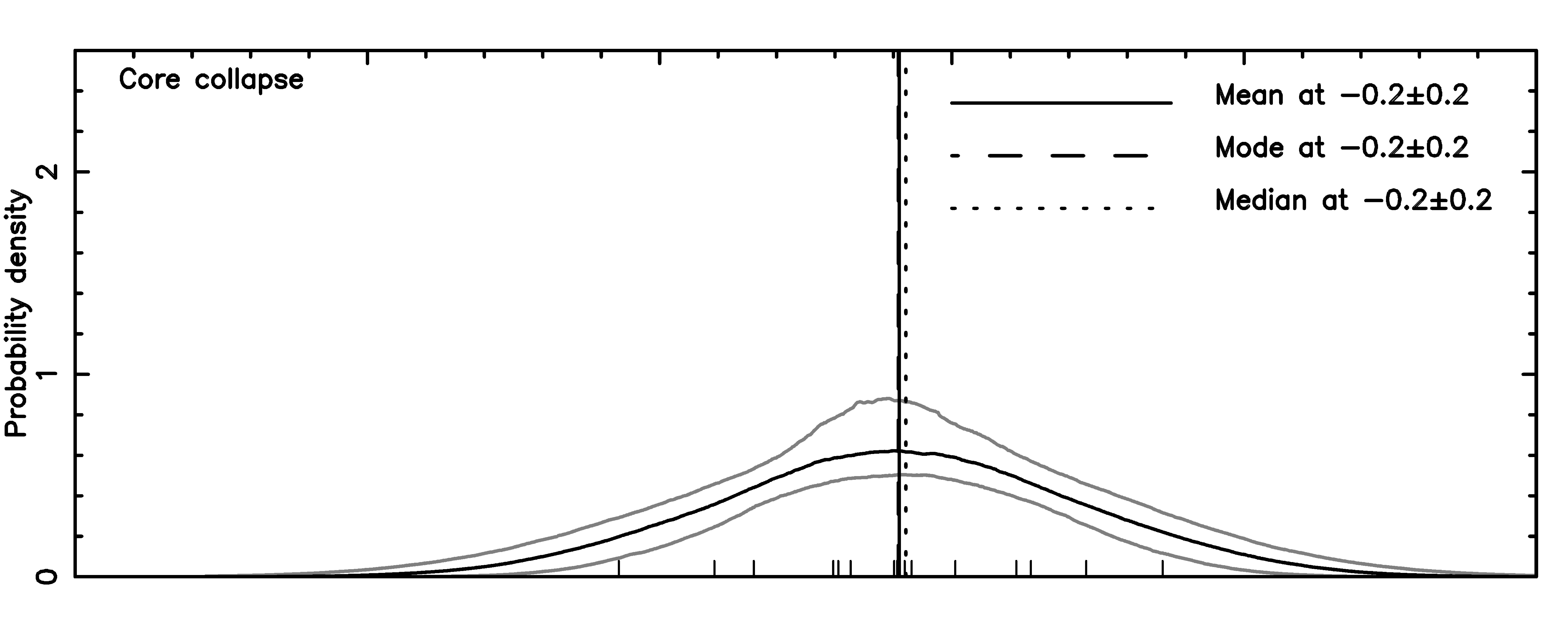}}\\
  \resizebox{1.6\columnwidth}{!}{\includegraphics[trim = 0 0 0 60, clip]{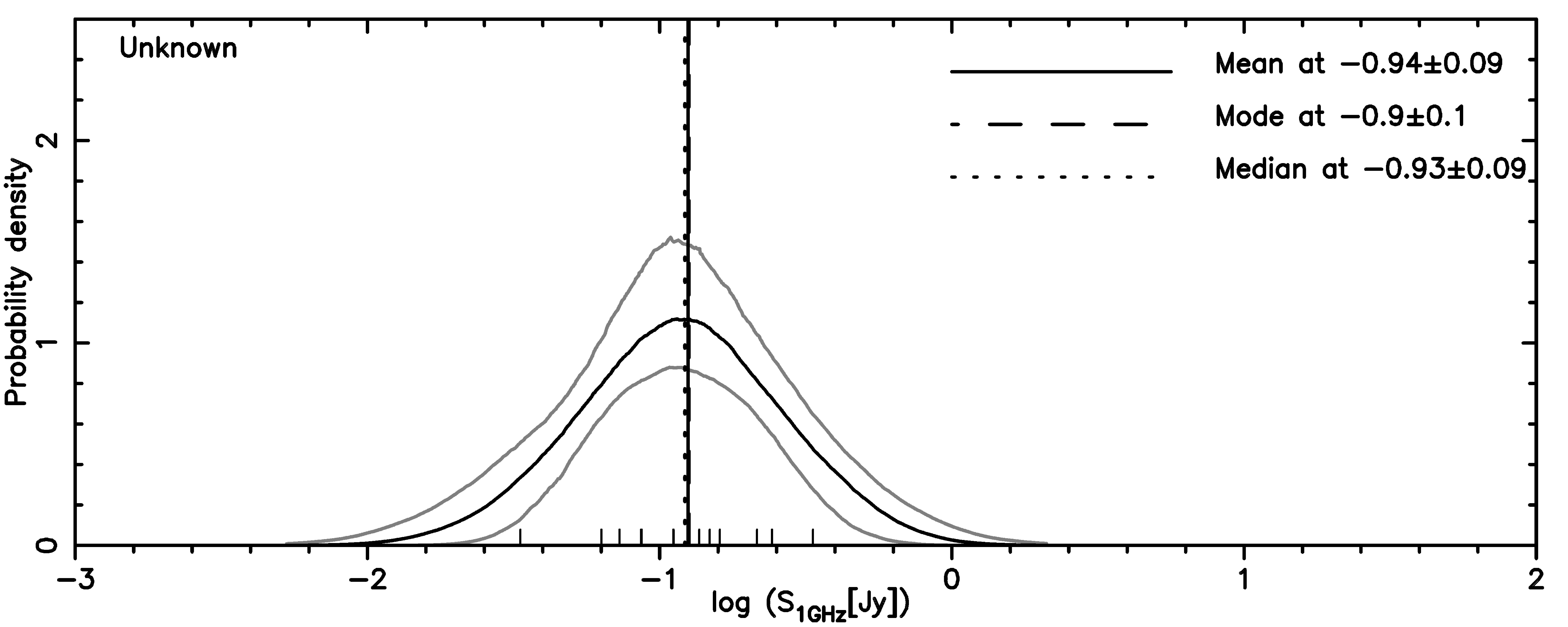}}\\
  \caption{ Smoothed density distribution for 1~GHz flux density data (as in Figure~\ref{fig:fluxhist}), for sub-samples based on progenitor type (designated on each panel). The optimal smoothing bandwidths for the analysed data samples were found to be ${0.369}$, ${0.458}$ and ${0.265}$ for type~Ia, CC and unknown sub-samples, respectively. Unlike the distribution in Figure~\ref{fig:fluxhist} these ones were smoothed on log $S_\mathrm{1\,GHz}$ scale.}
  \label{dvsid2}
 \end{center}
\end{figure*}

{ Similar to the SNR spherical symmetry tests (Section~\ref{Sect_SS}), the independence of the flux density sub-samples presented in Figure~\ref{dvsid2} was tested with the {AD} two sample test. For the CC vs. type~Ia SNRs, the Anderson-Darling variable test value is $6.4$ which is significantly higher than the $3.857$ { \citep[critical value for $\alpha = 0.01$ level from;][]{10.2307/2335097} indicating} that that CC and type~Ia remnants belong to vastly separate populations of objects. Similar results are found for CC vs. Unknown SN type with even higher confidence since the value of Anderson-Darling variable is 7.2. This implies that SNRs with undetermined progenitor type are more likely to originate from type~Ia than CC progenitors. This is in agreement with the remaining case (type~Ia vs. Unknown) where the Anderson-Darling variable is $0.73$ which is well below $1.933$ {(critical value at $\alpha = 0.1$ level,} as also used in Table~\ref{table_ad_test} for testing ovality based sub-samples), indicating that the tested samples are likely sampled from the same underlying distribution. We also emphasise that the second youngest \citep[310~yr old;][]{2014MNRAS.440.3220B,2015ApJ...809..119H} SNR in the LMC -- MCSNR\,J0509-6731 -- is a well-known type Ia and a relatively weak radio emitter with $S_\mathrm{1\,GHz}$ of 97.4~mJy.}

\begin{figure*}[]
 \begin{center}
  \resizebox{2\columnwidth}{!}{\includegraphics{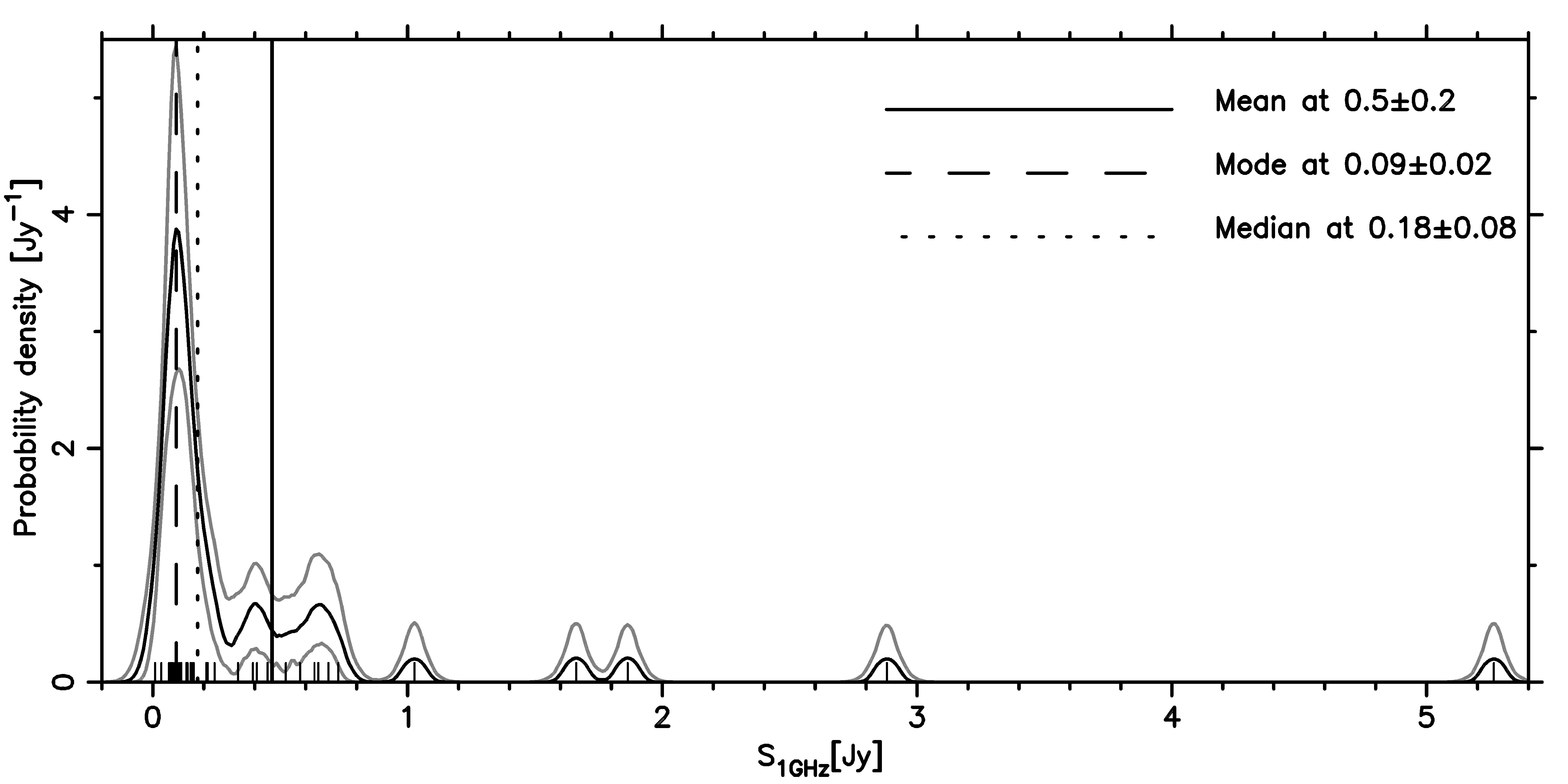}}
  \caption{The 1~GHz flux density smoothed density distribution (Section~\ref{kds}) for the sample of \spccountsn\ LMC SNRs with measured flux densities. The round-up values for mean, mode and median are also presented. The data points are marked with vertical dashes on the horizontal axis. Distribution parts with $S_\mathrm{1\,GHz}<0$ have no physical meaning and are plotted for completeness. A confidence interval of $75\%$ was used for estimating the uncertainties (grey lines). The optimal smoothing bandwidth for the analysed data sample was calculated to be $0.0491$~Jy.
  \label{fig:fluxhist}}
 \end{center}
\end{figure*}

The flux density distribution for the \spccountsn\ LMC SNRs that have estimated flux densities are shown in Figs.~\ref{dvsid2} and \ref{fig:fluxhist}\footnote{{SNR\,1987A is excluded from the analysis because of its separation from the rest of the sample and different physical characteristics.}}. The flux density variable is, by its own nature, heavily biased with sensitivity selection. The faint objects are not usually detected in surveys alongside the majority of the objects but rather with specialised high sensitivity observations. This is reflected in the fact that only $\sim$70\% (40 out of \confcount) of known LMC SNRs have radio flux density measurements. When candidate SNRs are added, this number remain similar -- $\sim$67\% (49 out of \totalcount). The object DEM\,L71, which has the lowest flux value, is a well observed and studied object unlike much of the sample (see also Section~\ref{sec_radio_t_x}). Compared with other objects in the vicinity of the PDF mode (which is close to DEM\,L71), this object and its immediate neighbour [HP99]~460 appear to be rather well separated. This implies that the highest concentration of detected objects is very close to the sensitivity limits of the related observations and that surveys with sensitivities below 50~mJy should give a large number of new detections.

Although there are known LMC SNRs with 1~GHz flux densities less than 50~mJy, most of the sub 50~mJy sample is lacking reliable flux density estimates. This is because of confusion due to unassociated nearby emission, e.g., an \HII\ region or a nearby and strong background source. Also, some of the remnants are lacking flux density estimates because they are either too weak to be accurately measured or fall below the detection limit of the present generation of surveys. Future radio telescopes (such as Australia Square Kilometre Array Pathfinder, ASKAP) with higher sensitivity and resolution will be able to account for these remnants and provide a more complete sample.

\subsection{Radio to X-ray Flux Density Comparison}
 \label{sec_radio_t_x}
\begin{figure}[]
 \begin{center}
  \resizebox{1\columnwidth}{!}{\includegraphics{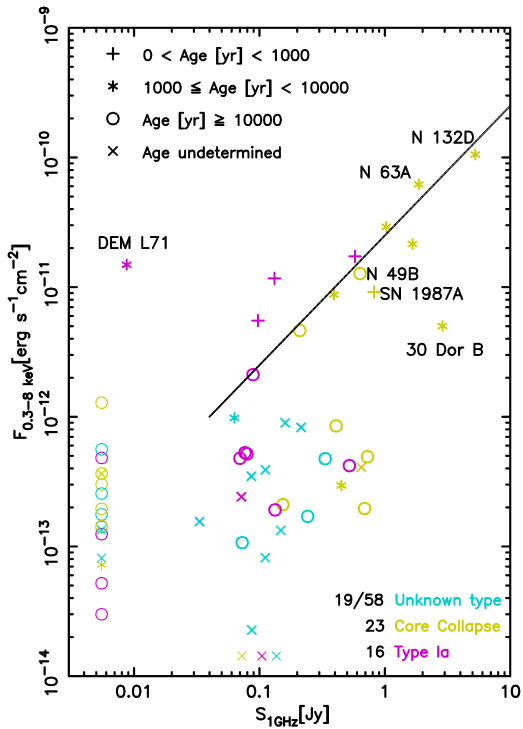}}
  \caption{Broad band X-ray flux in the 0.3-8~keV range versus radio flux density at 1~GHz for 58 LMC SNRs that have estimates for either one or both of these two frequencies. {Thin symbols} designate sources that only have estimates for X-ray flux or radio flux density but not both. The black solid line has the slope of 45 degrees and is plotted to indicate the 1--1 correspondence between the plotted variables for the objects with high values in both frequencies. The position of the { thin symbols} along the axis with the missing data is offset by 0.2~log scale from the faintest detection. Four symbols ($\times$, $\bigcirc$, $\ast$ and +) indicate SNR age based on Table~\ref{tab:lmcsnrs}. The pink colour symbols indicate type~Ia SN events while light green symbols indicate CC SN type. The light blue symbols represent unknown SN types.}
  \label{fig:fluxcomp}
 \end{center}
\end{figure}

We compared our estimated radio flux densities at 1~GHz (Table~\ref{tab:lmcsnrs}) and broad band X-ray flux in the 0.3-8 keV range from \citet{maggi16}. There are 58 known LMC SNRs with X-ray flux and/or radio flux density measurements (see Figure~\ref{fig:fluxcomp}). Only one confirmed LMC SNR (J0521-6542) has no measurements at either frequency.

At present, the SNR with the faintest measured radio flux density is DEM\,L71. As it is a well studied SNR (and one of the brightest X-ray SNRs in the LMC), there are many deep radio observations available for DEM\,L71 that make its detection and radio flux density measurement easier than for most of the LMC sample. The colour-coded {symbols} in Figure~\ref{fig:fluxcomp} at $\sim$6~mJy indicating radio non-detections are not representative of true flux density limits. While the RMS noise will vary significantly across the LMC, the average radio sensitivity limit for the non-detected sample is $>$10~mJy. Likewise, the three X-ray non-detections (ticks at $\sim10^{-14}$~ergs~s$^{-1}$~cm$^{-2}$) stem from a lack of proper coverage (e.g. observations with $\xmm$ or $\cxo$), and it is premature to conclude that these sources are intrinsically X-ray fainter.

The two flux-bright sources (N\,63A and N\,49B) with $F_\mathrm{0.3-8 keV} > 1 \times10^{-11}$~ergs~s$^{-1}$~cm$^{-2}$ { have ages of 3\,500 and 10\,000~yr old, respectively. The other outlier is 30~Dor~B (also known as N\,157B)}, which has a bright/young cometary PWN {and is 5\,000~yr old}. The total X-ray flux (SNR+PWN, $F_\mathrm{0.3-8~keV}>5 \times10^{-12}$~ergs~s$^{-1}$~cm$^{-2}$) is given, though the thermal component (SNR only) is only $\sim 2 \times10^{-13}$~ergs~s$^{-1}$~cm$^{-2}$ as for the bulk of the sample. The radio flux density for this object ($\sim 3$~Jy) also includes a significant fraction of PWN emission, but is harder to separate as both have a non-thermal spectrum.

It appears that SNRs younger than $10\,000$~yr with higher flux and flux density values in X-rays and radio, respectively, show some correlation between these values. While it is difficult to quantify, we point to the possible correlation between young type~Ia and CC SNRs, though the latter appears somewhat brighter (in both, X-rays and radio) than the former. {Also, the data points near the plotted 1-1 correspondence line in Fig. \ref{fig:fluxcomp} imply that type~Ia objects might be somewhat younger than CC SNRs. For the $43$ objects that have age data (Table 1), the results of the AD two sample test (see sections \ref{Sect_SS} and \ref{sec_flux_density}) do not justify the assumption that CC (20 objects) and Type~Ia (15 objects) age data sub-samples are independent. However, the distribution of objects with undetermined progenitor type (8 objects) is distinct from the distributions of type~Ia { (at $\alpha=0.05$ level)} and CC objects { (at $\alpha=0.1$ level}. This is consistent with Fig. \ref{fig:fluxcomp}. The objects with undetermined progenitor types, unknown ages and ages $>10\,000$ yr have, in general,   $F_\mathrm{0.3-8~keV}<\times10^{-12}$~ergs~s$^{-1}$~cm$^{-2}$. This implies that objects with unknown ages are thus likely to be older remnants that have lost the signatures of their progenitors over time.}

There is also a number of radio non-detections that fall into the X-ray flux range of $F_\mathrm{0.3-8~keV}~<~1\times10^{-12}$~ergs~s$^{-1}$~cm$^{-2}$, which confirms the need for more sensitive observations of the LMC SNR population in radio.

\subsection{Radio Surface Brightness Evolution}
 \label{SigmaD_evo}
Following the theoretical work done initially by \citet{1960SvA.....4..243S}, the important relation connecting the radio surface brightness $\Sigma_{\nu}$ of a particular SNR at frequency $\nu$ and its diameter $D$ can be written in general form as:
\begin{equation}
\Sigma_{ \nu }(D) = AD^{-\beta}.
\label{Sigma-D}
\end{equation}
\noindent The parameter $A$ depends on the properties of both the SN explosion and the ISM (e.g. SN energy of explosion, ejecta mass, the density of the ISM, the magnetic field strength, etc.), while $\beta$ is thought to be independent of these properties \citep{2005MNRAS.360...76A} but explicitly depends on the spectral index $\alpha$ of the integrated radio emission from an SNR \citep{1960SvA.....4..243S}. Parameters $A$ and $\beta$ are obtained by fitting the data from the sample of SNRs with known distances. Despite of all the criticism of the $\Sigma-D$ relation \citep[e.g.][]{2005MmSAI..76..534G}, it remains an important statistical tool in estimating distances to an SNR from its observed, distance independent, radio surface brightness.

The best fit correlation (Eq.~\ref{Sigma-D}) is a straight line in the $\log{D}-\log{\Sigma}$ plane. However, explicit care has to be taken to use the appropriate form of regression. As concluded in \citet{2013ApJS..204....4P}, the slopes of the empirical $\Sigma-D$ relation should be determined by using orthogonal regression because of its robust nature and equal statistical treatment of both variables. Both variables suffer from significant scatter and it is not statistically justified to treat one of them as independent. Nevertheless, this is the usual practice and regressions that minimise over offsets along one variable while the other one is considered as independent, such as $\Sigma = f(D)$ or $D = f (\Sigma)$, are very often used due to their simplicity. In this work we used the more robust orthogonal regression since it minimises over orthogonal distance of the data points from the fit line and, consequently, both variables have the same significance without one of them being considered the function of the other and measured with infinite accuracy. The orthogonal fitting performs well, regardless of the regression slope for data sets with severe scatters  in both coordinates. However, for the  extragalactic samples of SNRs, distance (and hence SNR diameters) can be obtained with higher accuracy than for the Galactic remnants since they can be approximated to reside at the distance of the host galaxy. Even in such a case, the depth effect and intrinsic effects can cause significant scatter in $D$ and orthogonal fitting should be preferred.

We can see from extragalactic SNR samples that the intrinsic scatter still dominates the $\Sigma-D$ relation and errors in distance determination are not crucial, for both Galactic and extragalactic SNRs. Here, the intrinsic scattering originates from the modelling of diverse phenomena with just two parameters ($\Sigma$ and $D$) and not taking into account individual characteristics of SNRs such as different SN explosion energies, densities of ISM into which they expand, evolutionary stages, etc. Also, data sets made up of extragalactic SNRs do not suffer from the Malmquist bias because all SNRs are at the same distance, while they still suffer from other selection effects caused by limitations in sensitivity and resolution \citep{2005A&A...435..437U}.

\begin{figure}[]
\epsscale{1.15}
\plotone{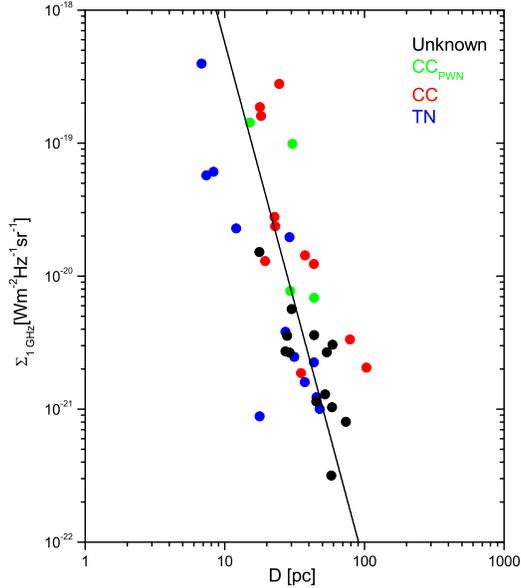}
\caption{Surface brightness versus diameter $\Sigma-D$ relation at 1~GHz for LMC SNRs. The solid black line represents the best orthogonal fit ($\beta = 3.92 \pm 1.20$). {SNR\,1987A, the smallest and brightest SNR in the LMC, is excluded from the graph because of its separation from the rest of the sample and different physical characteristics.}
\label{lmc-orth}}
\end{figure}

\begin{figure}[]
 \begin{center}
\resizebox{.9\columnwidth}{!}{\includegraphics{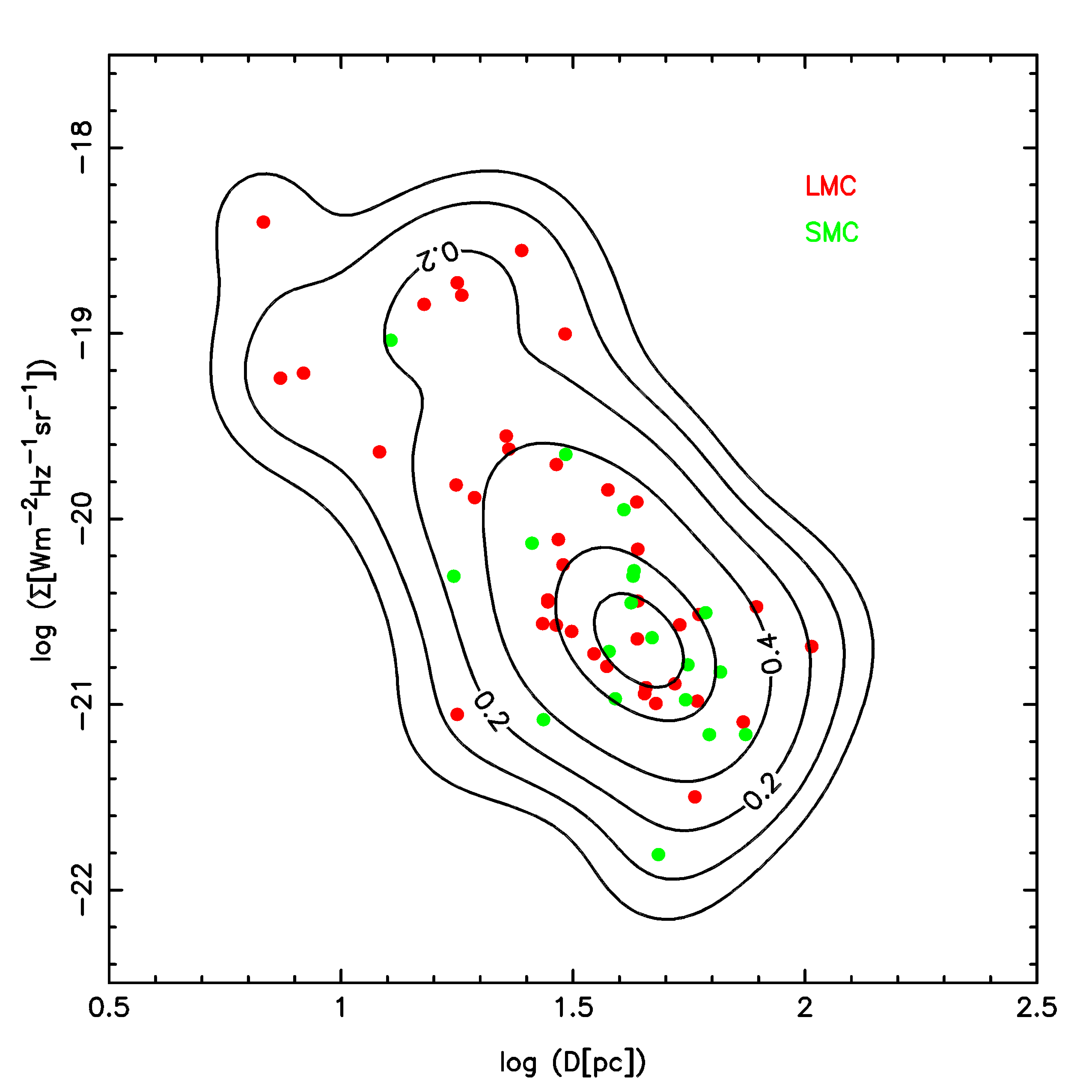}}
   \caption{Smoothed density distribution for the sample of LMC and SMC SNRs at 1~GHz, containing 59 SNRs. Red dots represent LMC SNRs while green dots represent SNRs from SMC. Contour levels are at ${0.05, 0.1, 0.2, 0.4, 0.8}$ and  ${1.0}$. The procedure for density smoothing is described in Section~\ref{kds_2d} with the relevant parameters given in Section~\ref{SigmaD_evo}. {As the smallest and brightest object of the sample, SNR\,1987A was not considered in this analysis.}
\label{lmc-track}}
 \end{center}
\end{figure}

\begin{figure*}[]
 \begin{center}
  \resizebox{1.6\columnwidth}{!}{\includegraphics[trim = 0 0 0 0]{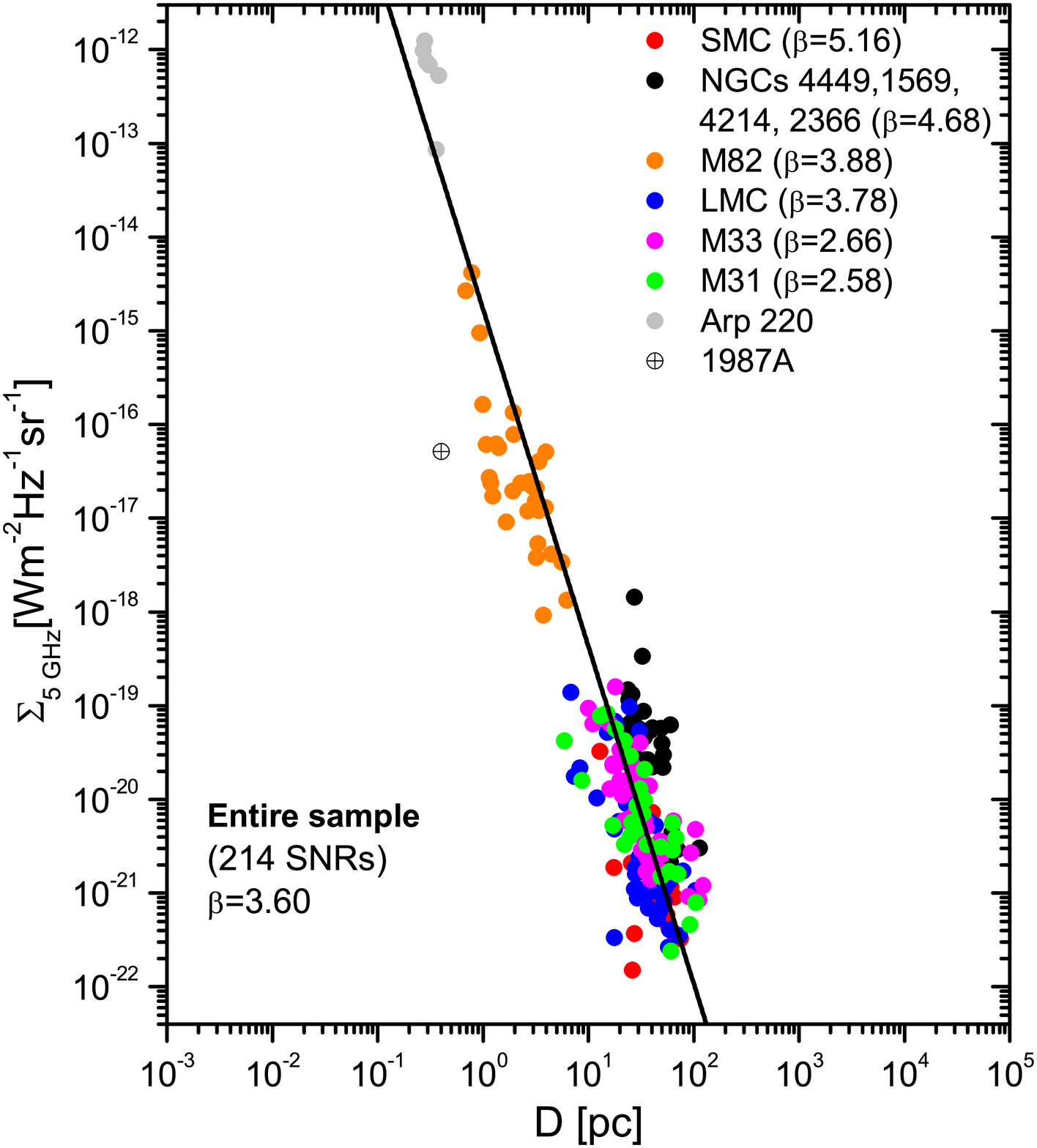}}
  \caption{The $\Sigma-D$ graph of LMC SNRs alongside other 9 nearby galaxies. This composite sample contains 214 SNRs. Orthogonal fitting has been applied to the 5~GHz data for the galaxies. {The youngest known SNR in the LMC, 1987A, is also shown (circled plus)
  but it is not included in the $\Sigma-D$ fit, being in the early free expansion phase of evolution.} {The solid black line
  represents the best orthogonal fit to the data ($\beta = 3.60 \pm 0.15$).}
  \label{entire215}}
 \end{center}
\end{figure*}

For the 40 LMC SNRs with measured flux densities, we estimated the radio surface brightness via:
\begin{equation}
  \Sigma_{ \nu } = 5.418 \times 10^{-16} \frac{S_{\nu}}{\theta^{2}},
 \label{calc}
\end{equation}
\noindent where $S_{\nu}$ is the integrated flux density in Jansky (Jy) and $\theta$ is the diameter in arc-seconds. Our data sample consisting of SNRs from the LMC is displayed in Figure~\ref{lmc-orth}, where there appears to be a correlation between $\Sigma$ and $D$, in confirmation of the theoretical models. Significant scattering is still present despite having more precise SNR diameters in comparison to the Galactic sample. This is expected due to an intrinsic scatter which dominates any errors arising from the measurement process. Furthermore, SNRs formed from a type~Ia explosion have lower surface-brightnesses than those arising from CC SN events. This is in agreement with the theoretical prediction that the larger ISM density produces greater synchrotron emission from an SNR, given in the form $\Sigma \propto \rho_{0}^{\eta} \propto n_{H}^{\eta}$ \citep{1986ApJ...301..308D,2004A&A...427..525B,2010A&A...509A..34B}, where $\rho_{0}$ and $n_{H}$ are the average ambient density and hydrogen number density, respectively. As the distance to the LMC of 50.0$\pm$1.3~kpc is determined to a very high accuracy \citep{2013Natur.495...76P}, the $\Sigma-D$ relation is in our case more important from a theoretical point of view, for comparison with Galactic and other extragalactic relations and also for comparing the environments into which SNRs expand.

Surface-brightnesses were calculated for the sample of 40 LMC SNRs at 5~GHz and orthogonal fitting was applied. The resulting fit shows a $\Sigma-D$ slope of $ \beta = 3.78 \pm 1.20$ which is very close to slope $\beta = 3.9 \pm 0.9$ obtained for the sample of 31 compact SNRs from the starburst galaxy M\,82, at the same frequency \citep{2010ApJ...719..950U}. The slope errors were calculated by using the bootstrap method. We have done $10^5$ bootstrap data re-samplings for each fit.

For the purpose of proving the universality of the $\Sigma-D$ law for SNRs, regardless of which samples are used, we constructed a composite sample containing available extragalactic SNR populations at the same frequency of 5~GHz (214 SNRs in total), which is shown in Figure~\ref{entire215}. Properties of the 10 extragalactic samples considered in this paper are listed in Table~\ref{table:galaxy-data}. The entire sample has a compact appearance for the presented variable range, with an overall slope of $\beta = 3.60 \pm 0.15$, very close to that of the LMC sample in this study. The resulting slope indicates that the observed extragalactic SNRs are mainly in the Sedov phase of evolution, as it was predicted by the values of slopes which are theoretically derived \citep{1986ApJ...301..308D,2004A&A...427..525B}.

\begin{table*}
  \begin{center}
 \caption{Parameters of available nearby galaxies data samples, using their 5~GHz radio fluxes}
 \vspace{3mm}
 \begin{tabular}{l c c c l}
\hline\hline
Name   & Number of SNRs & $\beta$    &$\Delta\beta$& Reference \\
\hline
Entire data sample & 214   &  3.60      & 0.15        & This work \\
LMC              & \p040   &  3.78      & 1.25        & This work \\
SMC              & \p019   &  5.16      & 5.76$^{1}$  & \citet{2005MNRAS.364..217F} \\
M\,82            & \p031   &  3.88      & 0.91        & \citet{2008MNRAS.391.1384F} \\
M\,31            & \p030   &  2.58      & 0.68        & Uro{\v s}evi{\'c} et al.~(2005) \\
\smallskip
M\,33            & \p051   &  2.66      & 0.85        & Uro{\v s}evi{\'c} et al.~(2005) \\
Arp\,220$^{2}$   &\p0\p06  & \nodata    & \nodata     & \citet{2011ApJ...740...95B} \\
NGC\,4449, NGC\,1569 &  \multirow{ 2}{*}{\p037}   & \multirow{ 2}{*}{4.68$^{3}$} & \multirow{ 2}{*}{2.58}  & \multirow{ 2}{*}{\citet{2009AJ....137.3869C}} \\
NGC\,4214, NGC\,2366 &     &       &          &   \\
\hline
\end{tabular}
\tablenotetext{\ }{ $^{1}$ A large bootstrap error for slope $\beta$ is expected for small samples like the SMC, containing only 19 SNRs.}
\tablenotetext{\ }{ $^{2}$ This sample contains only 6 SNRs and therefore calculating the $\Sigma-D$ slope does not make physical sense.}
\tablenotetext{\ }{ $^{3}$ The $\Sigma-D$ slope was calculated for the composite sample containing 37 SNRs in four galaxies
NGC\,4449, NGC\,1569, NGC\,4214 and NGC\,2366.}
 \label{table:galaxy-data}
 \end{center}
\end{table*}

\begin{figure*}[ht!]
 \begin{center}
  \resizebox{2\columnwidth}{!}{\includegraphics[trim = 0 0 0 0]{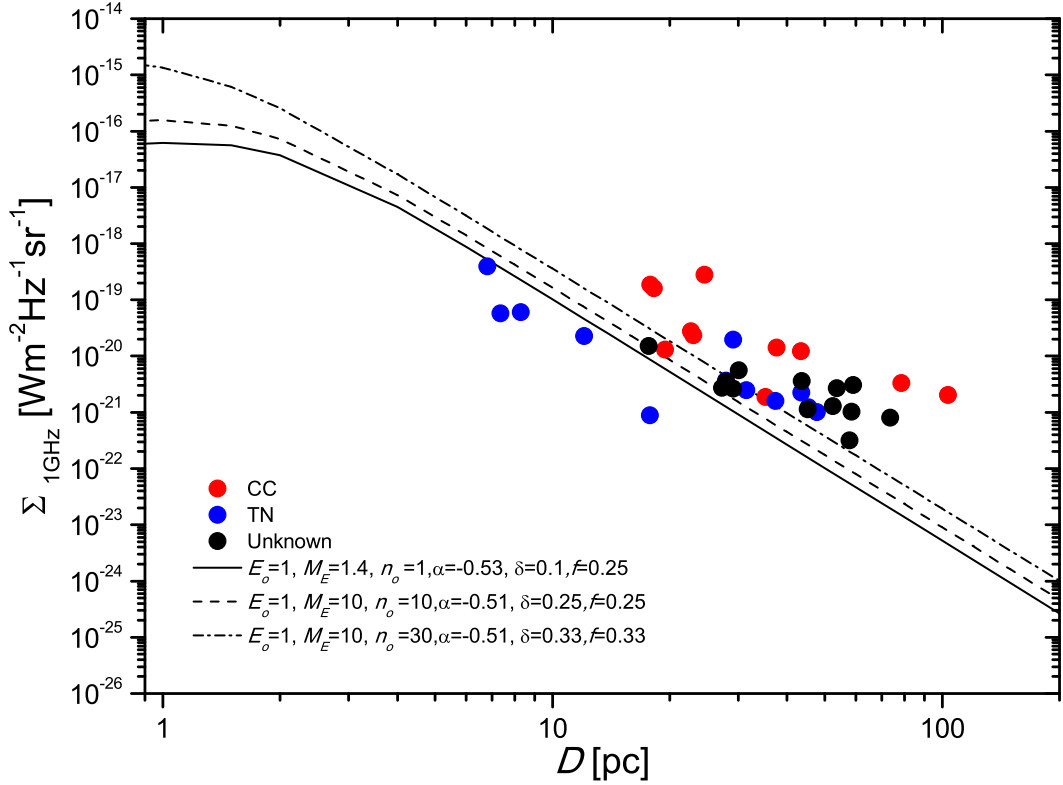}}
  \caption{An ``equipartition'' evolution model $ \epsilon_{\mathrm{CR}} \approx \frac{4}{\gamma +1}\epsilon_B  \sim \delta \rho  v^2, \ \ \ \epsilon_B = \frac{1}{8\pi}B^2 ,$ obtained by applying Eq.~\ref{eq22}, and assuming CRs proton to electron number 100:1. Spectral index is $\alpha$ and $f$ is the volume filling factor. The slope is in a range of $\beta = 4.265 - 4.295$.}
 \label{SDgraph}
 \end{center}
\end{figure*}

In Figure~\ref{SDgraph} we show theoretical ``equipartition'' models for radio evolution similar to the ones given by \citet{1981ApJ...245..912R} and \citet{2004A&A...427..525B}. The emissivity is defined as:
\begin{equation}
  \varepsilon _\nu = c_5 K (B \sin \Theta )^{(\gamma +1)/2}
  \Big(\frac{\nu}{2c_1}\Big)^{(1-\gamma )/2},
 \label{eq22}
\end{equation}
\noindent where $c_1, c_3$ and $c_5 = c_3 \Gamma (\frac{3\gamma -1}{12}) \Gamma (\frac{3\gamma +19}{12})/(\gamma +1)$ are from  \citet{1970ranp.book.....P}. The model assumes $K \propto  \epsilon_{\mathrm{CR}}$, $ \epsilon_{\mathrm{CR}} \approx \frac{4}{\gamma +1}\epsilon_B  \sim \delta \rho  v^2, \ \ \ \epsilon_B = \frac{1}{8\pi}B^2 $ \citep{2012ApJ...746...79A, 2013ApJ...777...31A}, where we applied Eq.~\ref{eq:ndlimits}, and assumed a CR proton-to-electron number of 100:1. The slope of the $\Sigma -D$ in the adiabatic phase is then given with approximately $\beta = \frac{-3\alpha +7}{2}$, where $\alpha$ is the spectral index. From Figure~\ref{SDgraph}, we also observe that, as expected, the type~Ia events could have overall lower surface brightness and smaller diameters then the known CC population. However, no definite conclusion can be drawn as there are large population of yet unknown SN types.

In addition to dependence on SNR size, the radio surface brightness of an SNR might also depend on the properties of the ambient medium into which the SNR expands. \citet{2005MNRAS.360...76A} argued that based on the progenitor type, SNRs are likely to be associated with an ambient medium of different density. Consequently, depending on the progenitor type (ambient density), SNRs should form a broad band corresponding to evolutionary tracks in $\Sigma-D$ plane. The dependence of $\Sigma-D$ evolution on ambient density is further explored in \citet{kostic15}. Under the assumption that SNRs expanding in an ambient medium of higher density have higher $\Sigma$, they argued in favour of the dependence of the slope of the $\Sigma-D$ relation on the fractal properties of the ISM density distribution in the areas crowded with molecular clouds. These clouds are denser than the surrounding ISM and therefore SNRs emit more synchrotron radiation while expanding within it. After reaching the edge of the cloud, SNR evolution continues outside the cloud and it emits less radiation due to the lower density of the ISM. This effect might result in different slopes of the $\Sigma-D$ relation.

To further analyse the density dependence, the probability density function of the data in the $\Sigma-D$ plane should be obtained. As described in \citet{2014MNRAS.440.2026V} this has many advantages compared to the standard fit parameters based analysis since all the information from the data sample is preserved and not just projected onto the parameters of the fit line. We calculated the $\Sigma-D$ PDF using kernel density smoothing described in Section~\ref{kds_2d}. To test if there are statistically significant data density features in the $\Sigma-D$ plane, in relation to $\Sigma(\rho)$ dependence, 2D kernel smoothing was performed on the $\Sigma-D$ LMC and SMC data sample (containing 40+19=59 SNRs; SN\,1987A not included in this sample). 100 smooth bootstrap re-samplings were applied in each step of the iterative procedure initialised with $h^{\log D}_0 = {0.129} $ and $h^{\log \Sigma}_0={0.389}$. The ranges over which the $BIMSE$ was calculated were $h^{\log D}=[0.01, 0.3]$ and $h^{\log \Sigma}=[0.1, 0.7]$, with steps of $0.01$ in both dimensions. We obtained optimal smoothing bandwidths at $h^{\log D} = {0.12}$ and $h^{\log \Sigma} = {0.34}$. The resulting $f_{h^{\log D}h^{\log \Sigma}}(\log D,\log \Sigma)$ in Figure~\ref{lmc-track} was calculated on a regular 100$\times$100 mapped on $\log D = (0.5, 2.5) $ and $\log \Sigma = (-22.5, -17.5)$ ranges.

From the given contour plot (Figure~\ref{lmc-track}) it is evident that no $\Sigma-D$ data groupings in parallel tracks are emergent{, or any other features possible indicative of SNRs} expanding out of molecular clouds or outgrowing the relevant density scale of the molecular clouds \citep[as analysed in][]{kostic15}. Further analysis and theoretical work is required on this matter in addition to thorough {surveys.}

%
%

\section{Supernova Remnants and Cosmic Rays}
 \label{Sect_SNRCR}

\citet{1934PNAS...20..259B} originally proposed that SNRs may be the primary site of {CR} acceleration. This initiated a debate as to the validity of this claim and to the extent of which SNRs accelerate CRs. \citet{2013Sci...339..807A} measured a gamma-ray spectrum which is better explained by a pion-decay origin rather than a leptonic origin in two galactic SNRs (IC\,443 and W\,44), providing direct evidence that CR protons are accelerated in SNRs. However, CR electrons are similarly important in this debate as they are accelerated in the SNR shock, as revealed by radio synchrotron emission and X-ray synchrotron filaments.

\begin{figure*}[]
 \begin{center}
  \resizebox{2\columnwidth}{!}{\includegraphics[trim = 0 0 0 0]{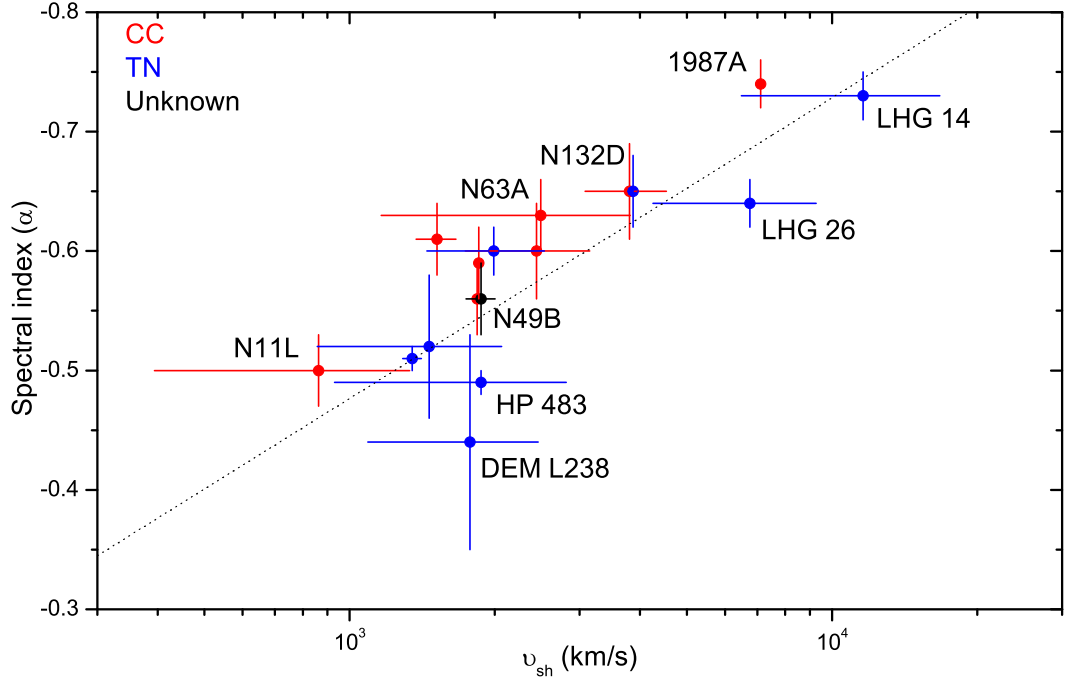}}
  \caption{Radio spectral index vs. shock velocity for {the sample of {17 young}} {LMC SNRs}. {These SNRs were selected as they exhibited a clear shell morphology, measured spectral index and age estimate.} The dotted line represents a fit to the entire sample resulting in ${\alpha={-0.73 - 0.25}}\;\text{log$_{10}$}(v_{\text{sh}}/10^4 \text{ km s}^{-1})$.}
\label{shockvelvssi}
\end{center}
\end{figure*}

The acceleration of {CRs} by strong shocks predicts a differential energy power-law spectrum $n(E)\text{d}E\,\propto\,E^{-\gamma}~\text{d}E$, with a spectral index of $\gamma=2$ \citep{1977SPhD...22..327K,1978MNRAS.182..147B,1978ApJ...221L..29B}. It was suggested that the acceleration of these {CR} may be due to repeatedly crossing the shock in a first-order Fermi process, gaining fractional energy $\Delta E/ E \propto u_s/c$ each crossover, where $u_s$ is the shock velocity. However, observationally, {CR} indices are found to significantly differ from such an index, e.g., the well known deviation found for relativistic shocks \citep{1989MNRAS.239..995K,1991MNRAS.249..551O,1993ApJ...409..327B,2001MNRAS.328..393A,2002PhRvD..66h3004K}. \citet{2011MNRAS.418.1208B} updated the theory of CR acceleration to explain deviations from $\gamma=2$ for shock velocities as low as 10\,000~km\,s$^{-1}$. For this, they plotted the SNR spectral index against its mean shock velocity, that is, the radius of the SNR divided by its age, in lieu of the momentary shock velocity. The reason for taking this approach is explained to be due to the spectrum being an addition of {CRs} that have been accelerated throughout the remnants lifetime. For their entire sample of SNe and SNRs, \citet{2011MNRAS.418.1208B} find a trend line of:
\begin{equation}\alpha=-0.7-0.3\text{ log$_{10}$}(v_{\text{sh}}/10^4 \text{ km s}^{-1})
\label{fig:bell}
\end{equation}
\noindent where $v_{\text{sh}}$ is the radius divided by the age of the SNRs. For a more reliable set of measurements, they also fit a trend line to only historic SNRs, resulting in:
\begin{equation}
 \alpha=-0.7-0.8\text{ log$_{10}$}(v_{\text{sh}}/10^4 \text{ km s}^{-1})
\end{equation}
\noindent Taking the same approach as in \citet{2011MNRAS.418.1208B}, we found:
\begin{equation}
{\alpha={-0.73b - 0.25}}\;\text{log$_{10}$}(v_{\text{sh}}/10^4 \text{ km s}^{-1})
\end{equation}
\noindent with fit quality $R^2$ = {0.80} ($R$ is correlation coefficient), {for the LMC subsample containing {17 young SNRs} with established shell morphology and readily available spectral index and age estimates. We selected only SNRs that are estimated to be younger than 10\,000 yr}. The best fit is represented as the dotted line in Figure~\ref{shockvelvssi}. This fit is close to the trend line, described by Equation~\ref{fig:bell} and found by \citet{2011MNRAS.418.1208B}, {for their entire sample of Galactic, historic and extragalactic SNRs and SNe}. Therefore, there appears to be a good alignment with the results between these two studies. Also, the LMC SNR sample confirms that there is a velocity dependent radio spectral index and that younger SNRs with higher shock velocities seem {to have steeper radio spectra}.

\begin{figure*}[]
 \begin{center}
  \resizebox{2\columnwidth}{!}{\includegraphics[trim = 0 0 0 0, clip]{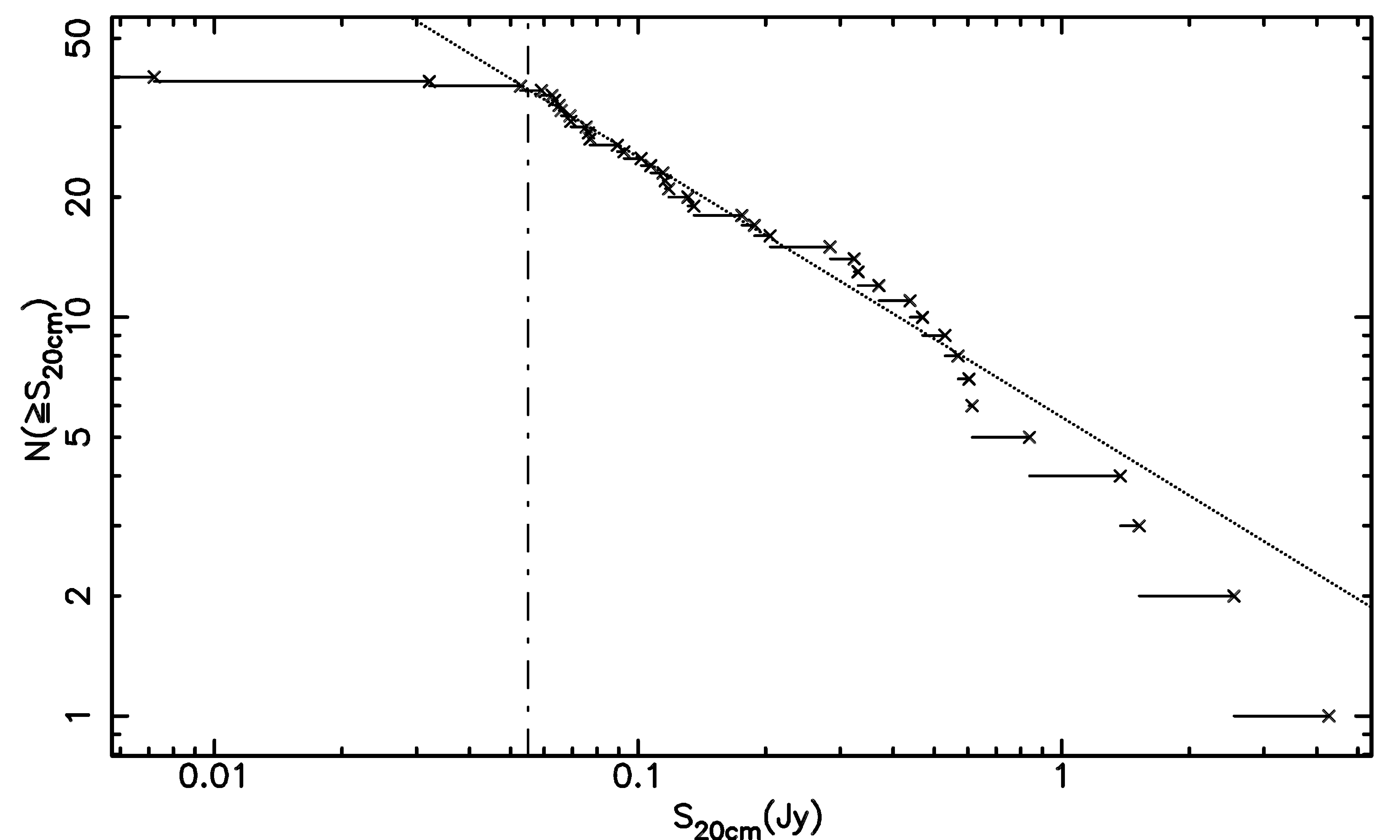}}
  \caption{ Luminosity function for a sample of 40 LMC SNRs at $\lambda=20$~cm, represented by a solid line. The dotted line represents the power-law fit to the component above the break at 55~mJy (dashed-dotted line). Crosses mark the data points obtained from the 1~GHz data scaled to 1.4~GHz (20~cm). }
  \label{lumfunc}
 \end{center}
\end{figure*}

The luminosity function can be used to test the hypothesis that SNRs are a primary site for CR acceleration. As SNRs only make up a fraction of the synchrotron radiation in the LMC, the electrons produced by SNRs must radiate well after the SNR has dissipated to account for the remaining majority of the synchrotron emission. To investigate this, a luminosity function needs to be created for the LMC SNR sample. To achieve this, we plot the cumulative luminosity function:
\begin{equation}
 N(s > S) \propto S^{-b}
\end{equation}
\noindent in Figure~\ref{lumfunc}, using integrated flux densities at 20~cm (deduced from the 1~GHz flux densities and the spectral indices shown in Table~\ref{tab:lmcsnrs}). As the LMC is located at a distance of $\sim50$~kpc, all objects that lie within the galaxy can be treated as approximately equidistant. Therefore, the luminosity and flux density plots will not differ and they are interchangeable in the text.

It can be seen that the spectrum breaks at $\sim55$~mJy which would imply that the sample is not complete. This was already known, as the sample used here contains only 40 (with SNR\,1987A also being excluded from the sample presented in Table \ref{tab:lmcsnrs}) out of the \confcount\ confirmed LMC SNRs, or 40 out of the \totalcount\ in the total sample including the candidates. Because of this, a simple power-law fit {on the whole S range} is not appropriate. Therefore, we fitted a power-law { (with the bootstrap procedure described in Section~\ref{sect:spcidxevo}) in the  $S > 55$~mJy interval (37 data points) resulting in }:
\begin{equation}
N({ \geqq S) =  (5.6\pm0.3) \cdot S^{-0.65\pm0.02} \text{~for~} 0.055 > S } \text{~Jy}
 \label{plflumfun}
\end{equation}
\noindent The total integrated flux density for the sample of 40 remnants was $\sim {17.5}$~Jy. In an attempt to account for the missing flux from the SNRs absent from the sample, we used the power-law-fit from Eq.~\ref{plflumfun} to integrate via:
\begin{equation}
 \int\limits_0^{{4.276}~\text{Jy}} {\frac{5.6}{(S)^{0.65}}}~dS = {26.7}~\text{Jy}
 \label{eqmaxflux}
\end{equation}
\noindent which is a better estimate of the total contribution of non-thermal emission from SNRs in the LMC. {As in \citet{1993A&AS...99..217D} we then use the equation}:
\begin{equation}
 \frac{\tau_e}{\tau_{\text{snr}}}= \frac{N_e(\text{LMC})}{N_e{\text{SNR}}} = \frac{S_{20}(\text{LMC})}{S_{20}(\text{SNR})} \frac{B_0(\text{SNR})^{1{-}\alpha(\text{SNR})}}{B_0(\text{LMC})^{1{-}\alpha(\text{LMC})}}
\end{equation}
\noindent where $\tau_e$ represents the residency time of electrons in the disk of the galaxy, $\tau_\text{snr}$ is the lifetime of the remnant and $S_{20}$ is the non-thermal flux density at 20~cm. { The equation has a similar form as in \citet{1993A&AS...99..217D} with the $\alpha$ sign being the only difference. \citet{1993A&AS...99..217D} used $S\propto\nu^{-\alpha}$ while here we adopted the $S\propto\nu^\alpha$ form.}

Along with the estimate of the total contribution of flux density from SNRs from Eq.~\ref{eqmaxflux}, we substituted the non-thermal flux density of the LMC at 20~cm given by \citet{2007MNRAS.382..543H}, who stated that the LMC exhibits a radio flux density of 426~Jy, where $\geq$20\% is thermal origin\footnote{{ We note here that radio flux from some SNRs might also have a significant thermal component \citep{2014Ap&SS.354..541U, 2013Ap&SS.346....3O} but to some degree it is compensated by using the lower limit for the fraction of the thermal radiation of the LMC.}}. The magnetic field strength of the LMC was taken to be{ $B_0 \approx 1~\mu$G} \citep{2005Sci...307.1610G} and a typical SNR would exhibit \mbox{$B_0$(SNR) = 40~$\mu$G} \citep{2012ApJ...746...79A,2013ApJ...777...31A} and ${\alpha = -0.52}$ {(Figure~\ref{spchist})}. The spectral index of the LMC is difficult to constrain and in the literature we found a range of values from  $-0.29,$ $-0.17$ \citep{2007MNRAS.382..543H}, $-0.56$ \citep{1989A&A...211..280K}, and --0.3 \citep{1991A&A...252..475H}. We note that the total contribution of SNRs to the non-thermal flux density of the LMC is much higher (\hilightwhite{$\sim${ 6}\%}) than that found by \citet{1998ApJS..117...89G} of M\,33 ($\approx2\%-3\%$).

Taking the value of $B_0(\text{LMC})$ to be $1~\mu$G, the spectral index of the galaxy does not affect the outcome, and results in:
\begin{equation}
 \frac{\tau_e}{\tau_{\text{snr}}}= {3\,603}.
\end{equation}
\noindent If the magnetic field of the LMC is varied, a wider range of values is possible. To investigate this, the  $\tau_e/\tau_{\text{snr}}$ values were measured for magnetic field strengths from $2-5~\mu$G and for spectral index values for the LMC of $\alpha= -0.17, -0.29, -0.30$ and --0.56, which can be found in Table~\ref{possvalues}.

\begin{table*}[]
\caption{Fraction of residency time of electrons to SNRs lifetime ($\tau_e/\tau_{\text{snr}}$) for different values of LMC magnetic field and LMC spectral index}
 \label{possvalues}
 \vspace{3mm}
 \centerline{\begin{tabular}{ccccc}
\hline\hline
$B_0~[\mu\mathrm{G}]$	&	$\alpha=-0.17$	&	$\alpha=-0.29$	&	$\alpha=-0.30$	&	$\alpha=-0.56$ \\
\hline
2 & 1601 & 1473 & 1463 & 1222  \\
3 & 996  & 873  & 864  & 649  \\
4 & 712  & 603  & 594  & 414  \\
5 & 548  & 452  & 445  & 293  \\ 	
\hline
 \end{tabular}}
\end{table*}

The measurement of $\alpha_{\text{LMC}}=-0.56$ was estimated from a much earlier paper; radio instruments have developed significantly since \citet{1989A&A...211..280K}. If measurements derived from this value of $\alpha$ are omitted, and the average SNR lifetime of $\tau_{\text{snr}}\approx10^4$ years is taken, this results in a residency time of ${4.4 - 16.0}$~million years. Although there is a large spread of possible values, this is in line with the residency time of CRs within the Galaxy of $15\pm1.6$~Myr \citep{1994ApJ...423..426L}, and with the value found by the same method for M\,33 of $7.5-10.1$~Myr \citep{1998ApJS..117...89G}. Consequently, these results are in agreement with the hypothesis that SNRs are a predominate source for CR acceleration in the LMC.

\section{Conclusions}

This work has provided an atlas of the \confcount\ currently confirmed LMC SNRs in addition to \candcount\ candidates from which 7 are shown here for the first time. Our statistical analysis of the available data has led to the following conclusions:

\begin{itemize}
\item While our LMC sample is the most complete sample of SNRs in any galaxy, the sensitivity detection level is sparse across the different electromagnetic domains. This leads to the conclusion that our sample is { under the influence} of various observational biases. Specifically, the sample with sizes $D>40$~pc still seems to be incomplete leaving room for a future detection of a mainly large (and older) LMC SNRs (Section~\ref{Sect_MFE}). 

\item We found evidence that the known \typeiacount\ type~Ia LMC SNRs are expanding in a somewhat lower density environment (mean=1.9$\times$10$^{21}$~atoms~cm$^{-2}$) while the \typecccount\ known CC type SNRs are expanding in a somewhat denser environment with a mean \HI\ column density of 2.4$\times$10$^{21}$~atoms~cm$^{-2}$. However, the uncertainties are on the larger side and these suggestions should be taken with caution (Section~\ref{Sect_SD}).

\item The LMC SNR population exhibits a mean diameter of { 39}~pc for the confirmed remnants and {41}~pc for the combined SNR and SNR candidates sample. These values are comparable to the mean diameters of the M\,31 (44~pc) and M\,33 (48~pc) {SNR} samples (Section~\ref{Sect_DSD}). 

\item The LMC SNRs spherical symmetry (or ovality -- as defined here) of the multi-frequency emission does not appear to correlate with the type of {SN} explosion. Also, no evidence for any type of correlation was found between the type of {SN} explosion, ovality or its known age (Section~\ref{Sect_SS}). 

\item The \ND\ relationship shows an exponent $a={0.96}$, which is close to \mbox{$a=1$}, even for the more complete sample given in this paper. Therefore, the earlier suggestion regarding randomised diameters readily mimicking such an exponent is probably the case here and not that the relation is indicative of the SNR population in the galaxy being in free-expansion (Section~\ref{sect:ND}).

\item The mean spectral index of the LMC SNRs ($\alpha$={--0.52}) is in line with the theoretically expected: $\alpha$=--0.5 (Section~\ref{Sect_RCSID}). However, our data show a clear flattening of the synchrotron spectral index as the remnant ages (Section~\ref{sect:spcidxevo}), at a rate of:
\[ \alpha = \hilightwhite{{0.18}}~\text{log}\left ( \frac{D}{pc} \right ) - \hilightwhite{{0.79}} \]

{\item Radio flux densities from CC and type~Ia remnants belong to the separate populations of objects. Namely, the CC type are distinctively brighter radio emitters than type~Ia remnants. Also, we conclude that the remaining population of presently unknown SNRs might be predominantly populated by type~Ia SN explosions (Section~\ref{sec_flux_density}).}

\item SNRs younger than $10\,000$~yr with higher flux and flux density values in X-rays and radio, respectively, show good correlation between these values. As expected, the young CC SNRs appear somewhat brighter in both X-ray and radio frequencies than the young type~Ia SNRs (Section~\ref{sec_radio_t_x}).

\item The 5~GHz \sigmaD\ relation for the LMC, with a slope of 3.78, is in line with the average for other nearby galaxies, where a slope of 3.60 was found (Section~\ref{SigmaD_evo}).

\item There is a clear relation between the shock velocity of an SNR and its synchrotron spectral index. {The trend for velocities of \textbf{17} LMC SNRs younger than 10\,000~yr shows}:
\[ \alpha={-0.73 - 0.25}\;\text{log}_{10}(v_{\text{sh}}/10^4 \text{ km s}^{-1})\]
\noindent which is in agreement with the slope found {for the sample of Galactic and extragalactic SNRs and SNe} (Section~\ref{Sect_SNRCR}).

\item The radio luminosity function has been used to find a CR electron residency time of \hilightwhite{${4.4 - 16.0}$~Myr}, which aligns well with the residency time of electrons in the Galaxy, and therefore, consistent with the suggestion that SNRs are the primary site for CR within a galaxy (Section~\ref{Sect_SNRCR}).

\end{itemize}

\acknowledgments We thank referee (Eric M. Schlegel) for his thorough review and highly appreciate his comments and suggestions, which significantly contributed to improving the quality of our paper. We also thank an anonymous statistical referee whose comments have greatly improved the data analysis and corresponding results of this work. {We thank Denis Leahy for discussions and comments.} The Australia Telescope Compact Array is part of the Australia Telescope National Facility, which is funded by the Commonwealth of Australia for operation as a National Facility managed by CSIRO. This paper includes archived data obtained through the Australia Telescope Online Archive (http://atoa.atnf.csiro.au). D.\,U., M.\,Z.\,P., B.\,V. and B.\,A. acknowledge financial support from the Ministry of Education, Science and Technological Development of the Republic of Serbia through the project \#176005 ``Emission nebulae: structure and evolution''. P.\,K. acknowledges support from the European Space Agency PRODEX Programme -- Contract Number 420090172. P.\,M. acknowledges support by the Centre National d'\'Etudes Spatiales (CNES). M.S.\ acknowledges support by the Deutsche Forschungsgemeinschaft through the research grant SA 2131/4-1 and the Heisenberg professor grant SA 2131/5-1. The MCELS was funded through the support of the Dean B. McLaughlin fund at the University of Michigan and through NSF grant 9540747.

{\it Facilities:} \facility{ATCA}, \facility{\xmm}.

\bibliographystyle{apj}
\bibliography{references}

\begin{thebibliography}{172}
\expandafter\ifx\csname natexlab\endcsname\relax\def\natexlab#1{#1}\fi

\bibitem[{{Achterberg} {et~al.}(2001){Achterberg}, {Gallant}, {Kirk}, \&
  {Guthmann}}]{2001MNRAS.328..393A}
{Achterberg}, A., {Gallant}, Y.~A., {Kirk}, J.~G., \& {Guthmann}, A.~W. 2001,
  \mnras, 328, 393

\bibitem[{{Ackermann} {et~al.}(2013){Ackermann}, {Ajello}, {Allafort},
  {Baldini}, {Ballet}, {Barbiellini}, {Baring}, {Bastieri}, {Bechtol},
  {Bellazzini}, {Blandford}, {Bloom}, {Bonamente}, {Borgland}, {Bottacini},
  {Brandt}, {Bregeon}, {Brigida}, {Bruel}, {Buehler}, {Busetto}, {Buson},
  {Caliandro}, {Cameron}, {Caraveo}, {Casandjian}, {Cecchi}, {{\c C}elik},
  {Charles}, {Chaty}, {Chaves}, {Chekhtman}, {Cheung}, {Chiang}, {Chiaro},
  {Cillis}, {Ciprini}, {Claus}, {Cohen-Tanugi}, {Cominsky}, {Conrad}, {Corbel},
  {Cutini}, {D'Ammando}, {de Angelis}, {de Palma}, {Dermer}, {do Couto e
  Silva}, {Drell}, {Drlica-Wagner}, {Falletti}, {Favuzzi}, {Ferrara},
  {Franckowiak}, {Fukazawa}, {Funk}, {Fusco}, {Gargano}, {Germani},
  {Giglietto}, {Giommi}, {Giordano}, {Giroletti}, {Glanzman}, {Godfrey},
  {Grenier}, {Grondin}, {Grove}, {Guiriec}, {Hadasch}, {Hanabata}, {Harding},
  {Hayashida}, {Hayashi}, {Hays}, {Hewitt}, {Hill}, {Hughes}, {Jackson},
  {Jogler}, {J{\'o}hannesson}, {Johnson}, {Kamae}, {Kataoka}, {Katsuta},
  {Kn{\"o}dlseder}, {Kuss}, {Lande}, {Larsson}, {Latronico}, {Lemoine-Goumard},
  {Longo}, {Loparco}, {Lovellette}, {Lubrano}, {Madejski}, {Massaro}, {Mayer},
  {Mazziotta}, {McEnery}, {Mehault}, {Michelson}, {Mignani}, {Mitthumsiri},
  {Mizuno}, {Moiseev}, {Monzani}, {Morselli}, {Moskalenko}, {Murgia},
  {Nakamori}, {Nemmen}, {Nuss}, {Ohno}, {Ohsugi}, {Omodei}, {Orienti},
  {Orlando}, {Ormes}, {Paneque}, {Perkins}, {Pesce-Rollins}, {Piron}, {Pivato},
  {Rain{\`o}}, {Rando}, {Razzano}, {Razzaque}, {Reimer}, {Reimer}, {Ritz},
  {Romoli}, {S{\'a}nchez-Conde}, {Schulz}, {Sgr{\`o}}, {Simeon}, {Siskind},
  {Smith}, {Spandre}, {Spinelli}, {Stecker}, {Strong}, {Suson}, {Tajima},
  {Takahashi}, {Takahashi}, {Tanaka}, {Thayer}, {Thayer}, {Thompson},
  {Thorsett}, {Tibaldo}, {Tibolla}, {Tinivella}, {Troja}, {Uchiyama}, {Usher},
  {Vandenbroucke}, {Vasileiou}, {Vianello}, {Vitale}, {Waite}, {Werner},
  {Winer}, {Wood}, {Wood}, {Yamazaki}, {Yang}, \&
  {Zimmer}}]{2013Sci...339..807A}
{Ackermann}, M., {Ajello}, M., {Allafort}, A., {et~al.} 2013, Science, 339, 807

\bibitem[{Anderson \& Darling(1954)}]{Anderson54}
Anderson, T.~W., \& Darling, D.~A. 1954, Journal of the American Statistical
  Association, 49, 765

\bibitem[{{Arbutina}(2005)}]{2005Arb}
{Arbutina}, B. 2005, Master's thesis, University of Belgrade, Serbia

\bibitem[{{Arbutina} \& {Uro{\v s}evi{\'c}}(2005)}]{2005MNRAS.360...76A}
{Arbutina}, B., \& {Uro{\v s}evi{\'c}}, D. 2005, \mnras, 360, 76

\bibitem[{{Arbutina} {et~al.}(2012){Arbutina}, {Uro{\v s}evi{\'c}},
  {Andjeli{\'c}}, {Pavlovi{\'c}}, \& {Vukoti{\'c}}}]{2012ApJ...746...79A}
{Arbutina}, B., {Uro{\v s}evi{\'c}}, D., {Andjeli{\'c}}, M.~M., {Pavlovi{\'c}},
  M.~Z., \& {Vukoti{\'c}}, B. 2012, \apj, 746, 79

\bibitem[{{Arbutina} {et~al.}(2013){Arbutina}, {Uro{\v s}evi{\'c}}, {Vu{\v
  c}eti{\'c}}, {Pavlovi{\'c}}, \& {Vukoti{\'c}}}]{2013ApJ...777...31A}
{Arbutina}, B., {Uro{\v s}evi{\'c}}, D., {Vu{\v c}eti{\'c}}, M.~M.,
  {Pavlovi{\'c}}, M.~Z., \& {Vukoti{\'c}}, B. 2013, \apj, 777, 31

\bibitem[{{Baade} \& {Zwicky}(1934)}]{1934PNAS...20..259B}
{Baade}, W., \& {Zwicky}, F. 1934, Proceedings of the National Academy of
  Science, 20, 259

\bibitem[{{Badenes} {et~al.}(2010){Badenes}, {Maoz}, \&
  {Draine}}]{2010MNRAS.407.1301B}
{Badenes}, C., {Maoz}, D., \& {Draine}, B.~T. 2010, \mnras, 407, 1301

\bibitem[{{Bandiera} \& {Petruk}(2010)}]{2010A&A...509A..34B}
{Bandiera}, R., \& {Petruk}, O. 2010, \aap, 509, A34

\bibitem[{{Baring} {et~al.}(1993){Baring}, {Ellison}, \&
  {Jones}}]{1993ApJ...409..327B}
{Baring}, M.~G., {Ellison}, D.~C., \& {Jones}, F.~C. 1993, \apj, 409, 327

\bibitem[{{Batejat} {et~al.}(2011){Batejat}, {Conway}, {Hurley}, {Parra},
  {Diamond}, {Lonsdale}, \& {Lonsdale}}]{2011ApJ...740...95B}
{Batejat}, F., {Conway}, J.~E., {Hurley}, R., {et~al.} 2011, \apj, 740, 95

\bibitem[{{Bell}(1978{\natexlab{a}})}]{1978MNRAS.182..147B}
{Bell}, A.~R. 1978{\natexlab{a}}, \mnras, 182, 147

\bibitem[{{Bell}(1978{\natexlab{b}})}]{1978MNRAS.182..443B}
---. 1978{\natexlab{b}}, \mnras, 182, 443

\bibitem[{{Bell} {et~al.}(2011){Bell}, {Schure}, \&
  {Reville}}]{2011MNRAS.418.1208B}
{Bell}, A.~R., {Schure}, K.~M., \& {Reville}, B. 2011, \mnras, 418, 1208

\bibitem[{{Berezhko} \& {V{\"o}lk}(2004)}]{2004A&A...427..525B}
{Berezhko}, E.~G., \& {V{\"o}lk}, H.~J. 2004, \aap, 427, 525

\bibitem[{{Berkhuijsen}(1987)}]{1987A&A...181..398B}
{Berkhuijsen}, E.~M. 1987, \aap, 181, 398

\bibitem[{{Blair} {et~al.}(2006){Blair}, {Ghavamian}, {Sankrit}, \&
  {Danforth}}]{2006ApJS..165..480B}
{Blair}, W.~P., {Ghavamian}, P., {Sankrit}, R., \& {Danforth}, C.~W. 2006,
  \apjs, 165, 480

\bibitem[{{Blandford} \& {Ostriker}(1978)}]{1978ApJ...221L..29B}
{Blandford}, R.~D., \& {Ostriker}, J.~P. 1978, \apjl, 221, L29

\bibitem[{{Boji{\v c}i{\'c}} {et~al.}(2007){Boji{\v c}i{\'c}}, {Filipovi{\'c}},
  {Parker}, {Payne}, {Jones}, {Reid}, {Kawamura}, \&
  {Fukui}}]{2007MNRAS.378.1237B}
{Boji{\v c}i{\'c}}, I.~S., {Filipovi{\'c}}, M.~D., {Parker}, Q.~A., {et~al.}
  2007, \mnras, 378, 1237

\bibitem[{{Borkowski} {et~al.}(2006{\natexlab{a}}){Borkowski}, {Hendrick}, \&
  {Reynolds}}]{2006ApJ...652.1259B}
{Borkowski}, K.~J., {Hendrick}, S.~P., \& {Reynolds}, S.~P. 2006{\natexlab{a}},
  \apj, 652, 1259

\bibitem[{{Borkowski} {et~al.}(2007){Borkowski}, {Hendrick}, \&
  {Reynolds}}]{2007ApJ...671L..45B}
---. 2007, \apjl, 671, L45

\bibitem[{{Borkowski} {et~al.}(2006{\natexlab{b}}){Borkowski}, {Williams},
  {Reynolds}, {Blair}, {Ghavamian}, {Sankrit}, {Hendrick}, {Long}, {Raymond},
  {Smith}, {Points}, \& {Winkler}}]{2006ApJ...642L.141B}
{Borkowski}, K.~J., {Williams}, B.~J., {Reynolds}, S.~P., {et~al.}
  2006{\natexlab{b}}, \apjl, 642, L141

\bibitem[{{Bozzetto} \& {Filipovi{\'c}}(2014)}]{2014Ap&SS.351..207B}
{Bozzetto}, L.~M., \& {Filipovi{\'c}}, M.~D. 2014, \apss, 351, 207

\bibitem[{{Bozzetto} {et~al.}(2010){Bozzetto}, {Filipovic}, {Crawford},
  {Bojicic}, {Payne}, {Medik}, {Wardlaw}, \& {de Horta}}]{2010SerAJ.181...43B}
{Bozzetto}, L.~M., {Filipovic}, M.~D., {Crawford}, E.~J., {et~al.} 2010,
  Serbian Astronomical Journal, 181, 43

\bibitem[{{Bozzetto} {et~al.}(2012{\natexlab{a}}){Bozzetto}, {Filipovic},
  {Crawford}, {De Horta}, \& {Stupar}}]{2012SerAJ.184...69B}
{Bozzetto}, L.~M., {Filipovic}, M.~D., {Crawford}, E.~J., {De Horta}, A.~Y., \&
  {Stupar}, M. 2012{\natexlab{a}}, Serbian Astronomical Journal, 184, 69

\bibitem[{{Bozzetto} {et~al.}(2012{\natexlab{b}}){Bozzetto}, {Filipovic},
  {Crawford}, {Payne}, {de Horta}, \& {Stupar}}]{2012RMxAA..48...41B}
{Bozzetto}, L.~M., {Filipovic}, M.~D., {Crawford}, E.~J., {et~al.}
  2012{\natexlab{b}}, \rmxaa, 48, 41

\bibitem[{{Bozzetto} {et~al.}(2015){Bozzetto}, {Filipovic}, {Haberl}, {Sasaki},
  {Kavanagh}, {Maggi}, {Urosevic}, \& {Sturm}}]{2015PKAS...30..149B}
{Bozzetto}, L.~M., {Filipovic}, M.~D., {Haberl}, F., {et~al.} 2015, Publication
  of Korean Astronomical Society, 30, 149

\bibitem[{{Bozzetto} {et~al.}(2012{\natexlab{c}}){Bozzetto}, {Filipovic},
  {Urosevic}, \& {Crawford}}]{2012SerAJ.185...25B}
{Bozzetto}, L.~M., {Filipovic}, M.~D., {Urosevic}, D., \& {Crawford}, E.~J.
  2012{\natexlab{c}}, Serbian Astronomical Journal, 185, 25

\bibitem[{{Bozzetto} {et~al.}(2014{\natexlab{a}}){Bozzetto}, {Filipovi{\'c}},
  {Uro{\v s}evi{\'c}}, {Kothes}, \& {Crawford}}]{2014MNRAS.440.3220B}
{Bozzetto}, L.~M., {Filipovi{\'c}}, M.~D., {Uro{\v s}evi{\'c}}, D., {Kothes},
  R., \& {Crawford}, E.~J. 2014{\natexlab{a}}, \mnras, 440, 3220

\bibitem[{{Bozzetto} {et~al.}(2012{\natexlab{d}}){Bozzetto}, {Filipovi{\'c}},
  {Crawford}, {Haberl}, {Sasaki}, {Uro{\v s}evi{\'c}}, {Pietsch}, {Payne}, {de
  Horta}, {Stupar}, {Tothill}, {Dickel}, {Chu}, \&
  {Gruendl}}]{2012MNRAS.420.2588B}
{Bozzetto}, L.~M., {Filipovi{\'c}}, M.~D., {Crawford}, E.~J., {et~al.}
  2012{\natexlab{d}}, \mnras, 420, 2588

\bibitem[{{Bozzetto} {et~al.}(2013){Bozzetto}, {Filipovi{\'c}}, {Crawford},
  {Sasaki}, {Maggi}, {Haberl}, {Uro{\v s}evi{\'c}}, {Payne}, {De Horta},
  {Stupar}, {Gruendl}, \& {Dickel}}]{2013MNRAS.432.2177B}
---. 2013, \mnras, 432, 2177

\bibitem[{{Bozzetto} {et~al.}(2014{\natexlab{b}}){Bozzetto}, {Kavanagh},
  {Maggi}, {Filipovi{\'c}}, {Stupar}, {Parker}, {Reid}, {Sasaki}, {Haberl},
  {Uro{\v s}evi{\'c}}, {Dickel}, {Sturm}, {Williams}, {Ehle}, {Gruendl}, {Chu},
  {Points}, \& {Crawford}}]{2014MNRAS.439.1110B}
{Bozzetto}, L.~M., {Kavanagh}, P.~J., {Maggi}, P., {et~al.} 2014{\natexlab{b}},
  \mnras, 439, 1110

\bibitem[{{Brantseg} {et~al.}(2014){Brantseg}, {McEntaffer}, {Bozzetto},
  {Filipovic}, \& {Grieves}}]{2014ApJ...780...50B}
{Brantseg}, T., {McEntaffer}, R.~L., {Bozzetto}, L.~M., {Filipovic}, M., \&
  {Grieves}, N. 2014, \apj, 780, 50

\bibitem[{{Cajko} {et~al.}(2009){Cajko}, {Crawford}, \&
  {Filipovic}}]{2009SerAJ.179...55C}
{Cajko}, K.~O., {Crawford}, E.~J., \& {Filipovic}, M.~D. 2009, Serbian
  Astronomical Journal, 179, 55

\bibitem[{{Callingham} {et~al.}(2016){Callingham}, {Gaensler}, {Zanardo},
  {Staveley-Smith}, {Hancock}, {Hurley-Walker}, {Bell}, {Dwarakanath},
  {Franzen}, {Hindson}, {Johnston-Hollitt}, {Kapi{\'n}ska}, {For}, {Lenc},
  {McKinley}, {Morgan}, {Offringa}, {Procopio}, {Wayth}, {Wu}, \&
  {Zheng}}]{2016MNRAS.462..290C}
{Callingham}, J.~R., {Gaensler}, B.~M., {Zanardo}, G., {et~al.} 2016, \mnras,
  462, 290

\bibitem[{{Chomiuk} \& {Wilcots}(2009)}]{2009AJ....137.3869C}
{Chomiuk}, L., \& {Wilcots}, E.~M. 2009, \aj, 137, 3869

\bibitem[{{Chu} \& {Kennicutt}(1988)}]{1988AJ.....96.1874C}
{Chu}, Y.-H., \& {Kennicutt}, Jr., R.~C. 1988, \aj, 96, 1874

\bibitem[{{Chu} {et~al.}(2000){Chu}, {Kim}, {Points}, {Petre}, \&
  {Snowden}}]{2000AJ....119.2242C}
{Chu}, Y.-H., {Kim}, S., {Points}, S.~D., {Petre}, R., \& {Snowden}, S.~L.
  2000, \aj, 119, 2242

\bibitem[{{Clark} \& {Caswell}(1976)}]{1976MNRAS.174..267C}
{Clark}, D.~H., \& {Caswell}, J.~L. 1976, \mnras, 174, 267

\bibitem[{{Crawford} {et~al.}(2008){Crawford}, {Filipovic}, {de Horta},
  {Stootman}, \& {Payne}}]{2008SerAJ.177...61C}
{Crawford}, E.~J., {Filipovic}, M.~D., {de Horta}, A.~Y., {Stootman}, F.~H., \&
  {Payne}, J.~L. 2008, Serbian Astronomical Journal, 177, 61

\bibitem[{{Crawford} {et~al.}(2010){Crawford}, {Filipovi{\'c}}, {Haberl},
  {Pietsch}, {Payne}, \& {de Horta}}]{2010A&A...518A..35C}
{Crawford}, E.~J., {Filipovi{\'c}}, M.~D., {Haberl}, F., {et~al.} 2010, \aap,
  518, A35

\bibitem[{{Crawford} {et~al.}(2014){Crawford}, {Filipovi{\'c}}, {McEntaffer},
  {Brantseg}, {Heitritter}, {Roper}, {Haberl}, \& {Uro{\v
  s}evi{\'c}}}]{2014AJ....148...99C}
{Crawford}, E.~J., {Filipovi{\'c}}, M.~D., {McEntaffer}, R.~L., {et~al.} 2014,
  \aj, 148, 99

\bibitem[{{Davies} {et~al.}(1976){Davies}, {Elliott}, \&
  {Meaburn}}]{1976MmRAS..81...89D}
{Davies}, R.~D., {Elliott}, K.~H., \& {Meaburn}, J. 1976, \memras, 81, 89

\bibitem[{{de Horta} {et~al.}(2012){de Horta}, {Filipovi{\'c}}, {Bozzetto},
  {Maggi}, {Haberl}, {Crawford}, {Sasaki}, {Uro{\v s}evi{\'c}}, {Pietsch},
  {Gruendl}, {Dickel}, {Tothill}, {Chu}, {Payne}, \&
  {Collier}}]{2012A&A...540A..25D}
{de Horta}, A.~Y., {Filipovi{\'c}}, M.~D., {Bozzetto}, L.~M., {et~al.} 2012,
  \aap, 540, A25

\bibitem[{{De Horta} {et~al.}(2014){De Horta}, {Sommer}, {Filipovi{\'c}},
  {O'Brien}, {Bozzetto}, {Collier}, {Wong}, {Crawford}, {Tothill}, {Maggi}, \&
  {Haberl}}]{2014AJ....147..162D}
{De Horta}, A.~Y., {Sommer}, E.~R., {Filipovi{\'c}}, M.~D., {et~al.} 2014, \aj,
  147, 162

\bibitem[{{Desai} {et~al.}(2010){Desai}, {Chu}, {Gruendl}, {Dluger}, {Katz},
  {Wong}, {Chen}, {Looney}, {Hughes}, {Muller}, {Ott}, \&
  {Pineda}}]{2010AJ....140..584D}
{Desai}, K.~M., {Chu}, Y.-H., {Gruendl}, R.~A., {et~al.} 2010, \aj, 140, 584

\bibitem[{{Dickel} {et~al.}(2010){Dickel}, {Gruendl}, {McIntyre}, \&
  {Amy}}]{2010AJ....140.1511D}
{Dickel}, J.~R., {Gruendl}, R.~A., {McIntyre}, V.~J., \& {Amy}, S.~W. 2010,
  \aj, 140, 1511

\bibitem[{{Dopita} {et~al.}(2010){Dopita}, {Blair}, {Long}, {Mutchler},
  {Whitmore}, {Kuntz}, {Balick}, {Bond}, {Calzetti}, {Carollo}, {Disney},
  {Frogel}, {O'Connell}, {Hall}, {Holtzman}, {Kimble}, {MacKenty}, {McCarthy},
  {Paresce}, {Saha}, {Silk}, {Sirianni}, {Trauger}, {Walker}, {Windhorst}, \&
  {Young}}]{2010ApJ...710..964D}
{Dopita}, M.~A., {Blair}, W.~P., {Long}, K.~S., {et~al.} 2010, \apj, 710, 964

\bibitem[{{Duric} \& {Seaquist}(1986)}]{1986ApJ...301..308D}
{Duric}, N., \& {Seaquist}, E.~R. 1986, \apj, 301, 308

\bibitem[{{Duric} {et~al.}(1993){Duric}, {Viallefond}, {Goss}, \& {van der
  Hulst}}]{1993A&AS...99..217D}
{Duric}, N., {Viallefond}, F., {Goss}, W.~M., \& {van der Hulst}, J.~M. 1993,
  \aaps, 99, 217

\bibitem[{Efron \& Tibshirani(1993)}]{tEFR93a}
Efron, B., \& Tibshirani, R.~J. 1993, An Introduction to the Bootstrap (New
  York, NY: Chapman \& Hall)

\bibitem[{Faraway \& Jhun(1990)}]{Faraway:1990:BCB}
Faraway, J.~J., \& Jhun, M. 1990, Journal of the American Statistical
  Association, 85, 1119

\bibitem[{Feigelson \& Babu(2012)}]{Feigelsonbabu2012astrostatistics}
Feigelson, E., \& Babu, G. 2012, Modern Statistical Methods for Astronomy: With
  R Applications (Cambridge University Press)

\bibitem[{{Fenech} {et~al.}(2008){Fenech}, {Muxlow}, {Beswick}, {Pedlar}, \&
  {Argo}}]{2008MNRAS.391.1384F}
{Fenech}, D.~M., {Muxlow}, T.~W.~B., {Beswick}, R.~J., {Pedlar}, A., \& {Argo},
  M.~K. 2008, \mnras, 391, 1384

\bibitem[{{Ferri{\`e}re}(2001)}]{2001RvMP...73.1031F}
{Ferri{\`e}re}, K.~M. 2001, Reviews of Modern Physics, 73, 1031

\bibitem[{{Filipovi{\'c}} \& {Bozzetto}(2016)}]{2016arXiv160401458F}
{Filipovi{\'c}}, M.~D., \& {Bozzetto}, L.~M. 2016, ArXiv e-prints

\bibitem[{{Filipovic} {et~al.}(1998){Filipovic}, {Haynes}, {White}, \&
  {Jones}}]{1998A&AS..130..421F}
{Filipovic}, M.~D., {Haynes}, R.~F., {White}, G.~L., \& {Jones}, P.~A. 1998,
  \aaps, 130, 421

\bibitem[{{Filipovi{\'c}} {et~al.}(2001){Filipovi{\'c}}, {Jones}, \&
  {Aschenbach}}]{2001AIPC..565..267F}
{Filipovi{\'c}}, M.~D., {Jones}, P.~A., \& {Aschenbach}, B. 2001, in American
  Institute of Physics Conference Series, Vol. 565, Young Supernova Remnants,
  ed. S.~S. {Holt} \& U.~{Hwang}, 267--270

\bibitem[{{Filipovi{\'c}} {et~al.}(2005){Filipovi{\'c}}, {Payne}, {Reid},
  {Danforth}, {Staveley-Smith}, {Jones}, \& {White}}]{2005MNRAS.364..217F}
{Filipovi{\'c}}, M.~D., {Payne}, J.~L., {Reid}, W., {et~al.} 2005, \mnras, 364,
  217

\bibitem[{{Filipovi{\'c}} {et~al.}(2008){Filipovi{\'c}}, {Haberl}, {Winkler},
  {Pietsch}, {Payne}, {Crawford}, {de Horta}, {Stootman}, \&
  {Reaser}}]{2008A&A...485...63F}
{Filipovi{\'c}}, M.~D., {Haberl}, F., {Winkler}, P.~F., {et~al.} 2008, \aap,
  485, 63

\bibitem[{{Finke} \& {Dermer}(2012)}]{2012ApJ...751...65F}
{Finke}, J.~D., \& {Dermer}, C.~D. 2012, \apj, 751, 65

\bibitem[{{Gaensler} {et~al.}(2005){Gaensler}, {Haverkorn}, {Staveley-Smith},
  {Dickey}, {McClure-Griffiths}, {Dickel}, \& {Wolleben}}]{2005Sci...307.1610G}
{Gaensler}, B.~M., {Haverkorn}, M., {Staveley-Smith}, L., {et~al.} 2005,
  Science, 307, 1610

\bibitem[{{Gal-Yam} {et~al.}(2009){Gal-Yam}, {Mazzali}, {Ofek}, {Nugent},
  {Kulkarni}, {Kasliwal}, {Quimby}, {Filippenko}, {Cenko}, {Chornock},
  {Waldman}, {Kasen}, {Sullivan}, {Beshore}, {Drake}, {Thomas}, {Bloom},
  {Poznanski}, {Miller}, {Foley}, {Silverman}, {Arcavi}, {Ellis}, \&
  {Deng}}]{2009Natur.462..624G}
{Gal-Yam}, A., {Mazzali}, P., {Ofek}, E.~O., {et~al.} 2009, \nat, 462, 624

\bibitem[{{Galvin} \& {Filipovic}(2014)}]{2014SerAJ.189...15G}
{Galvin}, T.~J., \& {Filipovic}, M.~D. 2014, Serbian Astronomical Journal, 189,
  15

\bibitem[{{Galvin} {et~al.}(2012){Galvin}, {Filipovi{\'c}}, {Crawford}, {Wong},
  {Payne}, {De Horta}, {White}, {Tothill}, {Dra{\v s}kovi{\'c}}, {Pannuti},
  {Grimes}, {Cahall}, {Millar}, \& {Laine}}]{2012Ap&SS.340..133G}
{Galvin}, T.~J., {Filipovi{\'c}}, M.~D., {Crawford}, E.~J., {et~al.} 2012,
  \apss, 340, 133

\bibitem[{{Galvin} {et~al.}(2014){Galvin}, {Filipovi{\'c}}, {Tothill},
  {Crawford}, {O'Brien}, {Seymour}, {Pannuti}, {Kosakowski}, \&
  {Sharma}}]{2014Ap&SS.353..603G}
{Galvin}, T.~J., {Filipovi{\'c}}, M.~D., {Tothill}, N.~F.~H., {et~al.} 2014,
  \apss, 353, 603

\bibitem[{{Ghavamian} {et~al.}(2003){Ghavamian}, {Rakowski}, {Hughes}, \&
  {Williams}}]{2003ApJ...590..833G}
{Ghavamian}, P., {Rakowski}, C.~E., {Hughes}, J.~P., \& {Williams}, T.~B. 2003,
  \apj, 590, 833

\bibitem[{{Glushak}(1985)}]{1985SvAL...11..350G}
{Glushak}, A.~P. 1985, Soviet Astronomy Letters, 11, 350

\bibitem[{{Glushak}(1996)}]{1996A&AT...11..317G}
---. 1996, Astronomical and Astrophysical Transactions, 11, 317

\bibitem[{{Gordon} {et~al.}(1998){Gordon}, {Kirshner}, {Long}, {Blair},
  {Duric}, \& {Smith}}]{1998ApJS..117...89G}
{Gordon}, S.~M., {Kirshner}, R.~P., {Long}, K.~S., {et~al.} 1998, \apjs, 117,
  89

\bibitem[{{Green}(2005)}]{2005MmSAI..76..534G}
{Green}, D.~A. 2005, \memsai, 76, 534

\bibitem[{{Green}(2014)}]{2014BASI...42...47G}
---. 2014, Bulletin of the Astronomical Society of India, 42, 47

\bibitem[{{Grondin} {et~al.}(2012){Grondin}, {Sasaki}, {Haberl}, {Pietsch},
  {Crawford}, {Filipovi{\'c}}, {Bozzetto}, {Points}, \& {Smith}}]{grondin12}
{Grondin}, M.-H., {Sasaki}, M., {Haberl}, F., {et~al.} 2012, \aap, 539, A15

\bibitem[{{Haberl}(2014)}]{2014xru..confE...4H}
{Haberl}, F. 2014, in The X-ray Universe 2014, 4

\bibitem[{{Haberl} \& {Pietsch}(1999)}]{1999A&AS..139..277H}
{Haberl}, F., \& {Pietsch}, W. 1999, \aaps, 139, 277

\bibitem[{{Haberl} {et~al.}(2012){Haberl}, {Sturm}, {Ballet}, {Bomans},
  {Buckley}, {Coe}, {Corbet}, {Ehle}, {Filipovic}, {Gilfanov},
  {Hatzidimitriou}, {La Palombara}, {Mereghetti}, {Pietsch}, {Snowden}, \&
  {Tiengo}}]{2012A&A...545A.128H}
{Haberl}, F., {Sturm}, R., {Ballet}, J., {et~al.} 2012, \aap, 545, A128

\bibitem[{{Harris}(1962)}]{1962ApJ...135..661H}
{Harris}, D.~E. 1962, \apj, 135, 661

\bibitem[{{Haynes} {et~al.}(1991){Haynes}, {Klein}, {Wayte}, {Wielebinski},
  {Murray}, {Bajaja}, {Meinert}, {Buczilowski}, {Harnett}, {Hunt}, {Wark}, \&
  {Sciacca}}]{1991A&A...252..475H}
{Haynes}, R.~F., {Klein}, U., {Wayte}, S.~R., {et~al.} 1991, \aap, 252, 475

\bibitem[{{Hendrick} {et~al.}(2003){Hendrick}, {Borkowski}, \&
  {Reynolds}}]{2003ApJ...593..370H}
{Hendrick}, S.~P., {Borkowski}, K.~J., \& {Reynolds}, S.~P. 2003, \apj, 593,
  370

\bibitem[{{Hovey} {et~al.}(2015){Hovey}, {Hughes}, \&
  {Eriksen}}]{2015ApJ...809..119H}
{Hovey}, L., {Hughes}, J.~P., \& {Eriksen}, K. 2015, \apj, 809, 119

\bibitem[{{Hughes} {et~al.}(2007){Hughes}, {Staveley-Smith}, {Kim}, {Wolleben},
  \& {Filipovi{\'c}}}]{2007MNRAS.382..543H}
{Hughes}, A., {Staveley-Smith}, L., {Kim}, S., {Wolleben}, M., \&
  {Filipovi{\'c}}, M. 2007, \mnras, 382, 543

\bibitem[{{Hughes} {et~al.}(2006){Hughes}, {Rafelski}, {Warren}, {Rakowski},
  {Slane}, {Burrows}, \& {Nousek}}]{2006ApJ...645L.117H}
{Hughes}, J.~P., {Rafelski}, M., {Warren}, J.~S., {et~al.} 2006, \apjl, 645,
  L117

\bibitem[{{Hughes} {et~al.}(1995){Hughes}, {Hayashi}, {Helfand}, {Hwang},
  {Itoh}, {Kirshner}, {Koyama}, {Markert}, {Tsunemi}, \&
  {Woo}}]{1995ApJ...444L..81H}
{Hughes}, J.~P., {Hayashi}, I., {Helfand}, D., {et~al.} 1995, \apjl, 444, L81

\bibitem[{{Ingallinera} {et~al.}(2014){Ingallinera}, {Trigilio}, {Umana},
  {Leto}, {Agliozzo}, \& {Buemi}}]{2014MNRAS.445.4507I}
{Ingallinera}, A., {Trigilio}, C., {Umana}, G., {et~al.} 2014, \mnras, 445,
  4507

\bibitem[{{Jaskot} {et~al.}(2011){Jaskot}, {Strickland}, {Oey}, {Chu}, \&
  {Garc{\'{\i}}a-Segura}}]{2011ApJ...729...28J}
{Jaskot}, A.~E., {Strickland}, D.~K., {Oey}, M.~S., {Chu}, Y.-H., \&
  {Garc{\'{\i}}a-Segura}, G. 2011, \apj, 729, 28

\bibitem[{{Kavanagh} {et~al.}(2015{\natexlab{a}}){Kavanagh}, {Sasaki},
  {Bozzetto}, {Filipovi{\'c}}, {Points}, {Maggi}, \& {Haberl}}]{kavanagh15b}
{Kavanagh}, P.~J., {Sasaki}, M., {Bozzetto}, L.~M., {et~al.}
  2015{\natexlab{a}}, \aap, 573, A73

\bibitem[{{Kavanagh} {et~al.}(2015{\natexlab{b}}){Kavanagh}, {Sasaki},
  {Bozzetto}, {Points}, {Filipovi{\'c}}, {Maggi}, {Haberl}, \&
  {Crawford}}]{2015A&A...583A.121K}
---. 2015{\natexlab{b}}, \aap, 583, A121

\bibitem[{{Kavanagh} {et~al.}(2015{\natexlab{c}}){Kavanagh}, {Sasaki},
  {Whelan}, {Maggi}, {Haberl}, {Bozzetto}, {Filipovi{\'c}}, \&
  {Crawford}}]{kavanagh15}
{Kavanagh}, P.~J., {Sasaki}, M., {Whelan}, E.~T., {et~al.} 2015{\natexlab{c}},
  \aap, 579, A63

\bibitem[{{Kavanagh} {et~al.}(2013){Kavanagh}, {Sasaki}, {Points},
  {Filipovi{\'c}}, {Maggi}, {Bozzetto}, {Crawford}, {Haberl}, \&
  {Pietsch}}]{kavanagh13}
{Kavanagh}, P.~J., {Sasaki}, M., {Points}, S.~D., {et~al.} 2013, \aap, 549, A99

\bibitem[{{Kavanagh} {et~al.}(2016){Kavanagh}, {Sasaki}, {Bozzetto}, {Points},
  {Crawford}, {Dickel}, {Filipovi{\'c}}, {Haberl}, {Maggi}, \&
  {Whelan}}]{2016A&A...586A...4K}
{Kavanagh}, P.~J., {Sasaki}, M., {Bozzetto}, L.~M., {et~al.} 2016, \aap, 586,
  A4

\bibitem[{Kiefer(1953)}]{KieferFibonacciSearch}
Kiefer, J. 1953, Proceedings of the American Mathematical Society, 4, 502

\bibitem[{{Kim} {et~al.}(1998){Kim}, {Staveley-Smith}, {Dopita}, {Freeman},
  {Sault}, {Kesteven}, \& {McConnell}}]{1998ApJ...503..674K}
{Kim}, S., {Staveley-Smith}, L., {Dopita}, M.~A., {et~al.} 1998, \apj, 503, 674

\bibitem[{{Kirk} \& {Heavens}(1989)}]{1989MNRAS.239..995K}
{Kirk}, J.~G., \& {Heavens}, A.~F. 1989, \mnras, 239, 995

\bibitem[{{Klein} {et~al.}(1989){Klein}, {Wielebinski}, {Haynes}, \&
  {Malin}}]{1989A&A...211..280K}
{Klein}, U., {Wielebinski}, R., {Haynes}, R.~F., \& {Malin}, D.~F. 1989, \aap,
  211, 280

\bibitem[{{Klimek} {et~al.}(2010){Klimek}, {Points}, {Smith}, {Shelton}, \&
  {Williams}}]{2010ApJ...725.2281K}
{Klimek}, M.~D., {Points}, S.~D., {Smith}, R.~C., {Shelton}, R.~L., \&
  {Williams}, R. 2010, \apj, 725, 2281

\bibitem[{{Kobayakawa} {et~al.}(2002){Kobayakawa}, {Honda}, \&
  {Samura}}]{2002PhRvD..66h3004K}
{Kobayakawa}, K., {Honda}, Y.~S., \& {Samura}, T. 2002, \prd, 66, 083004

\bibitem[{{Kosti{\'c}} {et~al.}(2016){Kosti{\'c}}, {Vukoti{\'c}}, {Uro{\v
  s}evi{\'c}}, {Arbutina}, \& {Prodanovi{\'c}}}]{kostic15}
{Kosti{\'c}}, P., {Vukoti{\'c}}, B., {Uro{\v s}evi{\'c}}, D., {Arbutina}, B.,
  \& {Prodanovi{\'c}}, T. 2016, \mnras, 461, 1421

\bibitem[{{Krymskii}(1977)}]{1977SPhD...22..327K}
{Krymskii}, G.~F. 1977, Soviet Physics Doklady, 22, 327

\bibitem[{{Laki{\'c}evi{\'c}} {et~al.}(2015){Laki{\'c}evi{\'c}}, {van Loon},
  {Meixner}, {Gordon}, {Bot}, {Roman-Duval}, {Babler}, {Bolatto},
  {Engelbracht}, {Filipovi{\'c}}, {Hony}, {Indebetouw}, {Misselt}, {Montiel},
  {Okumura}, {Panuzzo}, {Patat}, {Sauvage}, {Seale}, {Sonneborn}, {Temim},
  {Uro{\v s}evi{\'c}}, \& {Zanardo}}]{2015ApJ...799...50L}
{Laki{\'c}evi{\'c}}, M., {van Loon}, J.~T., {Meixner}, M., {et~al.} 2015, \apj,
  799, 50

\bibitem[{{Lee} \& {Lee}(2014)}]{2014ApJ...786..130L}
{Lee}, J.~H., \& {Lee}, M.~G. 2014, \apj, 786, 130

\bibitem[{{Leonidaki} {et~al.}(2013){Leonidaki}, {Boumis}, \&
  {Zezas}}]{2013MNRAS.429..189L}
{Leonidaki}, I., {Boumis}, P., \& {Zezas}, A. 2013, \mnras, 429, 189

\bibitem[{{Long} {et~al.}(1990){Long}, {Blair}, {Kirshner}, \&
  {Winkler}}]{1990ApJS...72...61L}
{Long}, K.~S., {Blair}, W.~P., {Kirshner}, R.~P., \& {Winkler}, P.~F. 1990,
  \apjs, 72, 61

\bibitem[{{Long} {et~al.}(1981){Long}, {Helfand}, \&
  {Grabelsky}}]{1981ApJ...248..925L}
{Long}, K.~S., {Helfand}, D.~J., \& {Grabelsky}, D.~A. 1981, \apj, 248, 925

\bibitem[{{Long} {et~al.}(2010){Long}, {Blair}, {Winkler}, {Becker}, {Gaetz},
  {Ghavamian}, {Helfand}, {Hughes}, {Kirshner}, {Kuntz}, {McNeil}, {Pannuti},
  {Plucinsky}, {Saul}, {T{\"u}llmann}, \& {Williams}}]{2010ApJS..187..495L}
{Long}, K.~S., {Blair}, W.~P., {Winkler}, P.~F., {et~al.} 2010, \apjs, 187, 495

\bibitem[{{Lopez} {et~al.}(2011){Lopez}, {Ramirez-Ruiz}, {Huppenkothen},
  {Badenes}, \& {Pooley}}]{2011ApJ...732..114L}
{Lopez}, L.~A., {Ramirez-Ruiz}, E., {Huppenkothen}, D., {Badenes}, C., \&
  {Pooley}, D.~A. 2011, \apj, 732, 114

\bibitem[{{Lukasiak} {et~al.}(1994){Lukasiak}, {Ferrando}, {McDonald}, \&
  {Webber}}]{1994ApJ...423..426L}
{Lukasiak}, A., {Ferrando}, P., {McDonald}, F.~B., \& {Webber}, W.~R. 1994,
  \apj, 423, 426

\bibitem[{{Maggi} {et~al.}(2012){Maggi}, {Haberl}, {Bozzetto}, {Filipovi{\'c}},
  {Points}, {Chu}, {Sasaki}, {Pietsch}, {Gruendl}, {Dickel}, {Smith}, {Sturm},
  {Crawford}, \& {De Horta}}]{maggi12}
{Maggi}, P., {Haberl}, F., {Bozzetto}, L.~M., {et~al.} 2012, \aap, 546, A109

\bibitem[{{Maggi} {et~al.}(2014){Maggi}, {Haberl}, {Kavanagh}, {Points},
  {Dickel}, {Bozzetto}, {Sasaki}, {Chu}, {Gruendl}, {Filipovi{\'c}}, \&
  {Pietsch}}]{Maggi14}
{Maggi}, P., {Haberl}, F., {Kavanagh}, P.~J., {et~al.} 2014, \aap, 561, A76

\bibitem[{{Maggi} {et~al.}(2016){Maggi}, {Haberl}, {Kavanagh}, {Sasaki},
  {Bozzetto}, {Filipovi{\'c}}, {Vasilopoulos}, {Pietsch}, {Points}, {Chu},
  {Dickel}, {Ehle}, {Williams}, \& {Greiner}}]{maggi16}
---. 2016, \aap, 585, A162

\bibitem[{{Mathewson} \& {Clarke}(1973)}]{1973ApJ...180..725M}
{Mathewson}, D.~S., \& {Clarke}, J.~N. 1973, \apj, 180, 725

\bibitem[{{Mathewson} {et~al.}(1983){Mathewson}, {Ford}, {Dopita}, {Tuohy},
  {Long}, \& {Helfand}}]{1983ApJS...51..345M}
{Mathewson}, D.~S., {Ford}, V.~L., {Dopita}, M.~A., {et~al.} 1983, \apjs, 51,
  345

\bibitem[{{Mathewson} {et~al.}(1984){Mathewson}, {Ford}, {Dopita}, {Tuohy},
  {Mills}, \& {Turtle}}]{1984ApJS...55..189M}
---. 1984, \apjs, 55, 189

\bibitem[{{Mathewson} {et~al.}(1985){Mathewson}, {Ford}, {Tuohy}, {Mills},
  {Turtle}, \& {Helfand}}]{1985ApJS...58..197M}
{Mathewson}, D.~S., {Ford}, V.~L., {Tuohy}, I.~R., {et~al.} 1985, \apjs, 58,
  197

\bibitem[{{Mathewson} \& {Healey}(1964)}]{1964IAUS...20..283M}
{Mathewson}, D.~S., \& {Healey}, J.~R. 1964, in IAU Symposium, Vol.~20, The
  Galaxy and the Magellanic Clouds, ed. F.~J. {Kerr}, 283

\bibitem[{{Meixner} {et~al.}(2006){Meixner}, {Gordon}, {Indebetouw}, {Hora},
  {Whitney}, {Blum}, {Reach}, {Bernard}, {Meade}, {Babler}, {Engelbracht},
  {For}, {Misselt}, {Vijh}, {Leitherer}, {Cohen}, {Churchwell}, {Boulanger},
  {Frogel}, {Fukui}, {Gallagher}, {Gorjian}, {Harris}, {Kelly}, {Kawamura},
  {Kim}, {Latter}, {Madden}, {Markwick-Kemper}, {Mizuno}, {Mizuno}, {Mould},
  {Nota}, {Oey}, {Olsen}, {Onishi}, {Paladini}, {Panagia}, {Perez-Gonzalez},
  {Shibai}, {Sato}, {Smith}, {Staveley-Smith}, {Tielens}, {Ueta}, {van Dyk},
  {Volk}, {Werner}, \& {Zaritsky}}]{2006AJ....132.2268M}
{Meixner}, M., {Gordon}, K.~D., {Indebetouw}, R., {et~al.} 2006, \aj, 132, 2268

\bibitem[{{Millar} {et~al.}(2011){Millar}, {White}, {Filipovi{\'c}}, {Payne},
  {Crawford}, {Pannuti}, \& {Staggs}}]{2011Ap&SS.332..221M}
{Millar}, W.~C., {White}, G.~L., {Filipovi{\'c}}, M.~D., {et~al.} 2011, \apss,
  332, 221

\bibitem[{{Mills}(1983)}]{1983IAUS..101..551M}
{Mills}, B.~Y. 1983, in IAU Symposium, Vol. 101, Supernova Remnants and their
  X-ray Emission, ed. J.~{Danziger} \& P.~{Gorenstein}, 551--558

\bibitem[{{Mills} \& {Turtle}(1984)}]{1984IAUS..108..283M}
{Mills}, B.~Y., \& {Turtle}, A.~J. 1984, in IAU Symposium, Vol. 108, Structure
  and Evolution of the Magellanic Clouds, ed. S.~{van den Bergh} \& K.~S.~D.
  {de Boer}, 283--290

\bibitem[{{Mills} {et~al.}(1984){Mills}, {Turtle}, {Little}, \&
  {Durdin}}]{1984AuJPh..37..321M}
{Mills}, B.~Y., {Turtle}, A.~J., {Little}, A.~G., \& {Durdin}, J.~M. 1984,
  Australian Journal of Physics, 37, 321

\bibitem[{{Milne} {et~al.}(1980){Milne}, {Caswell}, \&
  {Haynes}}]{1980MNRAS.191..469M}
{Milne}, D.~K., {Caswell}, J.~L., \& {Haynes}, R.~F. 1980, \mnras, 191, 469

\bibitem[{{Ng} {et~al.}(2013){Ng}, {Zanardo}, {Potter}, {Staveley-Smith},
  {Gaensler}, {Manchester}, \& {Tzioumis}}]{2013ApJ...777..131N}
{Ng}, C.-Y., {Zanardo}, G., {Potter}, T.~M., {et~al.} 2013, \apj, 777, 131

\bibitem[{{Nishiuchi} {et~al.}(2001){Nishiuchi}, {Yokogawa}, {Koyama}, \&
  {Hughes}}]{2001PASJ...53...99N}
{Nishiuchi}, M., {Yokogawa}, J., {Koyama}, K., \& {Hughes}, J.~P. 2001, \pasj,
  53, 99

\bibitem[{{O'Brien} {et~al.}(2013){O'Brien}, {Filipovi{\'c}}, {Crawford},
  {Tothill}, {Collier}, {De Horta}, {Wong}, {Dra{\v s}kovi{\'c}}, {Payne},
  {Pannuti}, {Napier}, {Griffith}, {Staggs}, \& {Kotu{\v
  s}}}]{2013Ap&SS.347..159O}
{O'Brien}, A.~N., {Filipovi{\'c}}, M.~D., {Crawford}, E.~J., {et~al.} 2013,
  \apss, 347, 159

\bibitem[{{Oni{\'c}}(2013)}]{2013Ap&SS.346....3O}
{Oni{\'c}}, D. 2013, \apss, 346, 3

\bibitem[{{Ostrowski}(1991)}]{1991MNRAS.249..551O}
{Ostrowski}, M. 1991, \mnras, 249, 551

\bibitem[{{Pacholczyk}(1970)}]{1970ranp.book.....P}
{Pacholczyk}, A.~G. 1970, {Radio astrophysics. Nonthermal processes in galactic
  and extragalactic sources} (San Francisco: Freeman)

\bibitem[{{Pannuti} {et~al.}(2000){Pannuti}, {Duric}, {Lacey}, {Goss},
  {Hoopes}, {Walterbos}, \& {Magnor}}]{2000ApJ...544..780P}
{Pannuti}, T.~G., {Duric}, N., {Lacey}, C.~K., {et~al.} 2000, \apj, 544, 780

\bibitem[{{Pannuti} {et~al.}(2011){Pannuti}, {Schlegel}, {Filipovi{\'c}},
  {Payne}, {Petre}, {Harrus}, {Staggs}, \& {Lacey}}]{2011AJ....142...20P}
{Pannuti}, T.~G., {Schlegel}, E.~M., {Filipovi{\'c}}, M.~D., {et~al.} 2011,
  \aj, 142, 20

\bibitem[{{Pannuti} {et~al.}(2007){Pannuti}, {Schlegel}, \&
  {Lacey}}]{2007AJ....133.1361P}
{Pannuti}, T.~G., {Schlegel}, E.~M., \& {Lacey}, C.~K. 2007, \aj, 133, 1361

\bibitem[{{Park} {et~al.}(2012){Park}, {Hughes}, {Slane}, {Burrows}, {Lee}, \&
  {Mori}}]{2012ApJ...748..117P}
{Park}, S., {Hughes}, J.~P., {Slane}, P.~O., {et~al.} 2012, \apj, 748, 117

\bibitem[{{Park} {et~al.}(2003){Park}, {Hughes}, {Slane}, {Burrows}, {Warren},
  {Garmire}, \& {Nousek}}]{2003ApJ...592L..41P}
---. 2003, \apjl, 592, L41

\bibitem[{{Pavlovi{\'c}} {et~al.}(2013){Pavlovi{\'c}}, {Uro{\v s}evi{\'c}},
  {Vukoti{\'c}}, {Arbutina}, \& {G{\"o}ker}}]{2013ApJS..204....4P}
{Pavlovi{\'c}}, M.~Z., {Uro{\v s}evi{\'c}}, D., {Vukoti{\'c}}, B., {Arbutina},
  B., \& {G{\"o}ker}, {\"U}.~D. 2013, \apjs, 204, 4

\bibitem[{{Payne} {et~al.}(2008){Payne}, {White}, \&
  {Filipovi{\'c}}}]{2008MNRAS.383.1175P}
{Payne}, J.~L., {White}, G.~L., \& {Filipovi{\'c}}, M.~D. 2008, \mnras, 383,
  1175

\bibitem[{{Payne} {et~al.}(2007){Payne}, {White}, {Filipovi{\'c}}, \&
  {Pannuti}}]{2007MNRAS.376.1793P}
{Payne}, J.~L., {White}, G.~L., {Filipovi{\'c}}, M.~D., \& {Pannuti}, T.~G.
  2007, \mnras, 376, 1793

\bibitem[{Pettitt(1976)}]{10.2307/2335097}
Pettitt, A.~N. 1976, Biometrika, 63, 161

\bibitem[{{Pietrzy{\'n}ski} {et~al.}(2013){Pietrzy{\'n}ski}, {Graczyk},
  {Gieren}, {Thompson}, {Pilecki}, {Udalski}, {Soszy{\'n}ski}, {Koz{\l}owski},
  {Konorski}, {Suchomska}, {Bono}, {Moroni}, {Villanova}, {Nardetto},
  {Bresolin}, {Kudritzki}, {Storm}, {Gallenne}, {Smolec}, {Minniti}, {Kubiak},
  {Szyma{\'n}ski}, {Poleski}, {Wyrzykowski}, {Ulaczyk}, {Pietrukowicz},
  {G{\'o}rski}, \& {Karczmarek}}]{2013Natur.495...76P}
{Pietrzy{\'n}ski}, G., {Graczyk}, D., {Gieren}, W., {et~al.} 2013, \nat, 495,
  76

\bibitem[{{Reid} {et~al.}(2015){Reid}, {Stupar}, {Bozzetto}, {Parker}, \&
  {Filipovi{\'c}}}]{2015MNRAS.454..991R}
{Reid}, W.~A., {Stupar}, M., {Bozzetto}, L.~M., {Parker}, Q.~A., \&
  {Filipovi{\'c}}, M.~D. 2015, \mnras, 454, 991

\bibitem[{{Reynolds} \& {Chevalier}(1981)}]{1981ApJ...245..912R}
{Reynolds}, S.~P., \& {Chevalier}, R.~A. 1981, \apj, 245, 912

\bibitem[{{Rosado} {et~al.}(1993){Rosado}, {Laval}, {Le Coarer}, {Boulesteix},
  {Georgelin}, \& {Marcelin}}]{rosado93}
{Rosado}, M., {Laval}, A., {Le Coarer}, E., {et~al.} 1993, \aap, 272, 541

\bibitem[{{Sano} {et~al.}(2017){Sano}, {Yamane}, {Voisin}, {Fujii}, {Yoshiike},
  {Inaba}, {Tsuge}, {Babazaki}, {Mitsuishi}, {Yang}, {Aharonian}, {Rowell},
  {Filipovic}, {Mizuno}, {Tachihara}, {Kawamura}, {Onishi}, \&
  {Fukui}}]{2017arXiv170101962S}
{Sano}, H., {Yamane}, Y., {Voisin}, F., {et~al.} 2017, ArXiv e-prints

\bibitem[{{Sasaki} {et~al.}(2000){Sasaki}, {Haberl}, \&
  {Pietsch}}]{2000A&AS..143..391S}
{Sasaki}, M., {Haberl}, F., \& {Pietsch}, W. 2000, \aaps, 143, 391

\bibitem[{{Seok} {et~al.}(2013){Seok}, {Koo}, \& {Onaka}}]{2013ApJ...779..134S}
{Seok}, J.~Y., {Koo}, B.-C., \& {Onaka}, T. 2013, \apj, 779, 134

\bibitem[{{Seward} {et~al.}(2012){Seward}, {Charles}, {Foster}, {Dickel},
  {Romero}, {Edwards}, {Perry}, \& {Williams}}]{2012ApJ...759..123S}
{Seward}, F.~D., {Charles}, P.~A., {Foster}, D.~L., {et~al.} 2012, \apj, 759,
  123

\bibitem[{{Seward} {et~al.}(2006){Seward}, {Williams}, {Chu}, {Dickel},
  {Smith}, \& {Points}}]{2006ApJ...640..327S}
{Seward}, F.~D., {Williams}, R.~M., {Chu}, Y.-H., {et~al.} 2006, \apj, 640, 327

\bibitem[{{Seward} {et~al.}(2010){Seward}, {Williams}, {Chu}, {Gruendl}, \&
  {Dickel}}]{2010AJ....140..177S}
{Seward}, F.~D., {Williams}, R.~M., {Chu}, Y.-H., {Gruendl}, R.~A., \&
  {Dickel}, J.~R. 2010, \aj, 140, 177

\bibitem[{Sheather(2004)}]{Sheather04densityestimation}
Sheather, S.~J. 2004, Statistical Science, 588

\bibitem[{{Shklovskii}(1960)}]{1960SvA.....4..243S}
{Shklovskii}, I.~S. 1960, \sovast, 4, 243

\bibitem[{Silverman(1986)}]{silverman1986density}
Silverman, B. 1986, Density Estimation for Statistics and Data Analysis,
  Chapman \& Hall/CRC Monographs on Statistics \& Applied Probability (Taylor
  \& Francis)

\bibitem[{{Smith} {et~al.}(2000){Smith}, {Leiton}, \&
  {Pizarro}}]{2000ASPC..221...83S}
{Smith}, C., {Leiton}, R., \& {Pizarro}, S. 2000, in Astronomical Society of
  the Pacific Conference Series, Vol. 221, Stars, Gas and Dust in Galaxies:
  Exploring the Links, ed. D.~{Alloin}, K.~{Olsen}, \& G.~{Galaz}, 83

\bibitem[{{Smith} {et~al.}(1993){Smith}, {Kirshner}, {Blair}, {Long}, \&
  {Winkler}}]{1993ApJ...407..564S}
{Smith}, R.~C., {Kirshner}, R.~P., {Blair}, W.~P., {Long}, K.~S., \& {Winkler},
  P.~F. 1993, \apj, 407, 564

\bibitem[{{Spitzer}(1978)}]{1978ppim.book.....S}
{Spitzer}, L. 1978, {Physical processes in the interstellar medium} (New York
  Wiley-Interscience)

\bibitem[{{Staveley-Smith} {et~al.}(2005){Staveley-Smith}, {Manchester},
  {Gaensler}, {Kesteven}, {Tziournis}, {Bizunok}, \&
  {Wheaton}}]{2005coex.conf...89S}
{Staveley-Smith}, L., {Manchester}, R.~N., {Gaensler}, B.~M., {et~al.} 2005, in
  IAU Colloq. 192: Cosmic Explosions, On the 10th Anniversary of SN1993J, ed.
  J.-M. {Marcaide} \& K.~W. {Weiler}, 89

\bibitem[{{Stupar} {et~al.}(2005){Stupar}, {Filipovi{\'c}}, {Jones}, \&
  {Parker}}]{2005AdSpR..35.1047S}
{Stupar}, M., {Filipovi{\'c}}, M.~D., {Jones}, P.~A., \& {Parker}, Q.~A. 2005,
  Advances in Space Research, 35, 1047

\bibitem[{{Subramanian} \& {Subramaniam}(2010)}]{2010A&A...520A..24S}
{Subramanian}, S., \& {Subramaniam}, A. 2010, \aap, 520, A24

\bibitem[{{Turtle} \& {Mills}(1984)}]{1984PASAu...5..537T}
{Turtle}, A.~J., \& {Mills}, B.~Y. 1984, Proceedings of the Astronomical
  Society of Australia, 5, 537

\bibitem[{{Uro{\v s}evi{\'c}}(2014)}]{2014Ap&SS.354..541U}
{Uro{\v s}evi{\'c}}, D. 2014, \apss, 354, 541

\bibitem[{{Uro{\v s}evi{\'c}} {et~al.}(2005){Uro{\v s}evi{\'c}}, {Pannuti},
  {Duric}, \& {Theodorou}}]{2005A&A...435..437U}
{Uro{\v s}evi{\'c}}, D., {Pannuti}, T.~G., {Duric}, N., \& {Theodorou}, A.
  2005, \aap, 435, 437

\bibitem[{{Uro{\v s}evi{\'c}} {et~al.}(2010){Uro{\v s}evi{\'c}}, {Vukoti{\'c}},
  {Arbutina}, \& {Sarevska}}]{2010ApJ...719..950U}
{Uro{\v s}evi{\'c}}, D., {Vukoti{\'c}}, B., {Arbutina}, B., \& {Sarevska}, M.
  2010, \apj, 719, 950

\bibitem[{{van den Bergh} \& {Tammann}(1991)}]{1991ARA&A..29..363V}
{van den Bergh}, S., \& {Tammann}, G.~A. 1991, \araa, 29, 363

\bibitem[{{Vukoti{\'c}} {et~al.}(2014){Vukoti{\'c}}, {Jurkovi{\'c}}, {Uro{\v
  s}evi{\'c}}, \& {Arbutina}}]{2014MNRAS.440.2026V}
{Vukoti{\'c}}, B., {Jurkovi{\'c}}, M., {Uro{\v s}evi{\'c}}, D., \& {Arbutina},
  B. 2014, \mnras, 440, 2026

\bibitem[{{Warren} {et~al.}(2003){Warren}, {Hughes}, \&
  {Slane}}]{2003ApJ...583..260W}
{Warren}, J.~S., {Hughes}, J.~P., \& {Slane}, P.~O. 2003, \apj, 583, 260

\bibitem[{{Warth} {et~al.}(2014){Warth}, {Sasaki}, {Kavanagh}, {Filipovi{\'c}},
  {Points}, \& {Bozzetto}}]{warth14}
{Warth}, G., {Sasaki}, M., {Kavanagh}, P.~J., {et~al.} 2014, \aap, 567, A136

\bibitem[{Wasserman(2010)}]{Wasserman:2010:SCC:1965575}
Wasserman, L. 2010, All of Statistics: A Concise Course in Statistical
  Inference (Springer Publishing Company, Incorporated)

\bibitem[{{Westerlund} \& {Mathewson}(1966)}]{1966MNRAS.131..371W}
{Westerlund}, B.~E., \& {Mathewson}, D.~S. 1966, \mnras, 131, 371

\bibitem[{{Williams} \& {Chu}(2005)}]{2005ApJ...635.1077W}
{Williams}, R.~M., \& {Chu}, Y.-H. 2005, \apj, 635, 1077

\bibitem[{{Williams} {et~al.}(2005){Williams}, {Chu}, {Dickel}, {Gruendl},
  {Seward}, {Guerrero}, \& {Hobbs}}]{2005ApJ...628..704W}
{Williams}, R.~M., {Chu}, Y.-H., {Dickel}, J.~R., {et~al.} 2005, \apj, 628, 704

\bibitem[{{Williams} {et~al.}(2004){Williams}, {Chu}, {Dickel}, {Gruendl},
  {Shelton}, {Points}, \& {Smith}}]{2004ApJ...613..948W}
---. 2004, \apj, 613, 948

\bibitem[{{Williams} {et~al.}(1999){Williams}, {Chu}, {Dickel}, {Petre},
  {Smith}, \& {Tavarez}}]{1999ApJS..123..467W}
---. 1999, \apjs, 123, 467

\bibitem[{{Williams} {et~al.}(2006){Williams}, {Chu}, \&
  {Gruendl}}]{2006AJ....132.1877W}
{Williams}, R.~M., {Chu}, Y.-H., \& {Gruendl}, R. 2006, \aj, 132, 1877

\bibitem[{{Woltjer}(1972)}]{1972ARA&A..10..129W}
{Woltjer}, L. 1972, \araa, 10, 129

\bibitem[{{Zanardo} {et~al.}(2010){Zanardo}, {Staveley-Smith}, {Ball},
  {Gaensler}, {Kesteven}, {Manchester}, {Ng}, {Tzioumis}, \&
  {Potter}}]{2010ApJ...710.1515Z}
{Zanardo}, G., {Staveley-Smith}, L., {Ball}, L., {et~al.} 2010, \apj, 710, 1515

\end{thebibliography}

\end{document}